\documentclass[pdflatex,sn-mathphys-num]{sn-jnl}

\usepackage{graphicx}%
\usepackage{multirow}%
\usepackage{amsmath,amssymb,amsfonts}%
\usepackage{amsthm}%
\usepackage{mathrsfs}%
\usepackage[title]{appendix}%
\usepackage{xcolor}%
\usepackage{textcomp}%
\usepackage{manyfoot}%
\usepackage{booktabs}%
\usepackage{algorithm}%
\usepackage{algorithmicx}%
\usepackage{algpseudocode}%
\usepackage{listings}%

\usepackage{hyperref}
\usepackage{makecell}
\usepackage{pdflscape}
\usepackage{float}
\usepackage{caption}
\usepackage{adjustbox}

\theoremstyle{thmstyleone}%
\theoremstyle{thmstyletwo}%

\theoremstyle{thmstylethree}%

\begin{document}

\title[paper-title]{Residual neural networks to classify the high frequency emission in core-collapse supernova gravitational waves}


\author*[1]{\fnm{Manuel D.} \sur{Morales}}\email{manueld.morales@academicos.udg.mx}

\author[2]{\fnm{Javier M.} \sur{Antelis}}\email{mauricio.antelis@tec.mx}

\author[1]{\fnm{Claudia} \sur{Moreno}}\email{claudia.moreno@academicos.udg.mx}

\affil*[1]{\orgdiv{Departamento de Física}, \orgname{CUCEI, Universidad de Guadalajara}, \orgaddress{\city{Guadalajara}, \postcode{44430}, \state{Jalisco}, \country{México}}}

\affil[2]{\orgdiv{Tecnologico de Monterrey}, \orgname{Escuela de Ingeniería y Ciencias}, \orgaddress{\city{Monterrey}, \postcode{64849}, \state{Nuevo León}, \country{México}}}


\abstract{We present a new methodology to explore the morphology of the High Frequency Feature (HFF), i.e., the dominant, rising-frequency GW emission from a proto-neutron star in core-collapse supernovae (CCSNe). We used a residual neural network (ResNet50) to perform multi-class classification of image samples constructed from time–frequency Morlet wavelet scalograms. We defined a three-class problem by categorizing the HFF slope as Steep, Moderate, or Low, according to physically informed ranges. The ResNet50 model was optimized with phenomenological waveforms injected into real noise from the LIGO-Virgo O3b observing run and then tested with numerically simulated CCSN waveforms embedded in the same real noise. At galactic distances of 1 kpc and 5 kpc with H1 and L1 data and 1 kpc with V1 data, we obtained highly accurate results (test accuracies from 0.8933 to 0.9867), which show the feasibility of our methodology. For further distances, we observed declines in test accuracy until 0.8000 with H1 and L1 data at 10 kpc and until 0.5933 with V1 data at 10 kpc, which we attribute to limitations in the input datasets. Our methodology is sufficiently general to enable early-stage characterization of the HFF in real interferometric data.}

\keywords{gravitational waves, deep learning, residual neural networks, core-collapse supernovae, LIGO-Virgo detectors.}



\maketitle

\tableofcontents 

\section{Introduction}\label{sec:introduction}

Thanks to pioneering detections of gravitational-wave signals emitted by binary black holes~\cite{bA16} and binary neutron stars~\cite{bA17}, we entered an exciting new era of multi-messenger astronomy~\cite{bA17_2}. Moreover, the observational prospects now extend to other types of gravitational-wave (GW) signals, such as those emitted by core-collapse supernovae (CCSNe). These signals will be potentially detectable with the current generation of interferometric detectors~\cite{mS21}.

CCSNe are extremely violent astrophysical processes. They occur at the end of the lifetime of massive stars with initial masses of $9M_{\odot} \lesssim M \lesssim 25M_{\odot}$, and belonging to a metallicity range from metal-free stars to those with metallicity about the solar~\cite{kK06}. According to the basic overall picture of CCSNe~\cite{hB90,hJ12,eA22}, at the end of the life of massive stars with initial masses $M \geq 10M_{\odot}$, an iron core is formed, which in turn, after dissociation of its iron nuclei into alpha particles and free nucleons (due to collisions caused by ultra-energetic photons), undergoes a dynamical collapse. This collapse continues until the falling matter reaches supranuclear densities and a stiff proto-neutron star (PNS) is formed. In addition, as a result of the inner core bounce, a shock wave formed by the outer shells of the stellar envelope, with its own physical mechanism involved in its generation, revival, and propagation, is launched into the interstellar medium.

CCSNe emit GW signals because of the aspherical dynamics occurring in a strong gravity regime. From an inverse-problem perspective, these signals are expected to encode information about the physical processes driving the explosion, manifesting as observable features or signatures. This is pointed out in several numerical CCSNe simulations~\cite{sS08,bM13,tK16,dV23}, and also in studies about detection and characterization of CCSN GW signals~\cite{iH09,sS19,sS24,yY24,aM24}. Then, with this theoretical assumption at hand, the big challenge is to implement suitable algorithms to detect and estimate the parameters of these features.

Among the physical processes involved in CCSNe, those occurring in the proto-neutron star (PNS) contribute most to the emission of GW signals~\cite{kK06}. By simply exploring the time-frequency (TF) representation of simulated CCSN GW signals such as spectrograms, scalograms, etc., this emission can be visualized in many of the GW signal simulations (as those studied in \cite{mS21}). After core bounce, this emission exhibits a monotonically increasing frequency profile (approximately linear, first order), rising from about $100$ Hz to $1,000-2,000$ Hz. This emission usually is identified as the ``g-mode'' of the PNS, but it is not clear that it contains only information about the g-mode. Without a multimodal analysis, this emission cannot be unambiguously associated with a single \textit{g} as it may include both \textit{f} and \textit{g} modes \cite{vM18} or even multiple g-modes~\cite{kK06}. Despite this complexity, in the TF representation this emission appears as a dominant mode (i.e. the resonant frequency of the PNS) among weaker modes that are slightly or no visible, depending on the scale. In this work we focused on the dominant mode. Following the convention in~\cite{ac24}, we hereafter refer to this emission as the High-Frequency Feature (HFF).

Previous studies have aimed to characterize the HFF. For instance, Lin et al.~\cite{zL23} applied a chi-squared optimization to estimate the first-order slope and initial frequency of the HFF. It was performed with O3 LIGO (L1, H1), Virgo (V1) data and TF pixels of a single 3D CCSN waveform detected by the Coherent WaveBurst (cWB) pipeline. In Bruel et al.~\cite{tB23}, a polynomial fit was applied to estimate the HFF and, subsequently, to infer the evolution of the combination between the mass and the radius of the PNS. It drew on CCSN waveforms embedded in Gaussian colored noise. More recently, in Casallas-Lagos et al.~\cite{aC23}, an optimized regression deep neural network (fully connected) was applied to estimate the (first-order) slope of the HFF. They used O3 LIGO data and TF pixels of phenomenological and numerical simulated CCSN GW signals detected by cWB.

Rather than inferring parameters to fit a specific analytic model, we focus on understanding the morphology of the HFF. In this work, we present a new methodology to classify TF image samples (Morlet wavelet scalograms) containing the HFF, depending on the (first-order) HFF slope. We defined a three-class classification problem, categorizing the HFF slope as Class 1 (Steep), Class 2 (Moderate), and Class 3 (Low). These classes represent distinct morphological categories in the TF domain, with specific slope ranges provided in subsection~\ref{sec:phe_wf}.
For this, we drew on one of the state-of-the-art and open-source architectures in computer vision, namely Residual Neural Networks, in particular, ResNet50 \cite{kH16}.

The application of deep learning, particularly convolutional neural networks, has become increasingly prevalent in GW data analysis. Residual Neural Networks, in particular, have demonstrated excellent performance in tasks ranging from GW detection to parameter estimation on compact binary coalescences GW signals, processing 1D time series~\cite{pN23} and 2D TF representations~\cite{tS24}. Furthermore, recent works have begun to explore the potential of these architectures on CCSN signals, for instance, in characterizing the core-bounce feature via regression~\cite{sN24}. Our work builds directly upon this foundation by presenting a novel application of a ResNet50 architecture to the previously unaddressed problem of classifying the HFF based on its slope. By framing the task as a multi-class problem, we move beyond parameter estimation to establish clear, interpretable detectability thresholds for the morphological evolution of the HFF in the presence of real interferometric noise.

Our datasets consist of TF samples with real O3b noise plus GW signals, with the HFF being present. We used two types of waveforms: to optimize and train, we draw on phenomenological waveforms, and to test we draw on numerical simulated waveforms. To optimize our ResNet50 algorithm, we first draw on samples of phenomenological waveforms~\cite{pA18} injected into real noise data from L1, H1, and V1 detectors from the O3b run, which is freely available on the Gravitational Wave Open Science Center (gwosc.org). Then, to test the optimized ResNet50 model, we draw on a dataset formed by multidimensional numerical simulated CCSN waveforms (at distances $1$, $5$, and $10$ kpc) injected into real noise data from the L1, H1, and V1 detectors from the O3b run. This test is crucial because numerical simulated signals are the closest to what we expect in future detections of real CCSN GW signals.

In specific terms, we formulated the problem as a multi-class classification for three main reasons, one astrophysical and two methodological:

\begin{enumerate}
    \item \textit{Astrophysical interpretability and actionability}: The physical significance of specific HFF slope values in connection to core-collapse supernova (CCSN) is an on-going research area~\cite{aM24}. Without a precise physical model dictating specific values, classifying the slope into broad categories (Steep, Moderate, Low) provides a more actionable results for guiding rapid decision-making, namely follow-up studies, multi-messenger observations, and even theoretical inquiry. The boundaries between these categories can be directly interpreted as detectability thresholds for possible astrophysical scenarios to be discovered, in contrast to regression outputs that require over-interpretation and post-hoc thresholds.

    \vspace{.15cm}
    \item \textit{Robustness against data limitations}: Current numerical simulations of CCSNe are computationally expensive, resulting in a reduced number of available waveforms, and not all are even openly distributed. For effective training, classification models require less data (in quantity and diversity) compared to regression models, which can be prone to overfitting when predicting precise continuous values from a few examples. In classification, the model learns boundaries between regions of the feature space requiring limited data. Here complexity is associated to the discrimination of the model itself. Regression, on the other side, approximates a real function by introducing an explicit dependency between precision and the desired resolution to distinguish between functions. Greater accuracy requires more data, therefore complexity increase with resolution.
    
    \vspace{.15cm}
    \item \textit{Operational clarity in the low signal-to-noise (SNR) regime}: Given that the regression model's output is a continuous value, it involves a larger uncertainty in comparison to classification that work on decision boundaries. Therefore, when noise is predominant and wrong predictions increase, working with regression is difficult to interpret and act upon. A classification model provides a clear, discrete decision (e.g., 'Moderate slope detected'). This is more valuable for informing rapid astronomical decisions despite the inherent uncertainties of the detection, and easier to understand the rationale behind the prediction when wrong.
\end{enumerate}

Furthermore, even though current approaches to analyze the HFF are based in regression, classification is also commonly used in GW signal characterization for transient events (e.g., to distinguish between binary mergers, supernovae, and glitches). Therefore, we adhere to this established practice by extending to the HFF morphology.

From a theoretical point of view~\cite{aM24}, we know that the HFF is the dominant emission in CCSN gravitational-wave signals. Consequently, under typical detection pipeline conditions, we expect this emission to be visible in the vast majority of detected signals undergoing analyses in post-processing. The goal is to characterize the morphology of the HFF (its slope), not to determine its presence or absence. Nonetheless, in the final appendix, we include an analysis testing the optimized ResNet50. This test uses samples of pure noise (belonging to a ``ghost class'') and samples containing both noise and numerical simulated low SNR waveforms (at 50 kpc), performing a Kolmogorov–Smirnov comparative test to quantify statistical equivalence of both scenarios.

Within the above context, our choice of the classes and HFF slope ranges is practically motivated and physically-informed. As it is detailed in subsection~\ref{sec:phe_wf}, we designed a heuristic strategy in which, starting from the known broad range of HFF slopes in literature, we defined three adjacent ranges that, later, are validated by distinct astrophysical scenarios that are detailed in subsection~\ref{sec:num_wf}. Even though the physical significance of specific HFF slope values is an open research area~\cite{aM24}, target labels are associated with physical unknown configurations (for instance, related to different progenitor masses and rotation rates) to be discovered through future LIGO and Virgo observations. We need to anticipate discovery by implementing rapid, accurate, and robust computer pipelines to identify each of these configurations, even if their theoretical reasons are still unknown.

All computations were performed using open-source software: Python v3.9.7 (run locally), Python v3.10.12 (run on Google Colaboratory), TensorFlow v2.15.0, Keras v2.15.0, scikit-learn v1.2.2, SciKeras v0.12.0, Pillow v9.4.0, and PyCBC v2.3.0, among others. Following an open-science framework, we make the deep learning code freely available in the public GitHub repository \href{https://github.com/ManuelDMorales/resnet50-sngw-hff}{{\bf resnet50-sngw-hff}}, in addition to \href{https://github.com/ManuelDMorales/datagen-sngw-phen}{{\bf datagen-sngw-phen}} and \href{https://github.com/ManuelDMorales/datagen-sngw-genrel}{{\bf datagen-sngw-genrel}} containing the dataset generators with phenomenological and numerical simulated waveforms.

The paper is organized as follows. Section~\ref{sec:methodology} introduces the problem statement and the methodology, including dataset generation, injection procedure, pre-processing, and the applicatiom of the ResNet50. Later, section~\ref{sec:results} presents the results, separating them into those we obtained working with phenomenological waveforms (subsection~\ref{sec:results_phe_wf}) and with numerical simulated waveforms (subsection~\ref{sec:results_genrel_wf}). Finally, we conclude in section~\ref{sec:conclusions}, followed by an appendix.

\section{Methodology}\label{sec:methodology}

\subsection{Problem statement}\label{sec:prob_stat}

As a starting point, the raw strain time series is given by:
\begin{eqnarray}
    s_\mathrm{raw}{} (t) &=& \left[s(t_0), s(t_1), ..., s(t_{N_\mathrm{slice}-1}{})\right]^T ~. \label{eq:strain_raw}
\end{eqnarray}

Eq.~(\ref{eq:strain_raw}) represents the single-interferometer response signal, which contains non-Gaussian, non-stationary noise together with a number of embedded CCSN GW signals. Given a dataset of window time series previously extracted from Eq.~\ref{eq:strain_raw}, each containing a CCSN GW signal, our problem is to decide if each window belongs to class 1, 2, or 3, depending on the HFF slope-type of its GW signal. Subsections~\ref{sec:phe_wf}-\ref{sec:dataset_gen} describe the dataset generation process. Then, subsection~\ref{sec:resnet_model} is focused on the classification itself, which is addressed using an optimized ResNet50.



\subsection{Phenomenological waveforms}\label{sec:phe_wf}

Following the approach first proposed in~\cite{pA18} and later implemented in~\cite{cT23}, we used phenomenologically parameterized CCSN waveforms to optimize the ResNet50 model. These come from a simplified non-physical model that mimics one of the features that is common to all CCSN simulations, namely the HFF, that usually appears as an increasing arch in the time-frequency representation.

The phenomenological waveforms were generated using a damped-harmonic-oscillator equation with random forcing, expressed as follows~\cite{pA18}:
\begin{eqnarray}
    \frac{\partial^2{h}}{\partial t^2} + \frac{\omega(t)}{Q}\frac{\partial h}{\partial t} + \omega(t)^2 h &=& a(t) ~, \label{eq:phen_model}
\end{eqnarray}
where $h$ is the mimicked strain, $\omega(t)$ the excited eigenmode (angular frequency) of the PNS, $Q$ a constant quality factor, and $a(t)$ a random acceleration. The frequency $f(t)=\frac{\omega(t)}{2\pi}$ is modeled as a 2nd order polynomial:
\begin{eqnarray}
    f(t) &=& f_0 + f_1\left(t-t_\mathrm{ini}\right) + f_2\left(t-t_\mathrm{ini}\right)^2 ~,~
    t \in \left[t_\mathrm{ini},t_\mathrm{end}\right] ~. \label{eq:phen_freq}
\end{eqnarray}
where $t$ denotes the time, and $t_\mathrm{ini}$ and $t_\mathrm{end}$ the initial and end times of the signal, respectively. Given the external force (particularly $a(t)=a_n \delta (t-t_n)$ with $n=1,2,...,f_\mathrm{driver}(t_\mathrm{end}-t_\mathrm{ini})$ and $a_n$ randomly distributed on $[t_\mathrm{ini},t_\mathrm{end}]$ and $[0,a_\mathrm{max}]$, respectively), and rewriting Eq.~(\ref{eq:phen_freq}) with $f_1=f_\mathrm{1s}=f(t=1s)$, and $t_2=\text{argmax}\left[f(t)\right]$ ($t_2>t_\mathrm{end}$), this system can be numerically solved once we choose initial values for the $7$ free parameters: $t_\mathrm{ini}$, $t_\mathrm{end}$, $f_0$, $f_\mathrm{1s}$, $t_2$, $Q$, and $f_\mathrm{driver}$.

To categorize the waveforms, we defined three classes according to the numerical range of the HFF slope for each waveform:
\begin{itemize}
    \item \textbf{Class 1 (Steep)}: $1,620$ $\leq$ HFF slope $<4,990$ Hz/s

    \vspace{.15cm}
    \item \textbf{Class 2 (Moderate)}: $1,450$ $\leq$ HFF slope $<1,620$ Hz/s

    \vspace{.15cm}
    \item \textbf{Class 3 (Low)}: $950$ $\leq$ HFF slope $<1,450$ Hz/s \label{eq:wf_classes}
\end{itemize}

As previously mentioned, this choice is practically motivated, as the physical significance of specific HFF slope values remains an active area of research. To ensure the defined classes are physically relevant, our approach followed a three-step heuristic design. Firstly, we established a broad range of HFF slopes ($950$ Hz/s to $4990$ Hz/s) that encompasses typical values reported in the literature from multi-dimensional CCSN simulations. Next, within this range, we defined three adjacent, mutually exclusive intervals to create distinct morphological categories corresponding to low, moderate, and steep slopes as observed in the TF representation. And finally, we verified that these classes naturally map to distinct astrophysical scenarios, i.e. those detailed in the next subsection. This \textit{a posteriori} alignment with established numerical models provides direct astrophysical justification for the chosen class boundaries.

\begin{figure}[htb]
\begin{center}
  \includegraphics[width=13.0cm]{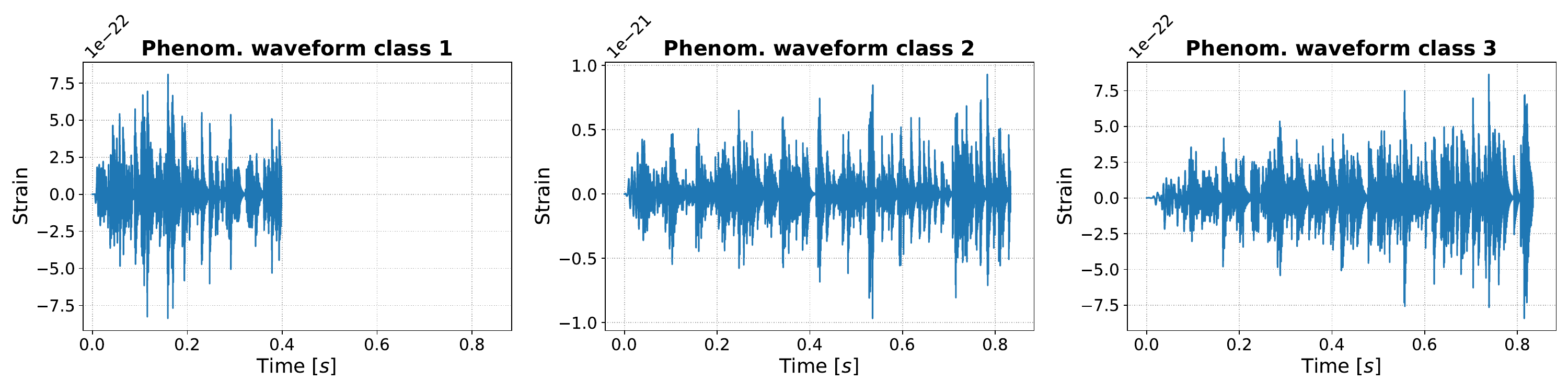}
  \includegraphics[width=13.2cm]{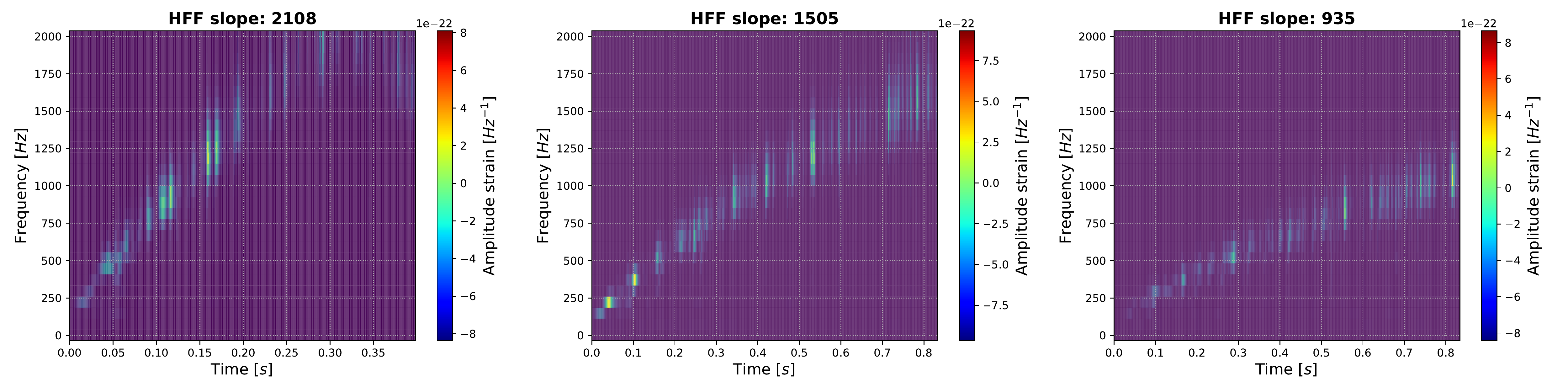}
  \caption{\label{fig:waveforms(phen)} Examples of phenomenological waveforms in the time domain (upper panel) and their Morlet wavelet scalograms (bottom panel). They belong to classes 1, 2, and 3. For HFF slope $2108$ Hz/s, $f_0 = 126.63$ Hz and $f_1 = 3416.08$ Hz; for slope $1505$ Hz/s, $f_0 = 108.14$ Hz and $f_1 = 1774.62$ Hz; and for slope $935$ Hz/s, $f_0 = 105.28$ Hz and $f_1 = 1148.24$ Hz. The time duration of the waveforms varies because of the nature of their model, which include a random force. Waveforms are shown in the absence of noise; however, as detailed in subsection~\ref{sec:dataset_gen}, they were injected into real LIGO and Virgo interferometric noise to generate the ResNet50 optimization dataset.}
\end{center}
\end{figure}

Fig.~\ref{fig:waveforms(phen)} shows three examples phenomenological waveforms, one of each class, plotted as strain time series with their respective time-frequency Morlet wavelet scalograms (in section~\ref{sec:dataset_gen} we briefly describe how these scalograms are generated). Notice from the strain time series that the duration of waveforms varies, which is expected because of the random force included in the the phenomenological model. We draw on $600$ phenomenological waveforms, $200$ per class, which will be randomly injected into the noise segments.

We stress that while this work have recourse to the phenomenological model~\cite{pA18} for its well-established representation of the dominant HFF, we note the recent development of more complex models (e.g., Cerda-Duran et al. 2025~\cite{pC25}) that incorporate additional physics. The pipeline presented here is well-suited for future application and retraining with such improved waveform models.

\subsection{Numerical simulated waveforms}\label{sec:num_wf}

\begin{figure}[htp]
 \begin{center}
  \includegraphics[width=4.25cm]{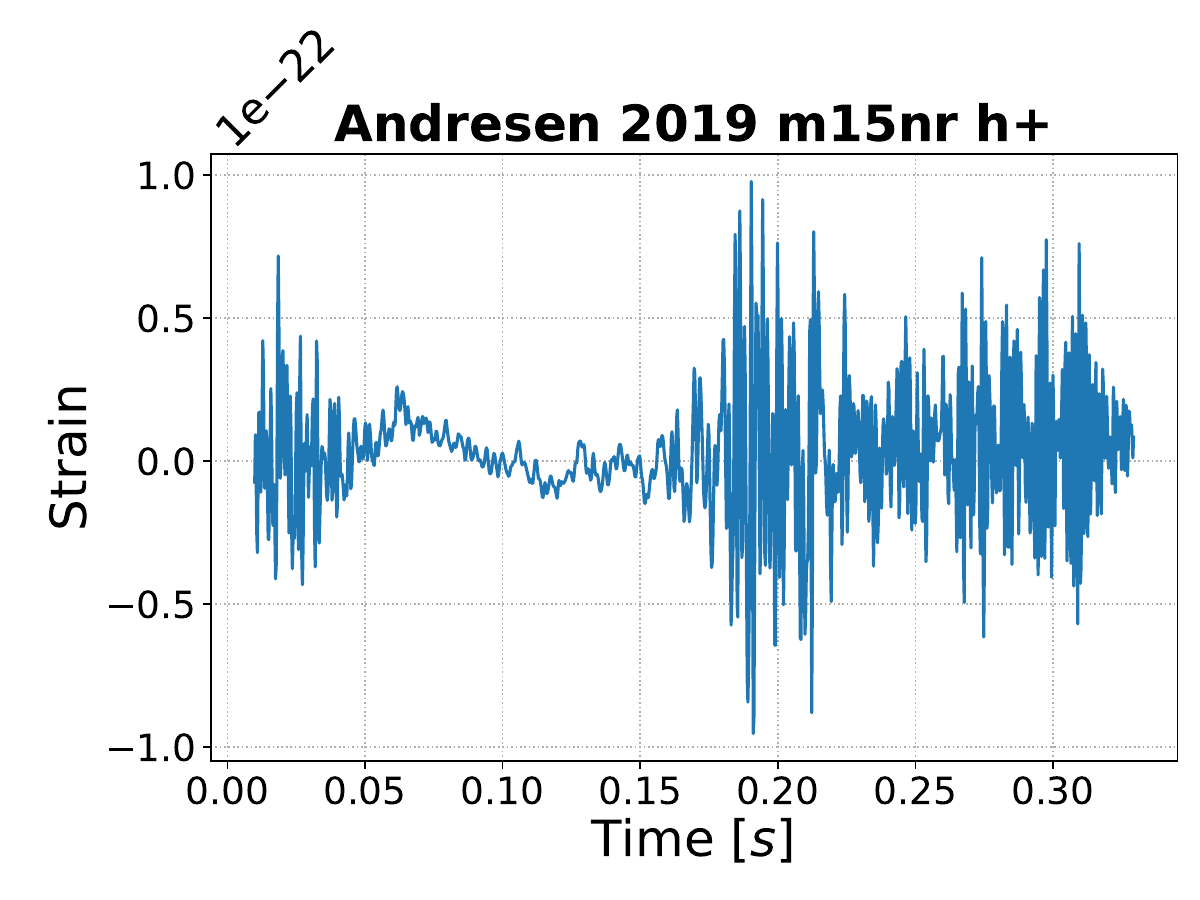}
  \includegraphics[width=4.25cm]{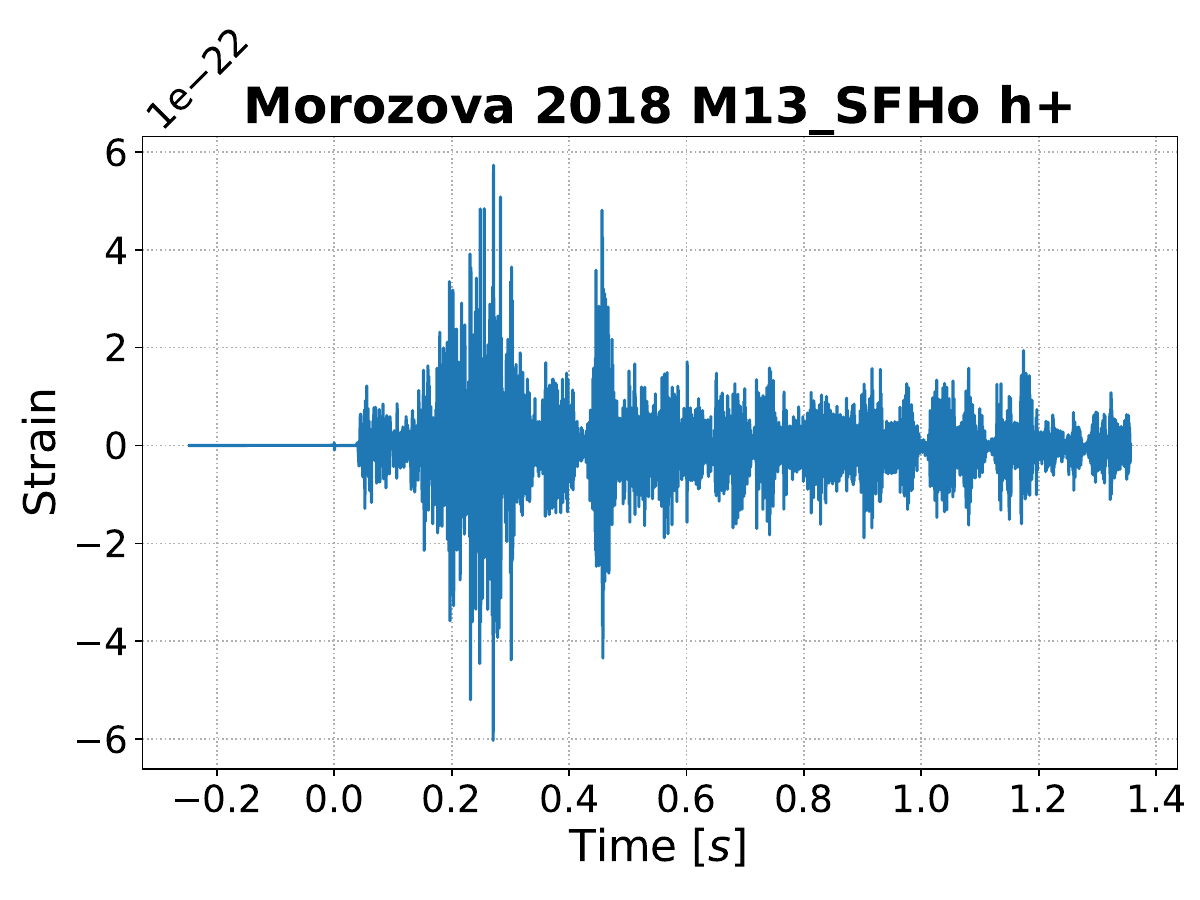}
  \includegraphics[width=4.25cm]{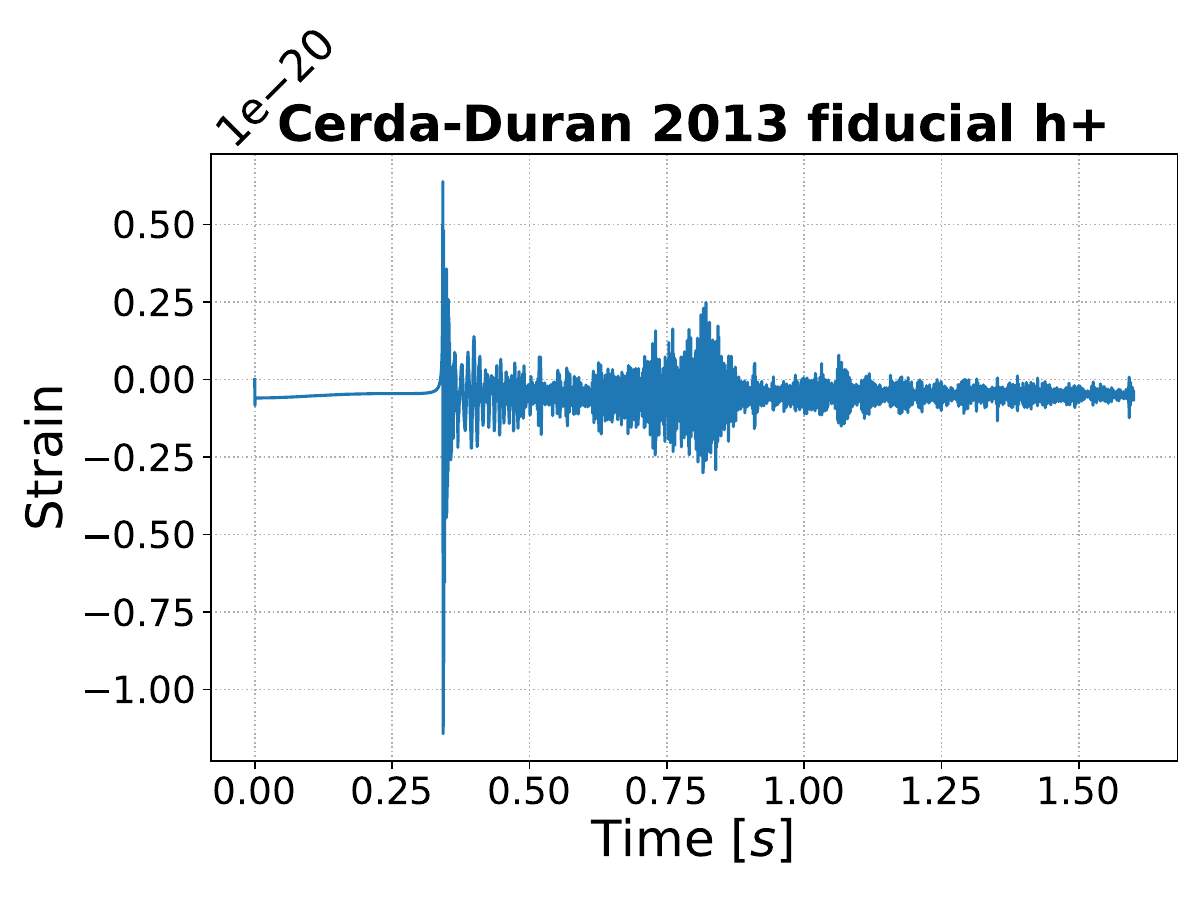}
  \\
  \includegraphics[width=4.43cm]{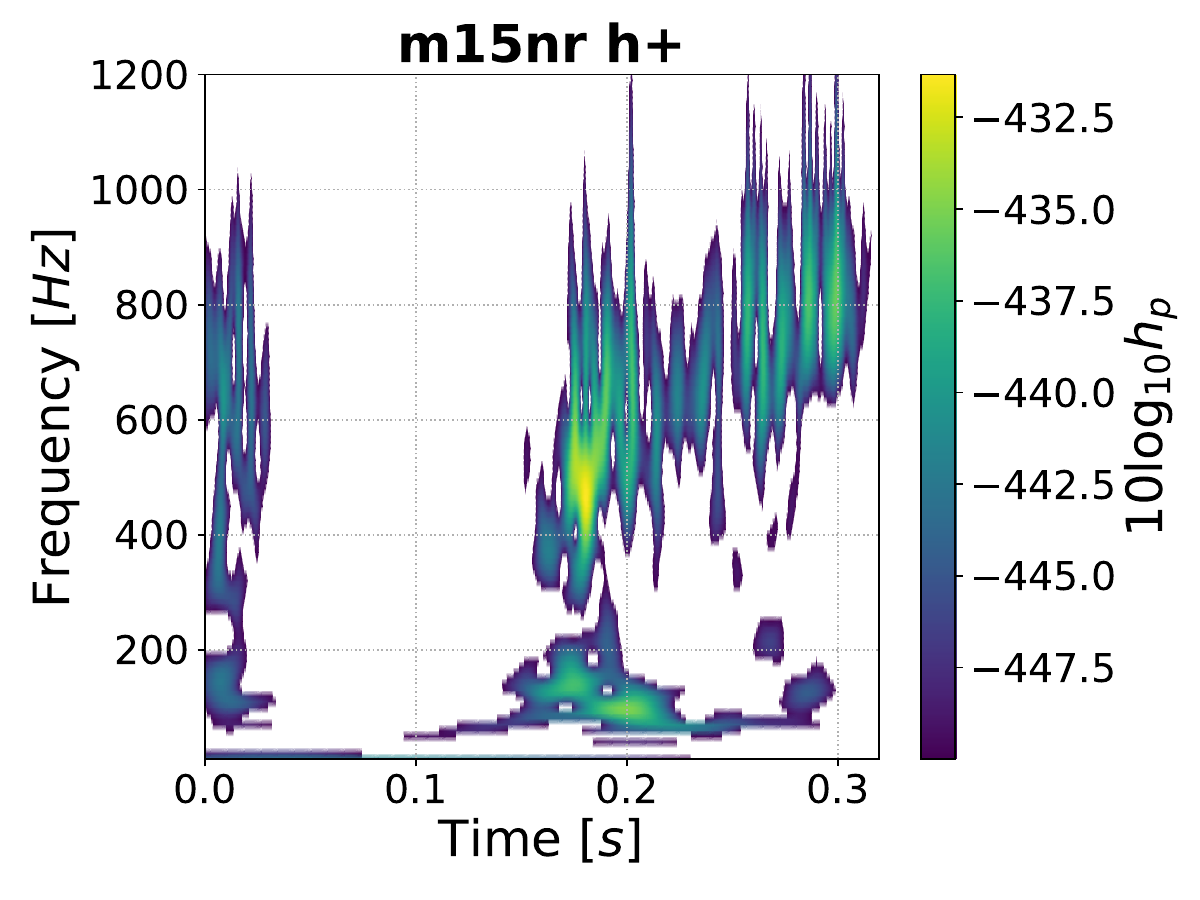}
  \hspace{-.3cm}
  \includegraphics[width=4.43cm]{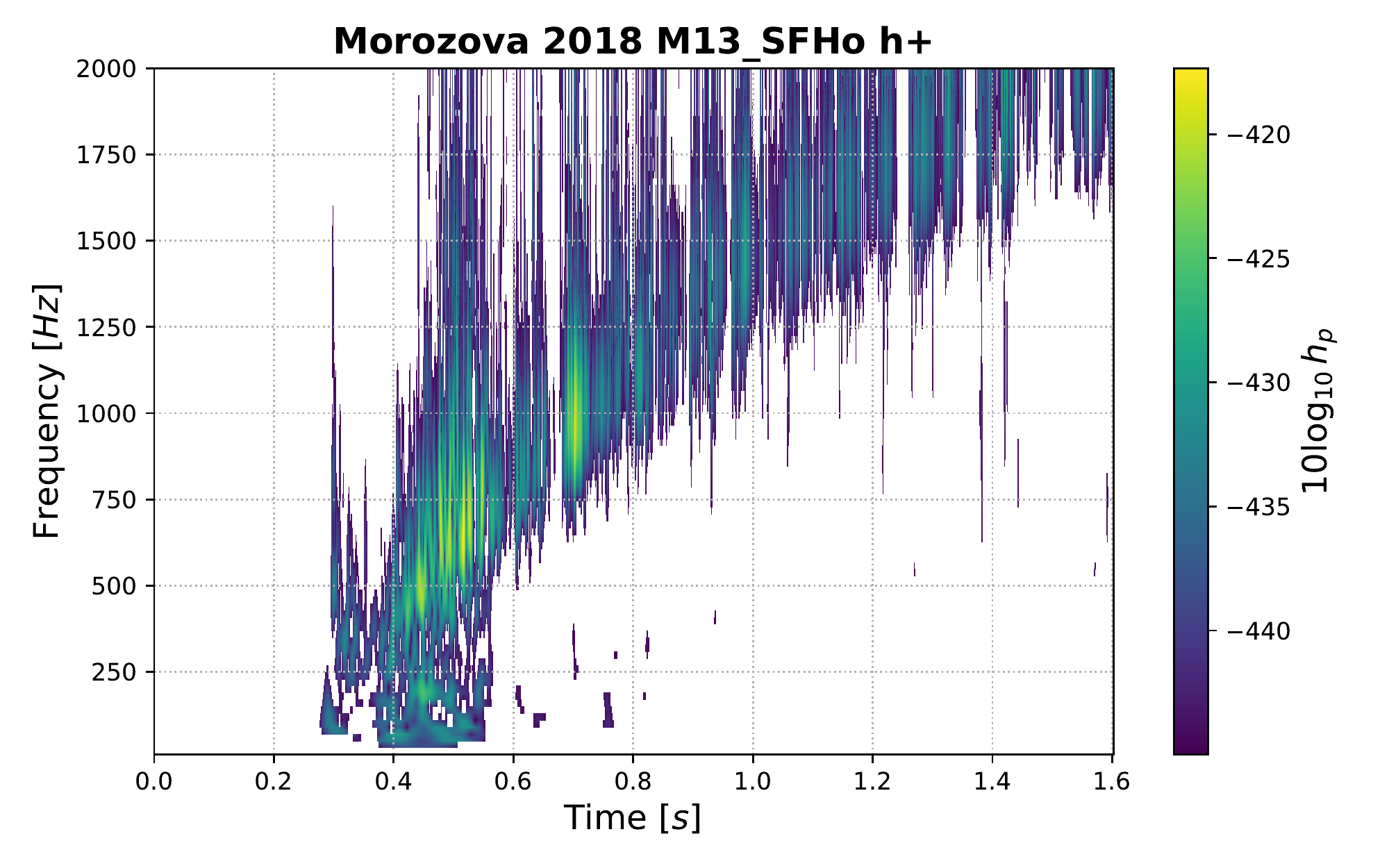}
  \hspace{-.3cm}
  \includegraphics[width=4.43cm]{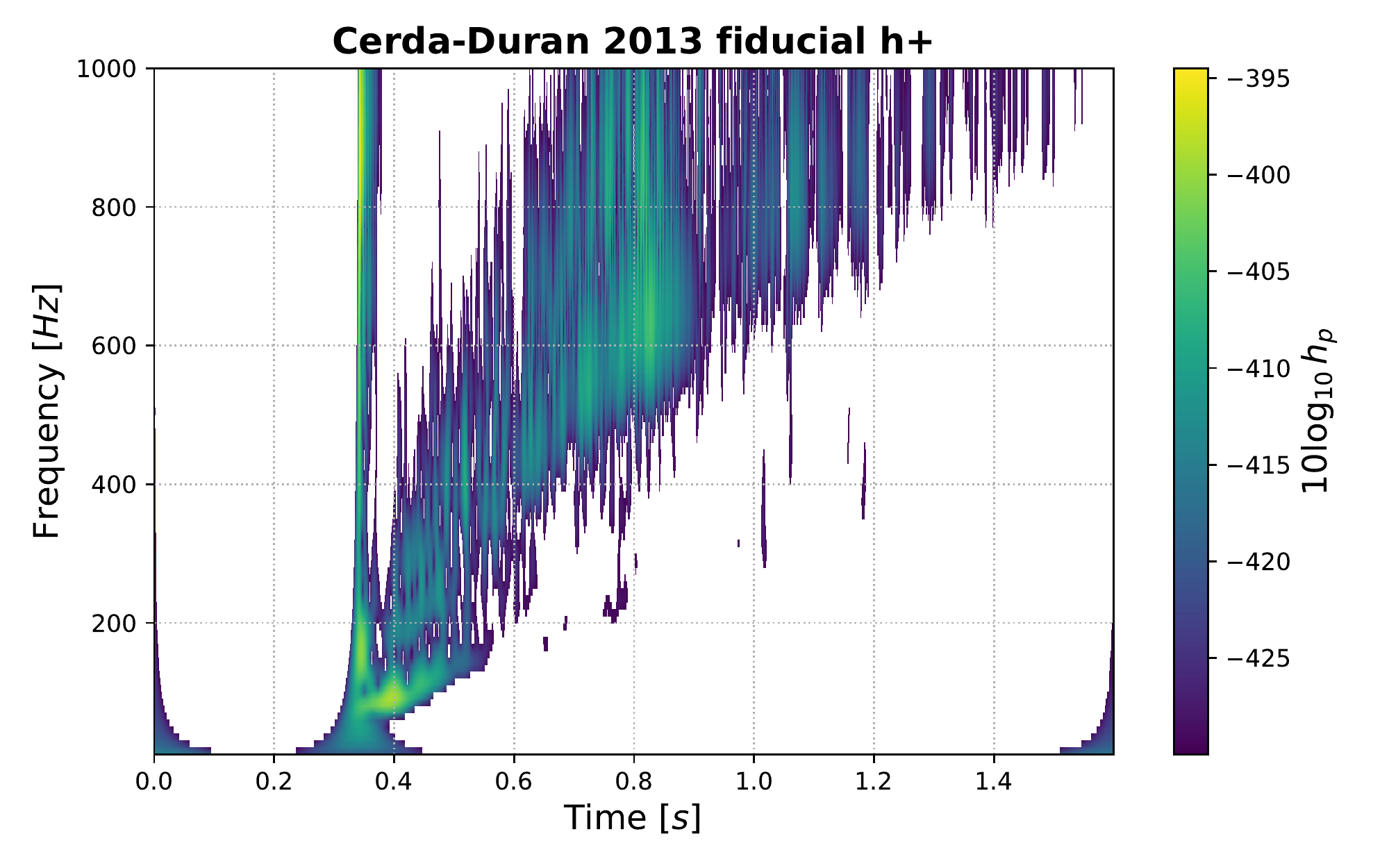}
  \caption{\label{fig:waveforms(num)}Numerical simulated CCSN waveforms simulations used in this work, both in the time domain (upper panel) and their Morlet wavelet scalograms (bottom panel). Andresen et al. 2019 $m15nr$ waveform comes from a 3D CCSN simulation. Morozova et al. 2018 $M13\_SFHo$ and Cerd\'a-Dur\'an et al. 2013 $fiducial$ al. waveforms come from 2D CCSN simulations. These waveforms are astrophysically realistic. Then, as detailed in subsection~\ref{sec:results_genrel_wf}, we used their HFF emission, injected into real LIGO and Virgo interferometric noise, to test the ResNet50 model.}
 \end{center}
\end{figure}

We considered three openly available CCSN waveforms. They have HFF slopes that belong to the classes defined in Eq.~(\ref{eq:wf_classes}) (one waveform per class), in addition to other features such as the standing accretion shock instability (SASI), and the prompt-convection feature.
\begin{itemize}
    \item Andresen et al. 2019 3D, $m15nr~h+$~\cite{hA19}. Model of a $15M_{\odot}$ non-rotating progenitor constrained by a Lattimer and Swesty equation of state (EoS) with nuclear compressibility $K = 220$MeV (LS220). Strong SASI activity dominated by the spiral mode, prompt convection, and the HFF are produced.

    \vspace{.15cm}
    \item Morozova et al. 2018 2D, $M13\_$SFHo~$h+$~\cite{vM18}. Model of a $13M_{\odot}$ non-rotating progenitor constrained by the Steiner (SFHo) EoS. Weak SASI activity, prompt-convection, and the HFF are produced.

    \vspace{.15cm}
    \item Cerd\'a-Dur\'an et al. 2013 2D, $fiducial$~\cite{pC13}. Model of a $35M_{\odot}$ rapidly rotating progenitor, constrained by the LS220 EoS. Weak SASI activity and the HFF are produced.
\end{itemize}
Fig.~\ref{fig:waveforms(num)} shows the three numerical simulated waveforms used in this work, both in the time domain (strain) and in the time-frequency domain (Morlet wavelet scalograms). Notice that SASI activity occurs in the frequency region below the early times of the HFF, while the prompt-convection feature appears within approximately $0.1$ s after core bounce.

To estimate the HFF slope of each waveform while minimizing ambiguity, we first isolated the HFF contribution. Starting from scalograms represented as pixel scatter plots, we applied a three-step procedure. Firstly, we removed all the pixels with intensity lower than the arithmetic mean of all wavelets in scalograms. Next, we removed the pixels from the initial time to $0.1$ s for the Andresen and Morozova waveforms and to $0.4$ s for Cerd\'a-Dur\'an waveform; these are the regions where the prompt-convection feature appears. Finally, we removed pixels located at frequencies less than $250$ Hz, which is the region where SASI appears.

\begin{figure}[htp]
\begin{center}
  \includegraphics[height=3.8cm]{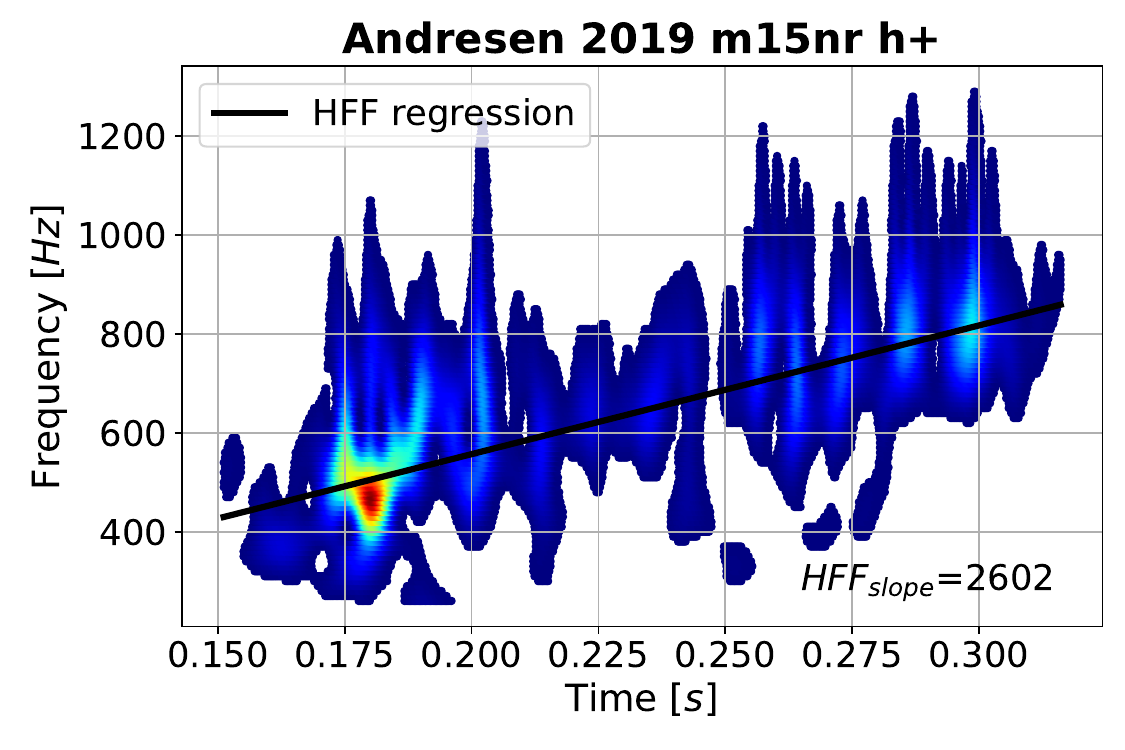}
  ~~~~
  \includegraphics[height=3.8cm]{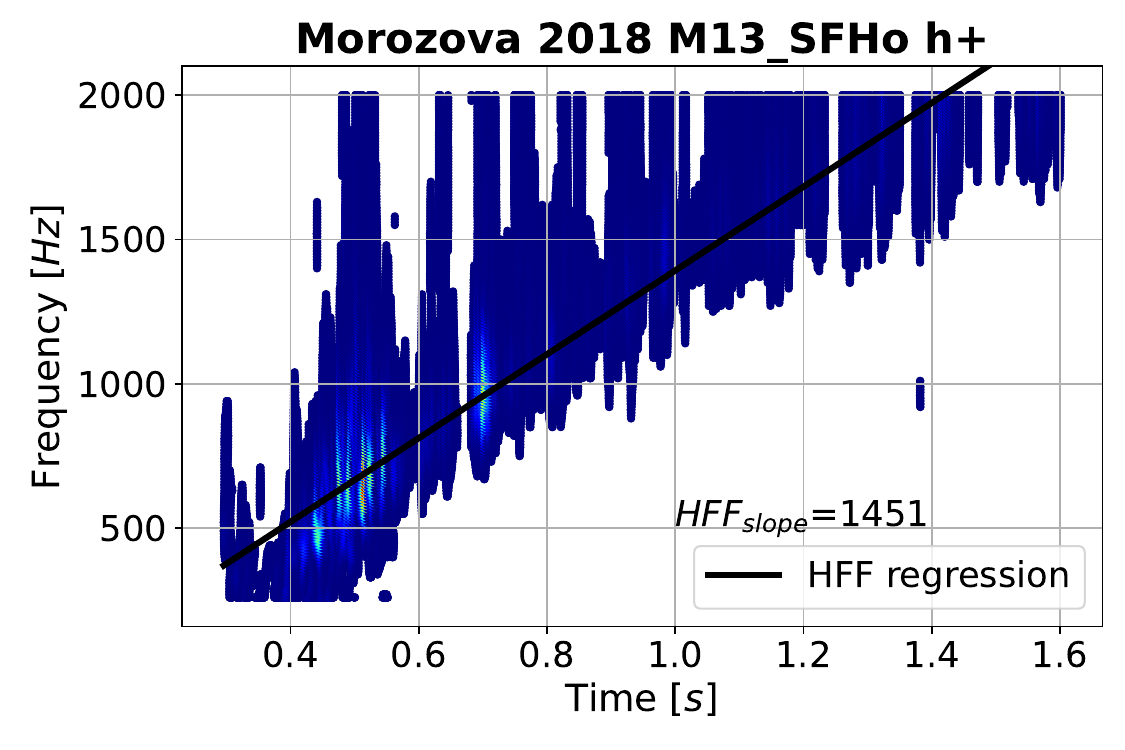}
  \includegraphics[height=3.8cm]{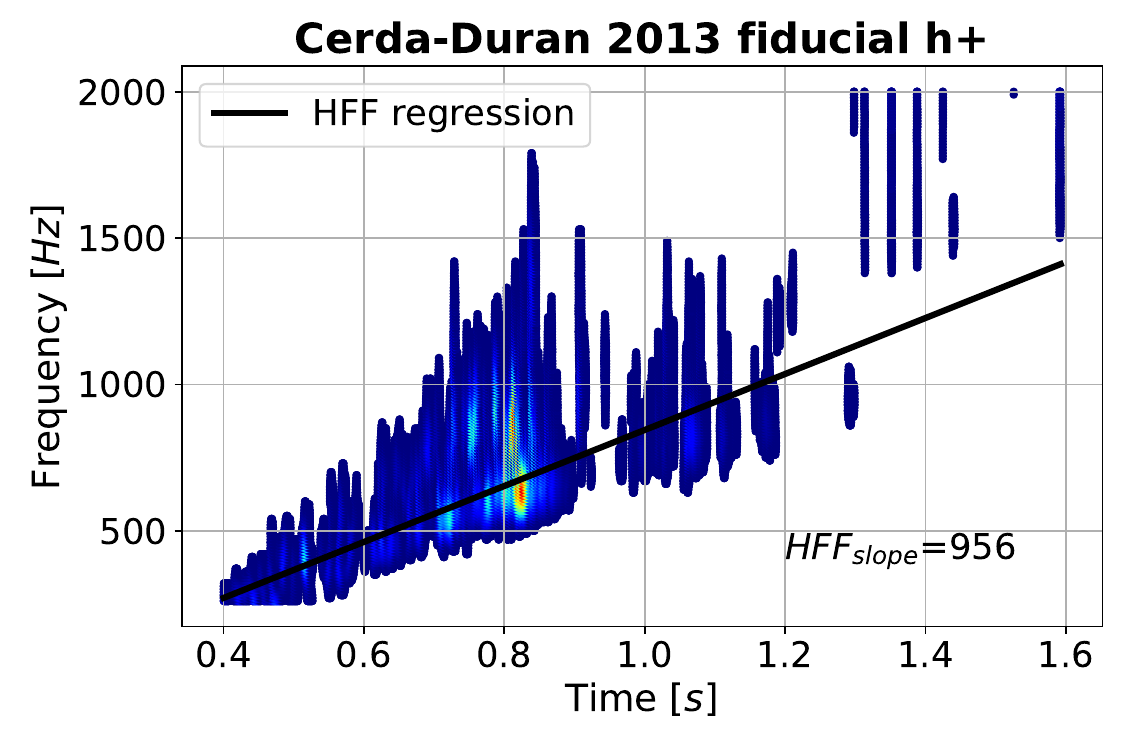}
  \caption{\label{fig:waveforms(num)_hffslope}HFF slope estimation for CCSNe numerical waveforms using linear regression. To unambiguously perform this fit, we previously removed wavelet contributions of the background (defined as pixels with intensity less than the arithmetic mean of all pixels) and features other than the HFF. This estimation was performed to assign, to each waveform, one of the previously defined target classes, as shown in Table~\ref{tab:HFFslopes}.}
\end{center}
\end{figure}

After previous removals, we estimated the slope of the HFF. For this, we apply a linear regression, considering the most energetic pixels each time. The results for the HFF slope estimation, for the three CCSN GW signals, are detailed in Fig.~\ref{fig:waveforms(num)_hffslope} and Table~\ref{tab:HFFslopes}.

\begin{table}[t]
\caption{\label{tab:HFFslopes}HFF slope estimates for the three numerical simulated CCSN waveforms used in this work. These results were computed using a linear regression in the absence of noise and after removing the contributions of wavelet features other than the HFF. Depending on having less or more frequency dispersion, the goodness of fit of the regression, $R^2$, is closer or farther to value 1, respectively.}
\captionsetup{width=1.0\textwidth}
\centering
\begin{tabular}{|c|c|c|c|c|}
 \hline
 \textbf{CCSN waveform} & \textbf{HFF slope [Hz/s]} & \textbf{HFF intercept [Hz]} & \textbf{$R^2$} & \textbf{Class} \\
 \hline\hline
 \makecell{Andresen 2019 \\ $m15nr~h+$} & 2601.99 & 36.4985 & 0.78906 & 1 \\ 
 \hline 
 \makecell{Morozova 2018 \\ $M13\_$SFHo~$h+$ } & 1451.37 & -58.9381 & 0.93960 & 2  \\
 \hline
 \makecell{Cerd\'a-Dur\'an 2013 \\ $fiducial~h+$ } & 956.261 & -112.200 & 0.77444 & 3 \\
 \hline
\end{tabular}
\end{table}

In phenomenological waveforms, the mimicked HFF is the only feature appearing, starting at the left-bottom corner of the scalogram, i.e., at the very beginning of the waveform. However, this is not the case for numerical simulated waveforms, which also include features other than the HFF. To be consistent between training and test datasets, we performed two applied steps before injecting numerical simulated waveforms into noise data. 

Firstly, we set the initial time of the numerical simulated waveforms such that the HFF begins approximately at the beginning of the waveform; that is to say, we remove its early times. For Andresen $m15nrh+$, we removed the first $0.16$ s; for Morozova $M13\_$SFHo~$h+$, the first $0.30$ s; and for Cerd\'a-Dur\'an $fiducial$, the first $0.38$ s. Next, we removed the contribution of features other than HFF. The prompt-convection feature was already removed with the early times cropping previously applied. Then, to remove the SASI feature we applied a Butterworth high-pass filter to the waveforms in the time domain, setting a critical frequency of $250$ Hz.

The pre-treatment applied to numerical waveforms, i.e., initial time cropping and high-pass filtering, was not applied to the phenomenological waveforms. This discrepancy is because the latter model only has the HFF and lacks other features. While this creates a difference in the data generation pipeline, it is a pragmatic approach: training a model on a clean representation of a feature before deploying it to identify that feature within more complex, pre-processed data. At the end, we note that high test accuracies achieved on numerical waveforms, especially at $1$ kpc and $5$ kpc, suggest that the model successfully learned the HFF morphology and was not biased by this pre-treatment difference.

\subsection{Dataset generation}\label{sec:dataset_gen}

\subsubsection{\label{sec:inj_cond}Injections and conditioning}

We used single-interferometer noise segments from the O3b run of LIGO (H1, L1) and Virgo (V1) detectors, with initial GPS times $1256783872$ and $1257050112$. Their time duration is $t_{N_\mathrm{slice}}=4096$ s, and we chose segments of sampling frequency $f_\mathrm{s}=4096$ Hz to minimize system resources.

To generate our datasets, we injected waveforms into the noise segments, conform an uniform injection grid except of a small jitter. The jitter was introduced to avoid the ResNet50 algorithm mislearns the HFF classification based on uniform fixed time locations of gravitational wave signals in noise data. Then, each $i$-th injection is:
  \begin{eqnarray}
   t_\mathrm{inj}{}^{(i)} = t_\mathrm{inj}{}^{(0)}+i\Delta t_\mathrm{inj} + t_\mathrm{jitter}{}^{(i)} ~, \label{eq:t_injections}
 \end{eqnarray}
where $t_\mathrm{jitter}{}^{(i)}=\mathrm{random}_i([-\mathrm{jitter}_\mathrm{lim},+\mathrm{jitter}_\mathrm{lim}])$ with $\mathrm{jitter}_\mathrm{lim}=0.01$ s, $t_\mathrm{inj}{}^{(0)}$ is the time location of the first injection, and $\Delta t_{\mathrm{inj}}=0.8$ s. We performed 511 injections of phenomenological waveforms belonging to a single class into a complete noise segment of $4096$ s. Given that we draw of two noise segments for three detectors and three classes, the resulting number of injections is $511 \times 2 \times 3 \times 3 = 9198$. For numerical simulated waveforms, we set the same input parameters for the injection procedure.

We stress that, as part of the aforementioned procedure, for each injected waveform we computed the SNR value, given by the following expression:
\begin{eqnarray}
    \mathrm{SNR} &=& \sqrt{ \int_{f_\mathrm{min}}^{f_\mathrm{max}} \frac{|\tilde{s}(f)|^2}{S_\mathrm{noise}(f)} df } ~,
\end{eqnarray}
where $\tilde{s}(f)$ denotes the one-sided Fourier transform of the strain time series $s(t)$, $S_\mathrm{noise}(f)$ is the power spectral density (PSD), in addition to $f_{\text{min}}$ and $f_{\text{max}}$ that are determined by the frequency range of the FFT.

After injections, we have raw strain data segments $s_\mathrm{raw}(t)$ as described in Eq.~(\ref{eq:strain_raw}). Next, we performed a conditioning that consists of two setps: i) whitening and ii) band-pass filtering. For this, we applied the PyCBC library~\cite{pyCBC}. Whitening approximates the detector noise as a Gaussian stochastic process; in particular, it attenuates the Fast Fourier Transform of $s_\mathrm{raw}(t)$ by its Amplitude Spectral Density (ASD). Here we used segments of $4$ s windowed with the Hanning function, overlapping by $2$ s, in a frequency domain from $0$ Hz to $2048$ Hz, with $0.25$ Hz of resolution. As whitening generates spurious frequencies on the edges of the data segment due to spectral leakage, we removed $5$ s of the data at each edge. Finally, we applied a high-pass and a low-pass Butterworth filter, both of $8th$ order, to remove frequency components lower than $100$ Hz and higher than $1800$ Hz, respectively.

\subsubsection{Extraction of window samples}

Starting from the conditioned data segments, we extracted window samples of duration $T_{\mathrm{win}}$. For phenomenological waveforms, each window sample was required to contain exactly one complete injected gravitational-wave signal, independent of $\mathrm{jitter}_\mathrm{lim}$, $\Delta t_\mathrm{inj}$, and time duration of injected waveforms $\Delta h^{(j)}$. To fulfill this, we imposed that:
\begin{eqnarray}
    T_\mathrm{win} &=& \Delta h^\mathrm{max} + \alpha ~~, \label{eq:Twin_set} \\
    T_\mathrm{win}^\mathrm{~start}{}^{(j)} &=& t_\mathrm{inj}{}^{(j)} + \frac{1}{2} \left(\Delta h^{(j)} - T_\mathrm{win}\right) - t_\mathrm{jitter}{}^{(j)} ~~, \label{eq:Twinstart_set}
\end{eqnarray}
where $\Delta h^\mathrm{max}$ is the duration of the longest injected phenomenological waveform, and $\alpha>2\mathrm{jitter}_\mathrm{lim}$ is an input parameter set as $\alpha=0.02001$. Moreover, if $t_\mathrm{jitter}{}^{(i)}=0 ~\forall i$ (due to $\mathrm{jitter}_\mathrm{lim}=0$), from Eq.~(\ref{eq:Twinstart_set}) we have that all injections are centered in their corresponding windows.

\subsubsection{Wavelet transform}

Each $p$-th strain-window sample ($s_\mathrm{win}{}^p$) was converted to its time–frequency (TF) scalogram using a Wavelet Transform (WT). In general, a WT requires a kernel (i.e., a localized ``mother wavelet''), which, varying its location and scale, is convolved with the window samples to scan frequency changes in the signal. We drew on a Morlet wavelet~\cite{rK87} as the kernel, which we also used in a previous study~\cite{mM21}:
\begin{equation}
  \psi(t_j,f_k) = \frac{1}{\sqrt{\sigma^t{}_j \sqrt{\pi}}}
              \exp\left[\frac{-t_j{}^2}{2 \left(\sigma^t{}_j\right){}^2}\right]
              \exp\left(2 i \pi f_k t_j \right)~, \label{eq:morlet_wavelet}
\end{equation}
which has a 2D Gaussian form, with standard deviations $\sigma^t$ and $\sigma^f$, related by $\sigma^t{}_j = 1 / \left(2 \pi \sigma^f{}_l \right)$ and $\sigma^f{}_k=f_l/\delta w$, where $\delta w$ is the wavelet's width (that we set beforehand in $7$ Hz) and $f_k$ its center in the frequency domain. Subsequently, to generate the TF scalogram of $s_\mathrm{win}{}^p$, we compute:
\begin{eqnarray}
  Ws_\mathrm{win}{}^p\left[t_n,f_j\right] &=& \sum_{m=0}^{N_\mathrm{win}-1} s_\mathrm{win}{}^p(t_m) \psi^*(t_{m-n},f_j)~, \label{eq:wavelet_transform}
\end{eqnarray}
where $s_\mathrm{win}{}^p(t_m)$ is the element $m$ of the vector $s_\mathrm{win}{}^p$. Besides, we have that $n=0,1,...,N_\mathrm{time}$ and $j=0,1,...,N_{\mathrm{freq}}$, where $N_\mathrm{time}$ and $N_\mathrm{freq}$ define the size of each 2D matrix generated by the WT, being $Ws^p\left[t_n,f_j\right]$ its $(n,j)$ element.

Notice from Eqs.~(\ref{eq:morlet_wavelet}) and~(\ref{eq:wavelet_transform}) that $\psi$ and $Ws_\mathrm{win}{}^p$ need to be defined over a finite grid containing discrete values of time and frequency. Here we defined a time vector taking values from $t_0=0$ to $t_{N_\mathrm{time}}=T_\mathrm{win}$ with $\delta t = 1/f$. We also set $f_\mathrm{ini}=10$ Hz, $f_\mathrm{end}=2000$ Hz, and $\delta f=10$ Hz for the initial frequency, the final frequency, and the frequency resolution, respectively. Then, the WT transform is such that its matrix has dimensions $N_\mathrm{time}=T_\mathrm{win}/\delta t = 3971$ and $N_\mathrm{freq}=(f_\mathrm{end}-f_\mathrm{ini})/\delta f + 1 = 200$.

We set the TF scalograms by applying a normalization such that magnitude, i.e., its intensity color bar, varies linearly between the minimum and maximum of its corresponding strain data.

\subsubsection{Image datasets}

To convert original WT matrices to pixelized images while minimizing resources, we apply the Pillow (PIL Fork) image module \cite{Pillow}. This generates pixelized images of dimensions $N_\mathrm{time}=64$ and $N_\mathrm{freq}=64$. Given that the ResNet50 architecture is designed to input RGB (red, green, blue) images in three channels by default, we decided to maintain this for practical purposes. Then, with Pillow we set the final dimensions of each TF image sample as $64 \times 64 \times 3$ by replicating the grayscale pixel values across the red, green, and blue channels. This implementation is general enough to be included in broader pipelines that, in a first stage of gravitational waves detection, the RGB input can be adapted to include simultaneous images of H1, L1, and V1 detectors (for instance, as proposed in \cite{pA18}), without modifying our ResNet50 architecture. The final dataset has the following form:
\begin{equation}
    D = \left\{ \boldsymbol{X}^{64 \times 64 \times 3}, \boldsymbol{y}_\mathrm{c} \right\}^{N_{\mathrm{s}}} ~~, \label{eq:final-phen_dataset}
\end{equation}
where $N_\mathrm{s}$ is the number of class $y_\mathrm{c}$ image samples. 

\begin{table}[t]
\caption{\label{tab:datasets}Size of datasets containing phenomenological and numerical simulated waveforms, and depending on detector noise and class. With numerical waveforms, we set galactic distances of $1$ kpc, $5$ kpc, and $10$ kpc.}
\captionsetup{width=1.0\textwidth}
\centering
\small
\begin{tabular}{|c|c|c|c|} 
 \hline
 $\begin{array}{c}
    \textbf{Datasets with}\\
      \textbf{phenomenological waveforms}
 \end{array}$ & \textbf{$y_\mathrm{1}$} & \textbf{$y_\mathrm{2}$} & \textbf{$y_\mathrm{3}$} \\
 \hline\hline
 L1 & 1,020 & 1,017 & 1,013 \\
 \hline
 H1 & 1,013 & 1,012 & 1,011 \\
 \hline
 V1 & 1,021 & 1,021 & 1,021 \\
 \hline
 $\begin{array}{c}
      \textbf{Datasets with}\\
      \textbf{numerical simulated waveforms}
 \end{array}$ & \textbf{$y_\mathrm{1}$} & \textbf{$y_\mathrm{2}$} & \textbf{$y_\mathrm{3}$} \\
 \hline\hline
 L1 or H1 or V1 (per distance)& 100 & 100 & 100 \\
 \hline
\end{tabular}
\end{table}

Table~\ref{tab:datasets} give the summary of window samples for each dataset used in this study. As it is described in section~\ref{sec:phen_pop}, for samples containing phenomenological waveforms, we used data from all interferometers, all together, to optimize the ResNet50 architecture. For samples containing numerical simulated waveforms, we set galactic distances of $1$ kpc, $5$ kpc, and $10$ kpc.

The execution time for generating the training dataset was approximately $10.5$ hours. Here, the conversion from window strain data to TF image samples was the most computationally intensive process, taking $1$ hour for single-class and single-detector samples. This scaled to $1 \times 3 \times 3 = 9$ hours (more than $90\%$ of the total generation time). Now, in terms of storage, each image sample (saved as a numpy file) was $12$ KB in size, giving a total of approximately $110$ MB.

\subsubsection{\label{sec:phen_pop}Population of optimization dataset}

\begin{figure}[htb]
\begin{center}
  \includegraphics[width=6.4cm]{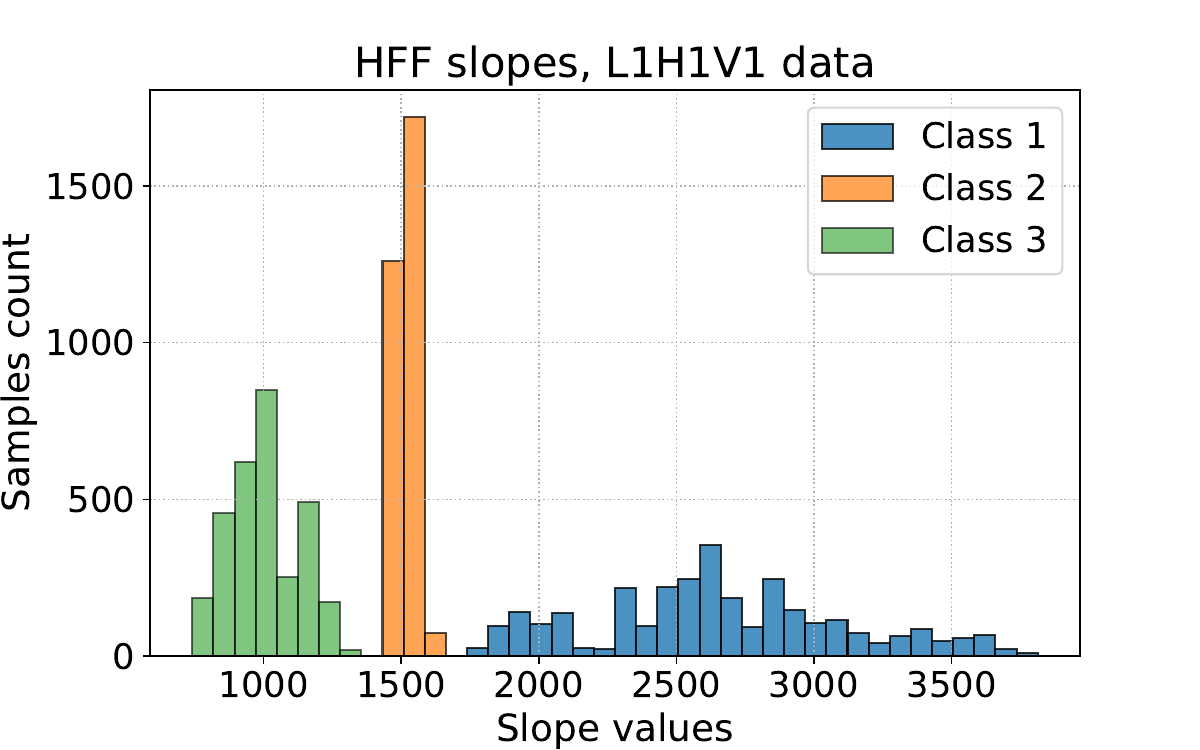}
  \includegraphics[width=6.4cm]{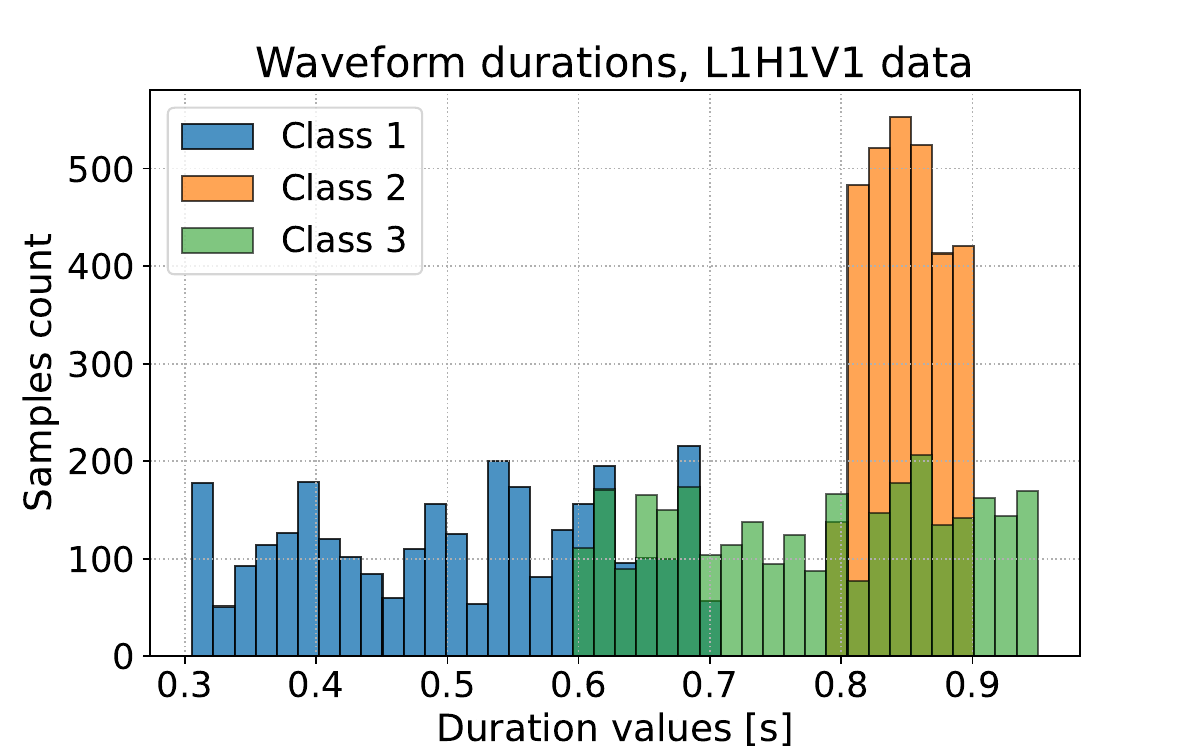}
  \includegraphics[width=6.4cm]{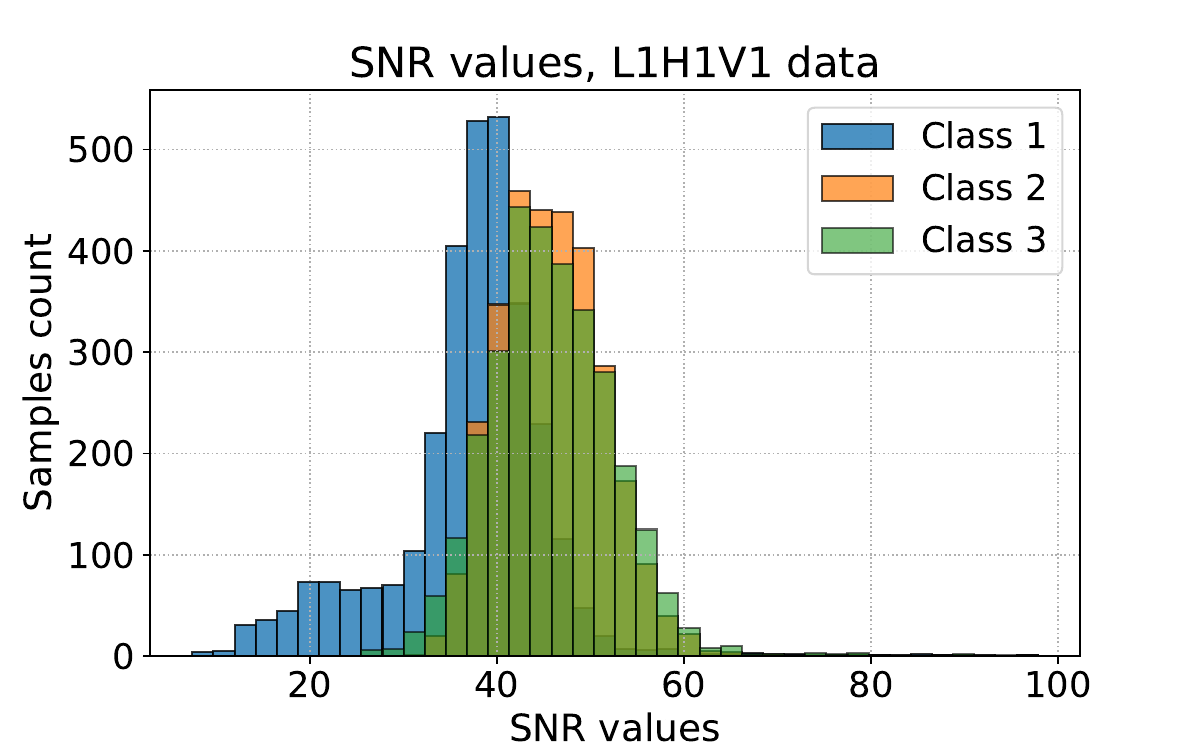}
  \caption{\label{fig:wf_phen_distribution}
  Population exploration of samples used to optimize the ResNet50 model. These contain phenomenological waveforms injected into LIGO-Virgo real O3b noise. Leveraging that these waveforms (by the nature of their model) do not codify the distance, amplitudes were chosen to create a high-SNR dataset to facilitate the learning process.. The SNR values distribution is shown in the bottom panel, in which class 1 samples are shifted to lower SNR values in comparison to SNR distributions for class 2 and 3 samples. Moreover, we have a significant overlap in the SNR distributions of class 2 and class 3 in the range of approximately 40-70, indicating that SNR alone could not be a reliable discriminator between these classes. On the other hand, the upper left and upper right panels show distributions (in the absence of noise) of HFF slope and waveform duration, respectively. Phenomenological waveforms were generated by a stochastic model, and their duration varies from $0.3$ s to just under $1.0$ s.}
\end{center}
\end{figure}

The dataset for the optimization consists of samples of noise plus phenomenological waveforms, with real noise data from the L1, H1, and V1 detectors, all together. The upper left and upper right panels of Fig.~\ref{fig:wf_phen_distribution} show the distribution, in absence of noise, of their HFF slopes and time durations, respectively. Notice that the HFF slopes (and therefore the class of GW signal samples) are clearly distinguishable, with only a reduced number of slopes around the boundary values.

Regarding waveform durations, we observe that their distributions overlap substantially. Moreover, variation of the duration is expected because of the stochasticity of our generating toy model. Class 1 waveforms tend to have shorter durations in comparison to class 2 and class 3 waveforms. To some extent, this difference is because class 1 waveforms have greater slope values, which implies that the HFF reaches the Nyquist frequency more rapidly in comparison to class 2 and class 3 waveforms. We used a sampling frequency of $4$ kHz to remain consistent with that of the interferometric noise segments.

\begin{figure}
\begin{center}
    \centering
    \includegraphics[width=0.65\linewidth]{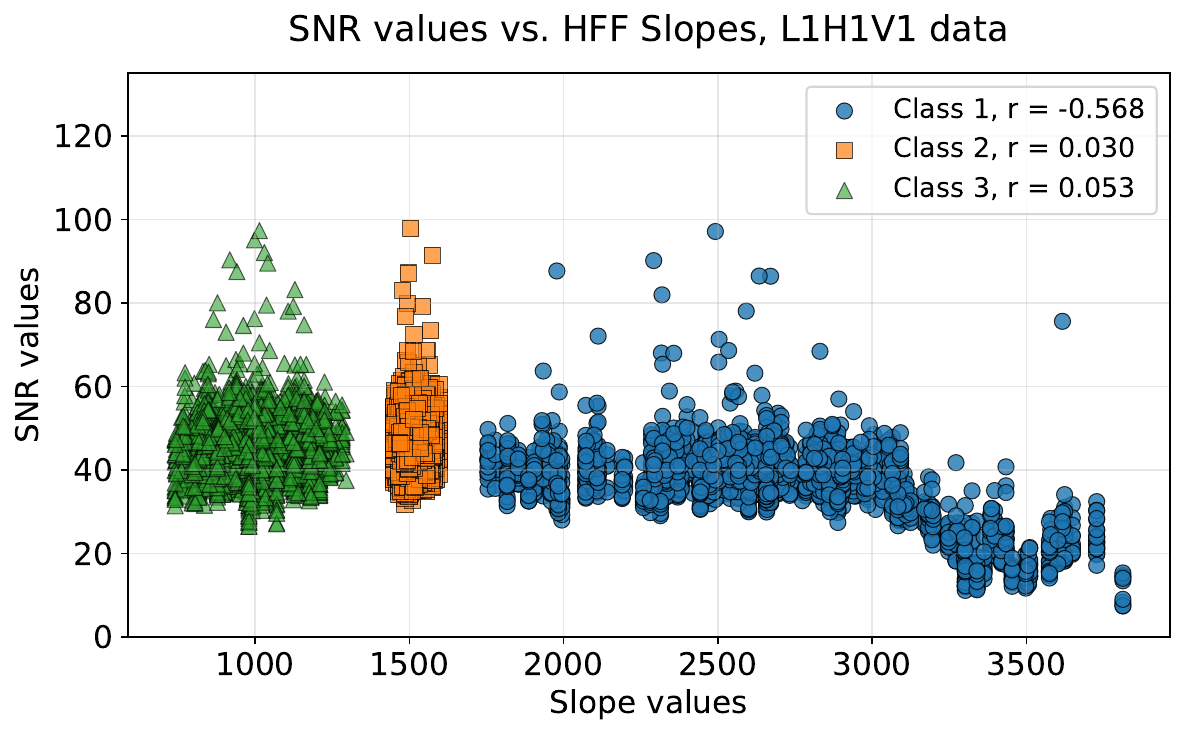}
    \caption{Correlation between HFF slopes and SNR values of samples of noise plus phenomenological waveform. The inverse relationship for class 1 samples arises because these reach the upper frequency limit more quickly, resulting in shorter durations and lower accumulated SNR for a given amplitude.}
    \label{fig:slope_snr_correlation}
\end{center}
\end{figure}

Furthermore, we have a negative correlation between SNR values and HFF slopes for class 1 phenomenological waveforms, as shown in Fig.~\ref{fig:slope_snr_correlation}. This physical effect is a consequence of shorter waveforms having the greatest HFF slopes, and this explains why the systematic shift of the class 1 SNR distribution to lower values, as seen in the bottom panel of Fig.~\ref{fig:wf_phen_distribution}.

Finally, the bottom panel of Fig.~\ref{fig:wf_phen_distribution} shows the distribution of waveform SNR values after being injected into L1, H1, and V1 noise data segments. Operating in the high-SNR regime served solely to facilitate learning of the HFF morphology, even though we discarded window samples with SNR values greater than $100$.

\subsection{ResNet50 architecture}\label{sec:resnet_model}

First introduced by He et al~\cite{kH16}, residual deep networks have shown greater performance results than those reached by shallower standard CNN architectures such as LeNet~\cite{yL98}, AlexNet~\cite{aK17}, and VGG~\cite{kS15}. Before the introduction of residual networks, conventional CNN architectures exhibited a degradation in training and testing accuracy as more and more layers were added (namely, surpassing a few tens of layers). Indeed, it is non-trivial for a stack of nonlinear layers to learn an identity mapping, which is the optimal solution for deeper networks. The residual framework explicitly addresses this problem by introducing skip connections that facilitate identity mappings, thereby solving the degradation problem and enabling the training of very deep CNN architectures.

\begin{figure}[htb]
    \begin{center}
        \includegraphics[height=8.0cm]{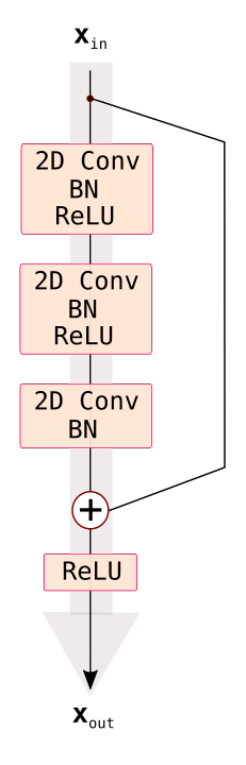} ~~~~~
        \includegraphics[height=8.0cm]{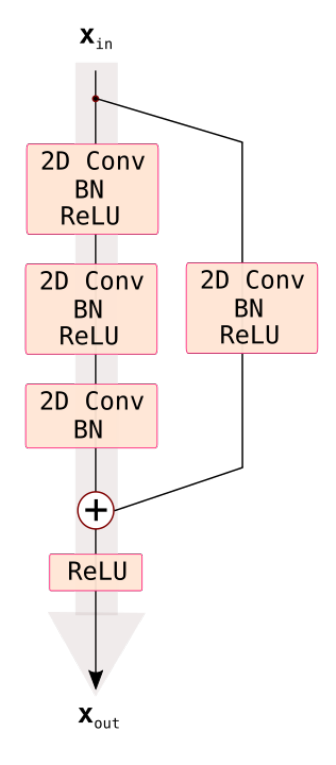}\caption{\label{fig:skip_conn} Basic residual blocks. Left panel: Identity Residual Block, in which its input data is directly added to the output of the third convolutional stack in the main path. Right panel: 2D Convolutional Residual Block, in which its input data passes through a Convolutional layer in the secondary path such that the dimensions before and after the first and third Convolutional layers of the main path, respectively, are the same.}
    \end{center}
\end{figure}

Let be stacked nonlinear layers that input data $\mathbf{x}_{\mathrm{in}}$, in the main pathway of a deep learning architecture. In both the left and right panels of Fig.~\ref{fig:skip_conn}, this main pathway consists of three 2D Convolutional stacks, including Batch Normalization (BN)~\cite{sI15} and a Rectified Linear Unit (ReLU) activation function~\cite{vN10}. Traditionally, the layers would learn the original mapping $\mathcal{M}\left(\mathbf{x}_{\mathrm{in}}\right)$. However, under this new benchmark, the layers learn instead the residual mapping $\mathcal{R}$, which related to the original mapping by:
\begin{eqnarray}
    \mathcal{M}\left(\mathbf{x}_{\mathrm{in}}\right) = p\left(\mathbf{x}_{\mathrm{in}}\right) + \mathcal{R} \left(\mathbf{x}_{\mathrm{in}}, \mathbf{W}\right) ~, \label{eq:skip-conn}
\end{eqnarray}
where $\mathbf{W}$ is the matrix codifying the weights and biases of the layers, and $p(\mathbf{x}_{\mathrm{in}})$ is a function that adds input data (either directly or after passing it through a convolutional stack) to the output of the third convolutional stack in the main pathway, as illustrated in Fig.~\ref{fig:skip_conn}. Consequently, the output of the stacked layers along the main path is as follows:
\begin{eqnarray}
    \mathbf{x}_\mathrm{out} = f_\mathrm{out} \left( \mathcal{M}\left(\mathbf{x}_{\mathrm{in}}\right) \right) ~, \label{eq:resnet-out}
\end{eqnarray}
where $f$ is the last activation function, which we choose beforehand as ReLU. Eq.~(\ref{eq:skip-conn}) is referred to as the ``skip connection'' and, more generally, Eqs.~(\ref{eq:skip-conn}) and~(\ref{eq:resnet-out}) together constitute the ``residual block'' or ``residual unit.''

Function $p(\mathbf{x}_{\mathrm{in}})$, present in Eq.~(\ref{eq:skip-conn}), is a projection used to match the dimensions of the Residual Block's input ($f_\mathrm{in}$ activation, where data $\mathbf{x}_\mathrm{in}$ come from before inputted into the residual block)) and output ($f_\mathrm{out}$ activation), to ensure that operations in the residual block are correctly defined. Let us call this the ``in-out dimensional'' condition. Particularly, in the left panel of Fig.~\ref{fig:skip_conn}, $p$ is simply the identity function, as the dimensions are preserved. However, in the right panel of Fig.~\ref{fig:skip_conn}, $p$ is a linear projection implemented by a 2D Convolution, including BN and ReLU, to adjust the number of feature maps. This follows the design of the original ResNet to handle dimension increases between stages.

Fig.~\ref{fig:skip_conn} shows two basic residual blocks, which are both formed by two paths. A main path is common to both blocks, which have three 2D Convolutional layers, including Batch Normalization (BN) and a ReLU layer. In addition, there is a secondary path describing the skip connection itself, which must satisfy the in-out dimensional condition. On the left panel, we have an identity skip connection, which is applied when $\mathbf{x}_\mathrm{in}$ data inputted by the first Convolutional layer in the main path have the same dimension as the data inputted to the third activation layer. We call this block ``Identity Block'' (Id Block). On the other hand, the right panel shows a 2D Convolutional skip connection, which is applied when the dimensions of the mentioned data are not the same. We call this block ``Convolutional Block'' (Conv Block).

\begin{figure}[htp]
\begin{center}
  \includegraphics[width=13.0cm]{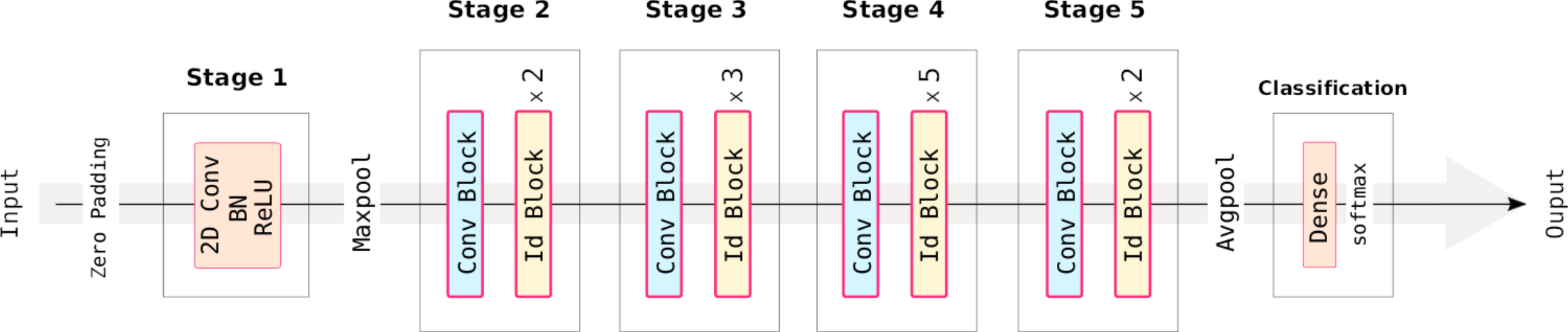}
  \caption{\label{fig:resnet50}ResNet50 architecture. This is formed by five stages in which identity blocks and/or convolutional blocks are applied. Then, a classification stage with a dense (fully connected) layer is applied, resulting in three output probabilistic scores per class, because of the application of a final softmax activation function. After stages 1 and 5, a max pooling and an average pooling, respectively, are applied to reduce the system resources. At the beginning, a zero padding is applied to the input data.}
\end{center}
\end{figure}

Having already explained the structure of the basic building residual blocks, we present in Fig.~\ref{fig:resnet50} the general architecture used in this work, namely ResNet50. This deep neural network has 50 parameter layers in the main path, and these layers are distributed in 5 stages in addition to the classification. Id Block and Conv Block, as shown in Fig.~\ref{fig:skip_conn}, have three parameter layers; therefore, we have 48 main parameter layers from stage 2 to stage 5, which are added to one layer in stage 1 and another layer in the classification.

\begin{table}[htp]
\caption{\label{tab:resnet50}Summary of the settings in the ResNet50 architecture that we used in this work. The total number of layers in the main path is $50$. In addition, the total number of parameters is $23,593,859$, occupying $90$MB in memory.}
\captionsetup{width=1.0\textwidth}
\centering
\begin{tabular}{|c|c|c|c|c|}
 \hline
  \textbf{Stage} & $\begin{array}{c} \textbf{Kernels dim }; \\ \textbf{$\#$ filters (per block)} \end{array}$ & $\begin{array}{c} \textbf{$\#$ layers in} \\ \textbf{main path} \end{array}$ & $\begin{array}{c} \textbf{$\#$ layers in} \\ \textbf{skip connect.} \end{array}$ & \textbf{$\#$ parameters} \\ 
  \hline
  \hline
  1 & $\begin{array}{ccc}
      ~ & 7 \times 7 ~; & 64
  \end{array}$ & 1 & --- & 9,728 \\
  \hline
  2 & $\begin{array}{ccc}
      ~ & 1 \times 1 ~; & 64\\
      ~ & 3 \times 3 ~; & 64\\
      ~ & 1 \times 1 ~; & 256
 \end{array}$ & 9 & 1 & 220,032 \\
  \hline
  3 & $\begin{array}{ccc}
      ~ & 1 \times 1~; & 128\\
      ~ & 3 \times 3~; & 128\\
      ~ & 1 \times 1~; & 512
 \end{array}$ & 21 & 1 & 1,230,336 \\
  \hline
  4 & $\begin{array}{ccc}
      ~ & 1 \times 1~; & 256\\
      ~ & 3 \times 3~; & 256\\
      ~ & 1 \times 1~; & 1024
 \end{array}$ & 18 & 1 & 7,129,088 \\
  \hline
  5 & $\begin{array}{ccc}
      ~ & 1 \times 1~; & 512\\
      ~ & 3 \times 3~; & 512\\
      ~ & 1 \times 1~; & 2048
 \end{array}$ & 9 & 1 & 14,998,528 \\
  \hline
  Classification & -- & 1 & --- & 6,147 \\
  \hline
 \end{tabular}
\end{table}

ResNet is a well-known (and well-tested) architecture in the Machine Learning community~\cite{mS22}, and the ResNet50 benchmark has a predefined configuration of hyperparameters. For this study, we considered that predefined configuration, which is shown in Table~\ref{tab:resnet50}. In addition, we consider a zero padding of dimension $3 \times 3$ after inputting data, a max pooling of $3 \times 3$ with stride $2$ after stage 1, and an averaging pooling of $2 \times 2$ after stage 5 of the ResNet50 model. Before the optimization, we initialized the weights of the ResNet50 by applying the Glorot uniform initializer~\cite{xG10}.

The ResNet50 implementation was carried out using the TensorFlow library~\cite{TensorFlow} with the Keras interface~\cite{Keras}.

\subsection{ResNet50 application}\label{sec:resnet_apply}

As a starting point, the input dataset of noise plus phenomenological waveforms was divided into two subsets: the testing set ($30\%$) and the optimization set ($70\%$). For training the ResNet50, we only drew on the optimization set.

Before applying global optimization (i.e., multiple trainings to find the optimal model), we perform a sanity check by running a single training. Given a set of hyperparameters, training refers to the iterative process in which the ResNet50 parameters are fitted in a finite number of epochs (one epoch is the time in which the entire training dataset of $N_\mathrm{train}$ samples passes once through ResNet50). Then, we randomly split the optimization set into two subsets: one for the training ($90\%$) and the other for validation ($10\%$). Validation is a mini-test to monitor the tranining in each epoch. We set an Adaptive Moment Estimation (Adam optimizer)~\cite{dK14} with a learning rate of $\alpha=0.0005$, batch size of $b_\mathrm{s}=50$, and $n_\mathrm{e}=40$ epochs. As detailed in subsection~\ref{sec:results_phe_wf}, this check was useful to set an appropriately early stopping strategy that was used in the training.

After the above sanity check, the next step is to optimize the ResNet50 by a global procedure. For this, sequential hyperparameter tunings (by repetitive training-validation processes) are performed. In this work, we follow a \texttt{GridSearchCV} strategy, which consists of repeatedly applying k-fold cross validation (CV)~\cite{Th09} to explore the performance of ResNet50 for different combinations of hyperparameters. These combinations were drawn from predefined dictionaries, and after the exploration, the best combination was selected. It should be stressed that, apart from model selection, k-fold CV allows us to address the inherent stochasticity of models by applying resampling and, therefore, to alleviate the artificiality introduced by class-balanced input datasets.

For each hyperparameter combination, we applied k-fold CV. This process first involves randomly splitting the training set into $k$ non-overlapping subsets (folds). Then, for each fold $i$ ($i=1,2,...,k$), the model is trained on the other $k-1$ folds and validated on the $i-$th fold. The model's performance is finally summarized by computing the average validation loss and accuracy across all $k$ trials. It is important to note that, unlike a single training process, validation in k-fold CV is performed after training on each fold is complete, not epoch-by-epoch.

As it was shown in Table~\ref{tab:resnet50}, the ResNet50 has several hyperparameter values that are specific to its architecture by design, i.e., they are predefined: number of layers, kernel dimensions, and number of neurons in specific layers, among others. We decided to maintain these hyperparameters as fixed. However, there are other relevant hyperparameters associated with training that we tuned with \texttt{GridSearchCV}, namely: the batch size $b_\mathrm{s}$, the number of training epochs $n_\mathrm{e}$, and its learning rate $\alpha$ and momentum $m$. Moreover, to take advantage of this tuning procedure, we included another element to tune, which, more than a single hyperparameter, is a process, namely the optimizer. While optimizers like Adam are often an effective default choice, their performance can depend on the problem. Therefore, to ensure a robust optimization of our model, we systematically evaluated a set of different optimizers as a key part of our hyperparameter tuning strategy. For a recent review of optimizers, see~\cite{rA23}.

Our global optimization procedure consisted of three \texttt{GridSearchCV} tunings. In each tuning, we monitor the performance of the ResNet50 by combining some hyperparameter values from one or more dictionaries. The best model in a single tuning is the one with the the greatest mean validation accuracy. The hyperparameter dictionaries are:

\begin{itemize}
    \item{\bf Tuning 1 dictionary}, for optimizer.
    
    \vspace{.15cm}
    \item[]$Opt=$ SGD, RMSprop, Adagrad, Adadelta, Adam, Adamax, Nadam.
    
    \vspace{.15cm}
    \item{\bf Tuning 2 dictionary}, for batch size and No. of epochs.
    
    \vspace{.15cm}
    \item[]$b_\mathrm{s}=25, 50, 75, 100.$ ~;~ $n_\mathrm{e}=20,30,40$.
    
    \vspace{.15cm}
    \item{\bf Tuning 2 dictionary}, for learning rate and momentum.
    
    \vspace{.15cm}
    \item[]$\alpha=0.0005, 0.001, 0.01, 0.1.$ ~;~ $m=0.0, 0.2, 0.4, 0.6.$
\end{itemize}

Before tuning 1, we set the initial hyperparameters $b_\mathrm{s}=50$, $n_\mathrm{e}=40$, $\alpha=0.0005$, and $m=0$. We also chose $k=5$ for CV, which gives us $n_\mathrm{fits}{}^{(1)} = 5 \times 7 = 35$ fits for tuning 1, $n_\mathrm{fits}{}^{(2)} = 5 \times 4 \times 3 = 60$ fits for tuning 2, and $n_\mathrm{fits}{}^{(3)} = 5 \times 4 \times 4 = 80$ fits for tuning 3. After completing the global optimization, we trained and tested the optimized ResNet50 on the full optimization set. Finally, our methodology includes a final test using data of noise plus numerical simulated waveforms.

\section{Results and Discussion}\label{sec:results}

\subsection{Analysis with phenomenological waveforms}\label{sec:results_phe_wf}

\subsubsection{Sanity check: Single-Training process}

\begin{figure}[ht]
\begin{center}
  \includegraphics[width=6.8cm]{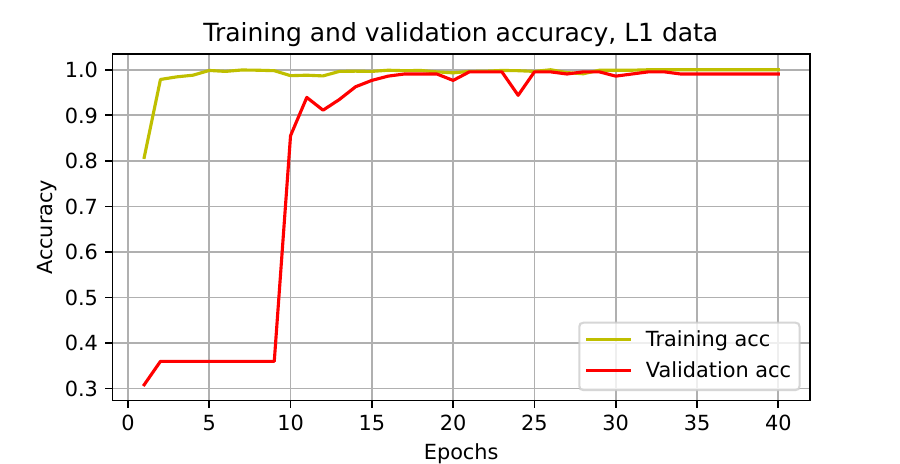}
  \hspace{-.8cm}
  \includegraphics[width=6.8cm]{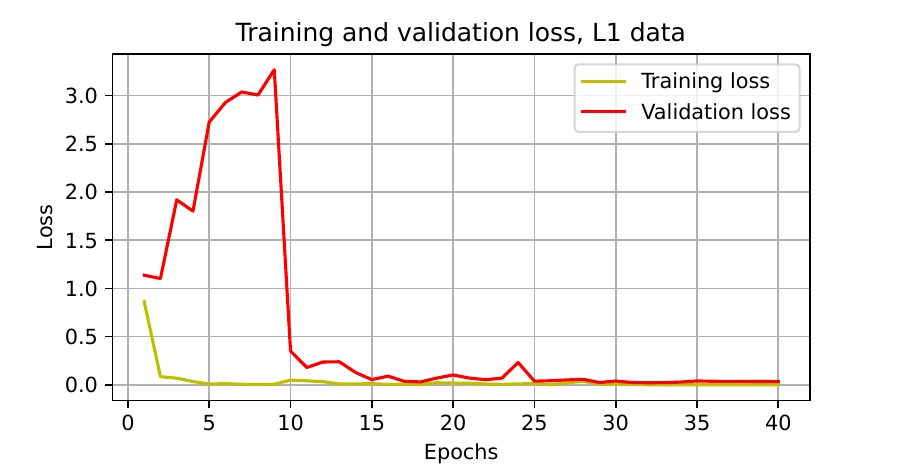} \\
  \includegraphics[width=6.8cm]{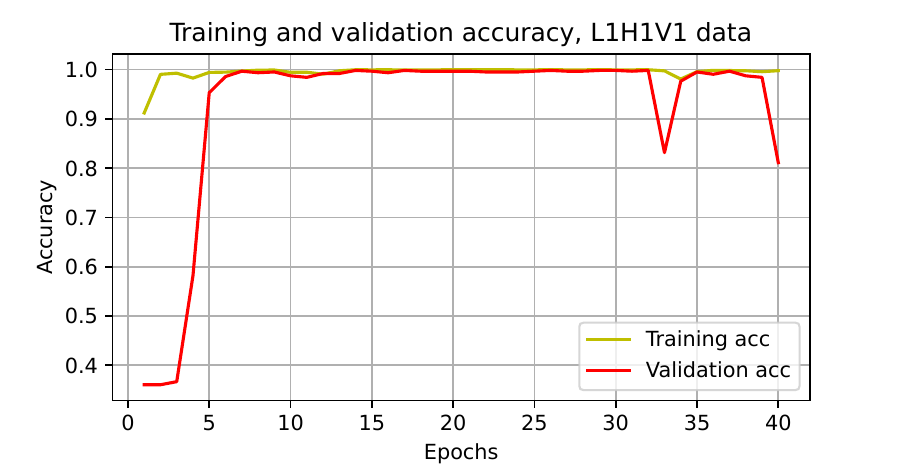}
  \hspace{-.8cm}
  \includegraphics[width=6.8cm]{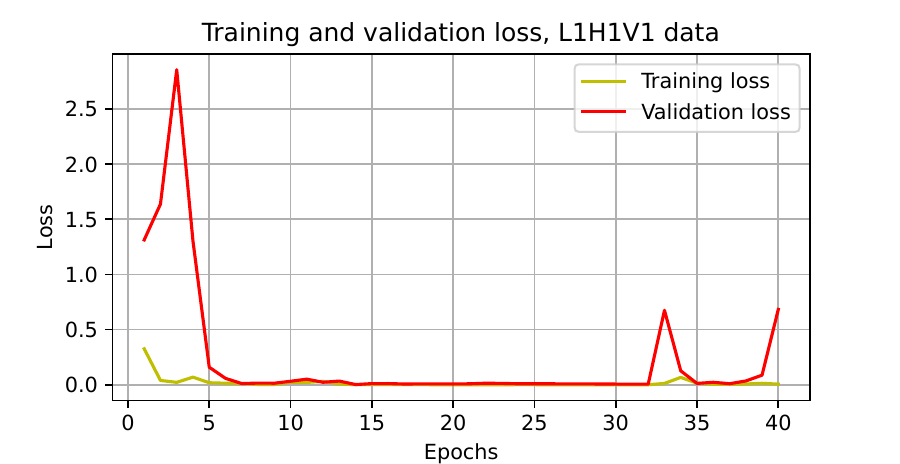}
  \caption{\label{fig:single_learning}Two representative single training process with image samples of phenomenological waveforms in L1 data (upper panel) and L1, H1, and V1 combined data (bottom panel). Accuracy (left panel) and categorical cross-entropy, i.e., the loss function (right panel), are shown during the training and validation processes.}
\end{center}
\end{figure}

Fig.~\ref{fig:single_learning} shows representative results from a single-training process using L1 data (upper panel) and combined L1, H1, and V1 data (lower panel). The results indicate no persistent perturbations arising from batch training, which confirms the stability of our estimates. In addition, we see that good performances are reaching very fast after the 5th to 15th epoch, which is good news for minimizing system resources.

However, the performance of the ResNet50 model does not necessarily improve with additional epochs. We can see from Fig.~\ref{fig:single_learning} that from the middle to the later epochs the improvement is minimal; even short detrimental effects to the performance can suddenly appear, as can be seen in the case of learning from L1, H1, V1 data. Then, it may happen that once the training is finished, the performance is not the best in comparison to previous epochs. To address this issue (as discussed in section~\ref{sec:final_traintest_phen}), we implemented a  \texttt{ModelCheckpoint} callback that saves, after training, the model achieving the maximum validation accuracy across all epochs.

\subsubsection{\texttt{GridSearchCV} tunings}

As noted above, the ResNet50 was optimized through three sequential hyperparameter tunings. For each tuning, we takes one value from a hyperparameter dictionary. Therefore, given our dictionary entries (detailed at the end of section~\ref{sec:resnet_apply}), we have single values in the first tuning and pairs of values in the second and third tunings. The results are shown in Table~\ref{tab:hyp_tuning}, in which we can immediately observe that, consecutively, (5-fold CV) mean accuracy increases as we advance to the next tuning. The mean fitting time was similar for tunings 1 and 2, with tuning 2 being slightly shorter. Here we used combined data from L1, H1, and V1 detectors.

In terms of computational resources, this three-stage tuning was by far the most demanding task. It was possible using a NVIDIA A100 Tensor Core GPU Architecture, available by upgrading Google Colaboratory, with a $83.5$ GB system RAM, a $40.0$ GB GPU RAM, and a disk space of $78.2$ GB. The total mean computing time (TMCT) was obtained by summing the values in the most-right column of the Table~\ref{tab:hyp_tuning}:
\begin{eqnarray}
  \text{TMCT}
  &=& \sum_{i=1}^{3} n_\mathrm{fits}{}^i \times \overline{t}_\mathrm{fit}{}^i \nonumber \\
  &=& (35 \times 115.1)~\text{s} + (60 \times 110.8)~\text{s} + (80 \times 101.5)~\text{s} \nonumber \\
  &\approx& 18796.5~\text{s} \approx 5.221~\text{hr} ~. \nonumber\\
\end{eqnarray}

\begin{table}[t]
\caption{\label{tab:hyp_tuning}Results of the \texttt{GridSearchCV tunings} applied to the training set, with L1, H1, and V1 combined data and phenomenological waveforms. The chosen model, sequentially, is better than the previous. We finally store the set of hyperparameters obtained in tuning 3, to be used in a final training.}
\captionsetup{width=1.0\textwidth}
\centering
\begin{tabular}{|c|c|c|c|c|c|} 
 \hline
  \textbf{Tuning} & \textbf{Best hyperparam} & $\begin{array}{c} \textbf{Mean acc.} \\ \pm ~ \textbf{std} \end{array}$ & \textbf{$\#$ fits $n_\mathrm{fits}$} & $\begin{array}{c} \textbf{Mean fit} \\ \textbf{time } \overline{t}_\mathrm{fit} \textbf{(s)}\end{array}$ \\
 \hline\hline
 1 & RMSprop & $0.9943 \pm 0.000884$ & $35$ & $115.1$ \\  
 \hline 
 2 & $b_\mathrm{s}=75, n_\mathrm{e}=40$ & $0.9960 \pm 0.001511 $ & $60$ & $110.8$ \\
 \hline
 3 & $\alpha=0.01, m=0.6$ & $0.9965 \pm 0.001550$ & $80$ & $101.5$\\
 \hline
 \end{tabular}
\end{table}

In summary, the \texttt{GridSearchCV} procedure identified the following optimal hyperparameter configuration: RMSprop optimizer with learning rate $\alpha=0.01$ and momentum $0.6$, batch size $b_\mathrm{s}=75$, and number of epochs $n_\mathrm{e}=40$.

\subsubsection{\label{sec:final_traintest_phen}Optimized ResNet50 model: Train and test}

\begin{figure}[htp]
\begin{center}
  \includegraphics[width=6.5cm]{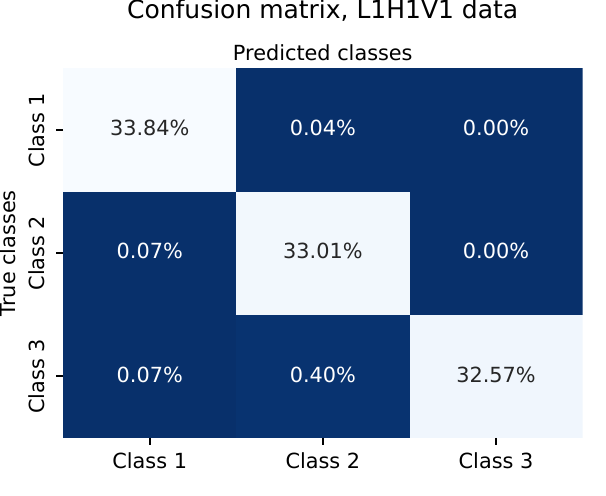}
  \caption{\label{fig:confu_matrix_phen_test}Confusion matrix obtained from testing the optimized ResNet50 model with samples with phenomenological waveforms in L1, H1, and V1 noise combined data. The matrix uses overall normalization, where the sum of all elements is $100\%$.}
\end{center}
\end{figure}

We next trained and tested the optimized model using the combined L1, H1, and V1 datasets.y In this case, the validation is still useful because, by implementing a \texttt{ModelCheckpoint} callback, we will save the model of maximum validation accuracy observed along all epochs once the training is finished. The resulting metrics are summarized in Table~\ref{tab:phenom_testtrain}, and Fig.~\ref{fig:confu_matrix_phen_test} presents the corresponding confusion matrix, demonstrating excellent classification performance across all three classes.

\begin{table}[htp]
\caption{\label{tab:phenom_testtrain}Metrics obtained after training, validating, and testing the ResNet50 using the best combination of hyperparameters found in \texttt{GridSearchCV} tunings. We drew on image samples with noise from detectors L1, H1, and V1 and injected phenomenological CCSN waveforms. We implemented a \texttt{ModelChekpoint} callback to choose the model with maximum validation accuracy across all training epochs.}
\captionsetup{width=1.0\textwidth}
\centering
\begin{tabular}{|c|c|c|c|c|} 
 \hline
  \textbf{Process} & \textbf{Accuracy} & \textbf{Loss} \\
 \hline\hline
 Train & $0.9990$ & $0.002400$ \\
 \hline
 Validation & $0.9984$ & $0.01700$ \\
 \hline
 Test & $0.9942$ & $0.03036$ \\
 \hline 
 \end{tabular}
\end{table}

Training was computationally efficient. We used an NVIDIA T4 Tensor Core GPU freely available from Google Colaboratory, with $12.7$GB system RAM, $16.00$GB GPU RAM, and a disk space of $78.2$GB. The total computing time was $344$ s $\approx 5.73$ min. Testing required only about $5$ s.

\subsection{Analysis with numerical simulated waveforms}\label{sec:results_genrel_wf}

\subsubsection{Input image samples}

We considered three different distances from the emitting numerical simulated CCSN, namely $1$ kpc, $5$ kpc, and $10$ kpc, corresponding to typical Galactic scales. Given a detector noise, we drew on $100$ image samples per class per distance, resulting in a total of $100 \times 3 \times 3 = 900$ TF image samples per detector. GW signals come from three CCSN multidimensional simulations: Andresen 2019 $m15nr~h+$ (3D model, class 1), Morozova 2018$M13\_$SFHo~$h+$ (2D model, class 2), and Cerd\'a-Dur\'an 2013 $fiducial~h+$ (2D model, class 3).

Fig.~\ref{fig:sample_images_genrel} shows some representative image samples, using L1 data, that were inputted to the trained ResNet50 classifier: one sample per class per distance. At a fixed distance, the samples exhibit different SNR values and varying HFF visibility. This happens for two physical reasons: the noise realization for each sample is not the same, and GW signals come from different CCSN models. For a given model (i.e., a single class), increasing the distance reduces the visibility of the HFF feature. This is expected because, for a given background noise data, the greater the distance from the emitting source, the smaller the strain magnitude of the signal.

\subsubsection{SNR distribution of samples}

\begin{landscape}
\begin{figure}[htp]
\begin{center}
  \includegraphics[width=6.0cm]{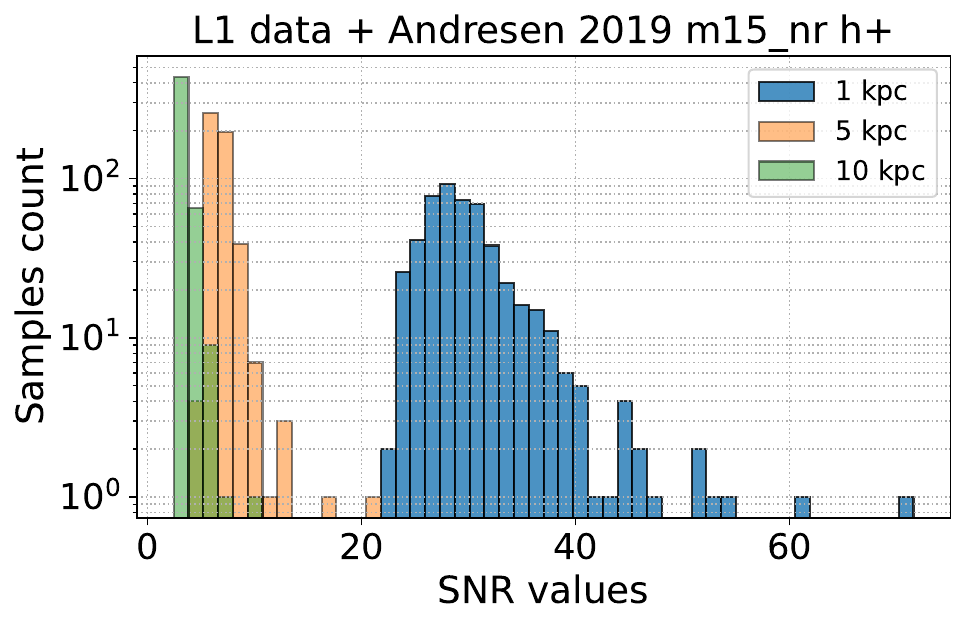}
  \includegraphics[width=6.0cm]{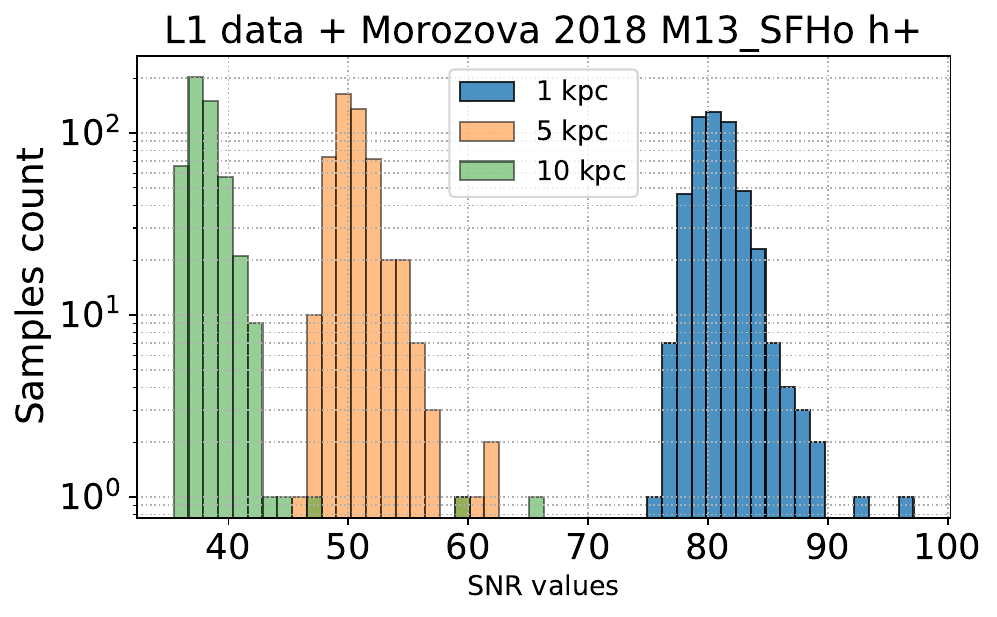}
  \includegraphics[width=6.0cm]{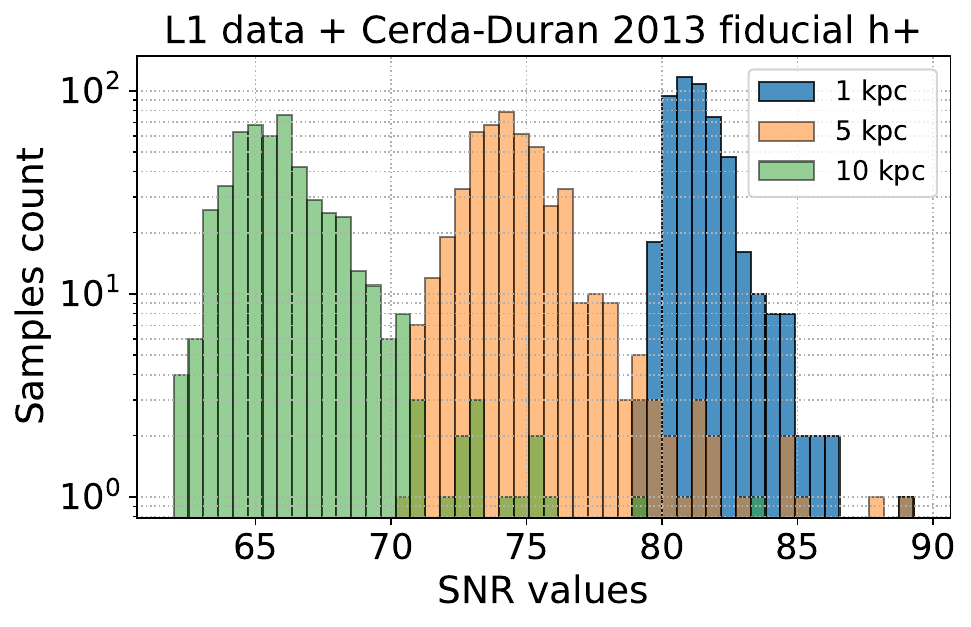}
  \\
  \includegraphics[width=6.0cm]{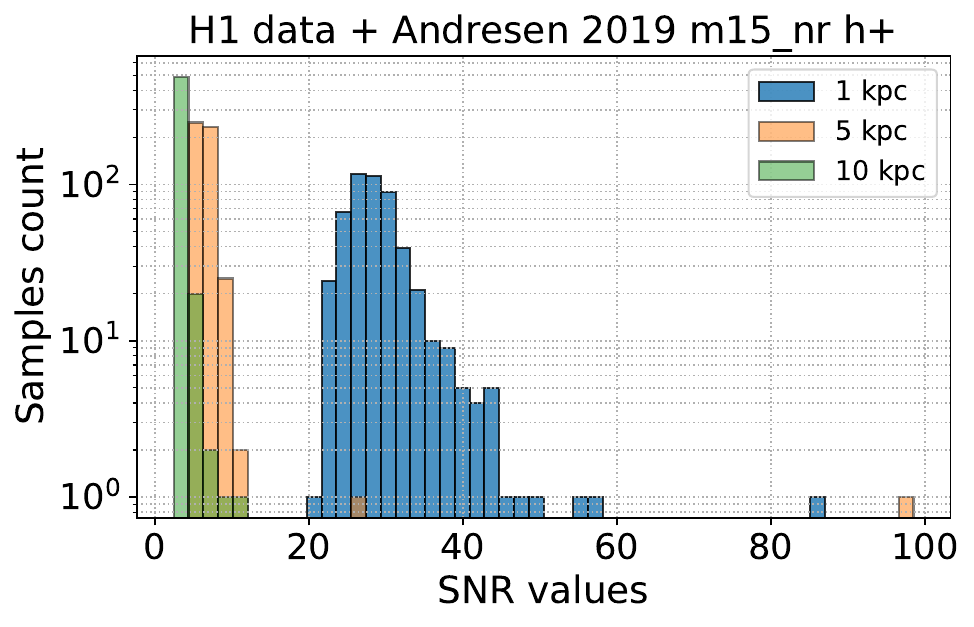}
  \includegraphics[width=6.0cm]{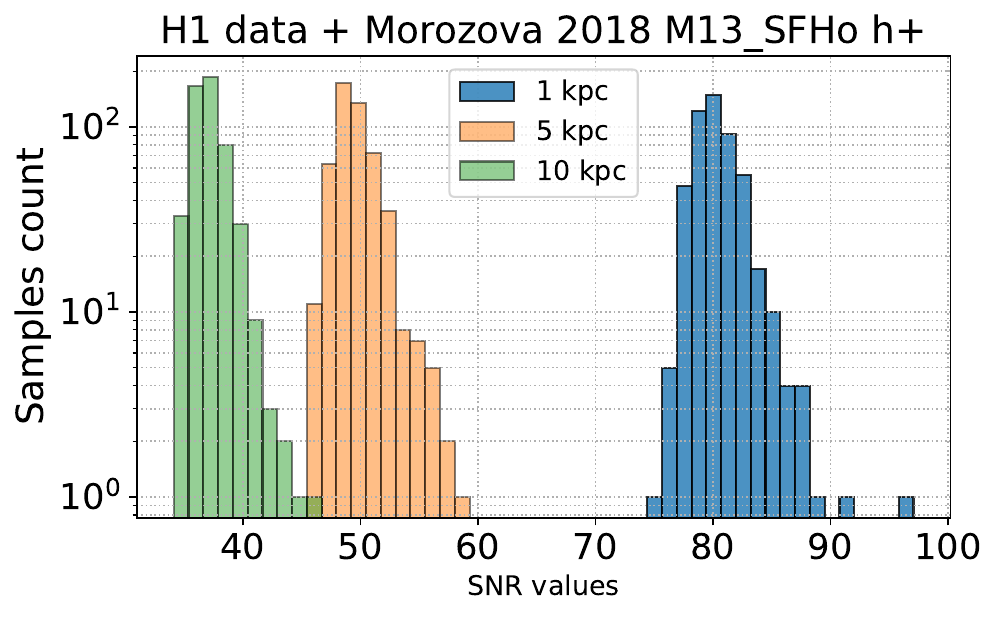}
  \includegraphics[width=6.0cm]{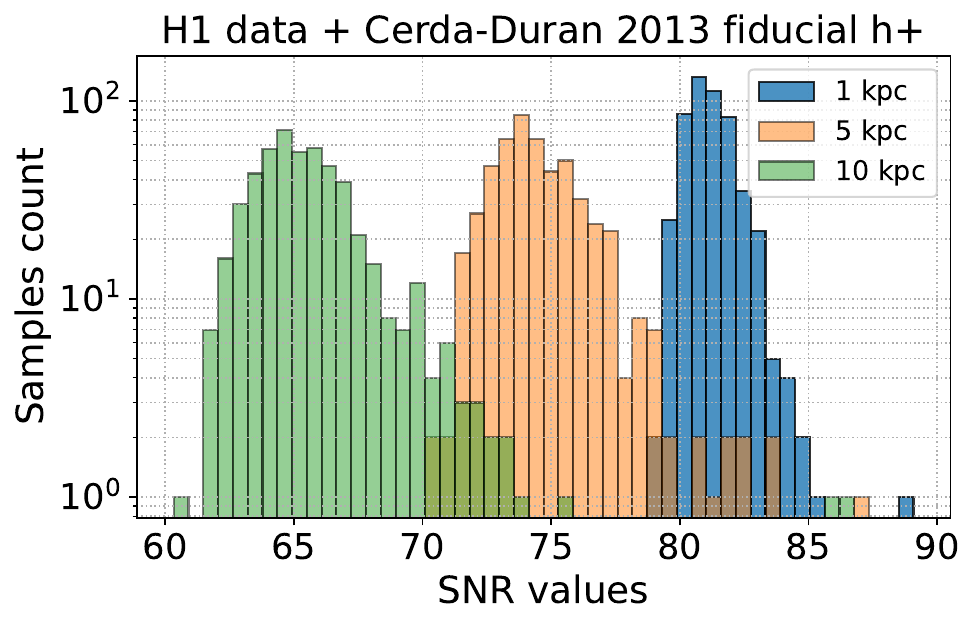}
  \\
  \includegraphics[width=6.0cm]{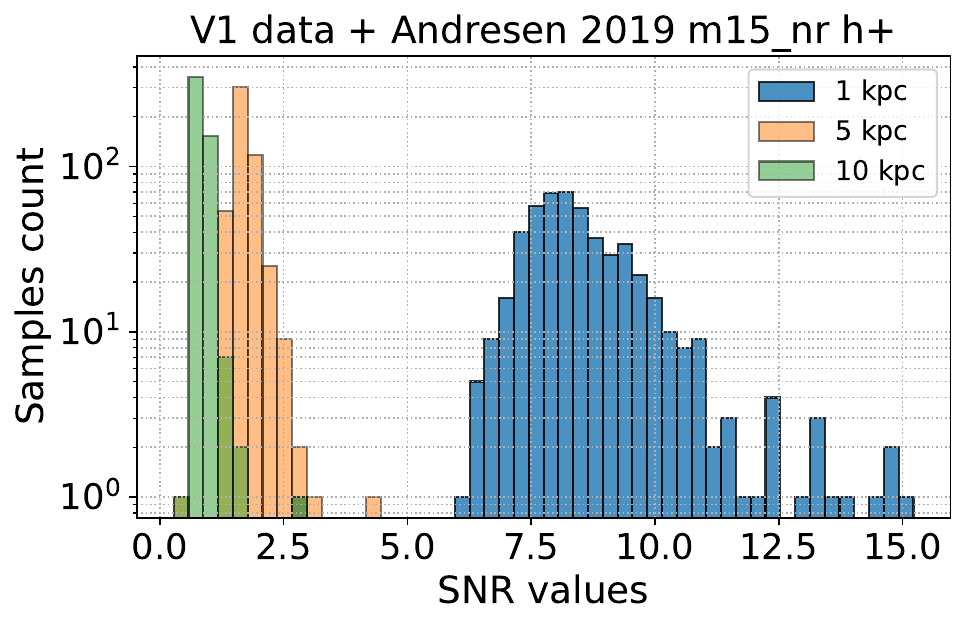}
  \includegraphics[width=6.0cm]{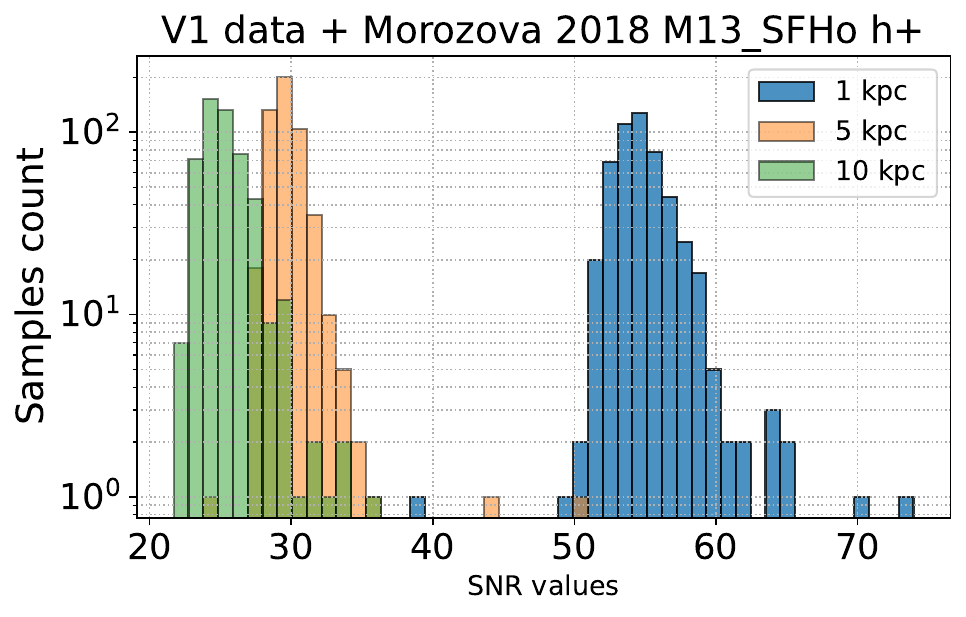}
  \includegraphics[width=6.0cm]{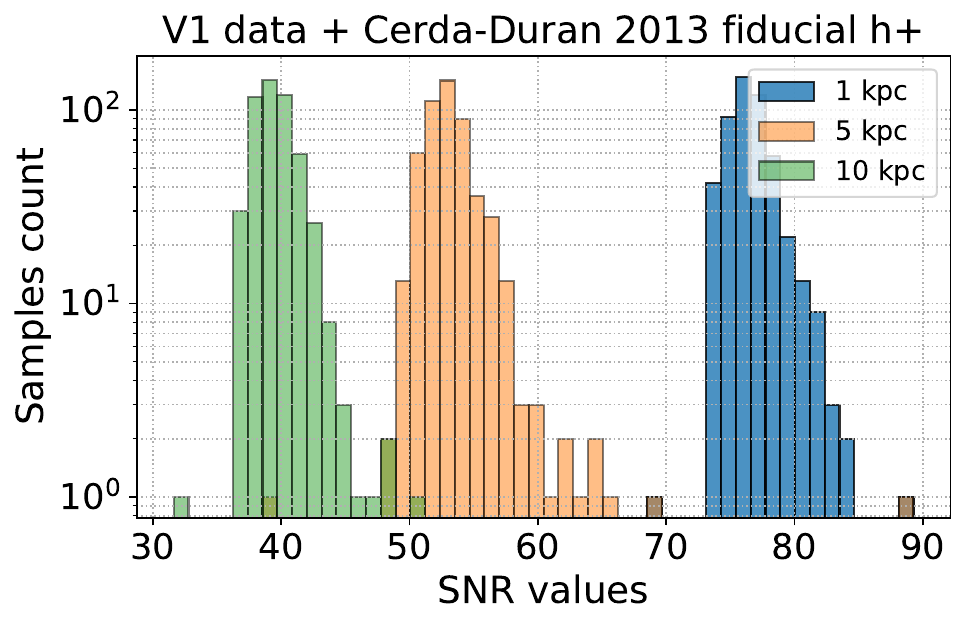}
  \caption{\label{fig:wf_snr_gr_distribution}SNR distributions of TF image samples containing interferometric noise plus numerical simulated CCSN GW models. Here, we identify three trends. Firstly, the greater the distance, the greater the occurrence of samples with lower SNR values than those at smaller distances. Besides, for a specific CCSN GW model and specific noise data, SNR histograms at larger distances tend to be more overlapped in a region of low SNR values. Finally, for a single CCSN GW model, we have that the relative dispersion of SNR distributions at different distances does not vary significantly with respect to the background noise. These statistical properties are crucial to interpret the classification performance of the ResNet50, shown in Fig.~\ref{fig:confu_matrix_genrel_test}. For instance, the strong overlap between distributions at $5$ kpc and $10$ kpc for the Andresen (class 1) and Morozova (class 2) samples with V1 data, underlies the increased misclassification rates at these distances. In contrast, the maintained separability of the Cerd\'a-Dur\'an (class 3) populations across distances correlates with its robust classification accuracy.}
\end{center}
\end{figure}
\end{landscape}

To provide a more comprehensive overview of the injected numerically simulated CCSN GW signals, Fig.~\ref{fig:wf_snr_gr_distribution} shows the distributions of their SNR values. These distributions are separated by classes (i.e., the CCSN emitting models) and distances ($1$ kpc, $5$ kpc, and $10$  kpc), considering data from the three detectors.

\begin{figure}[htp]
\begin{center}
  \includegraphics[width=12.9cm]{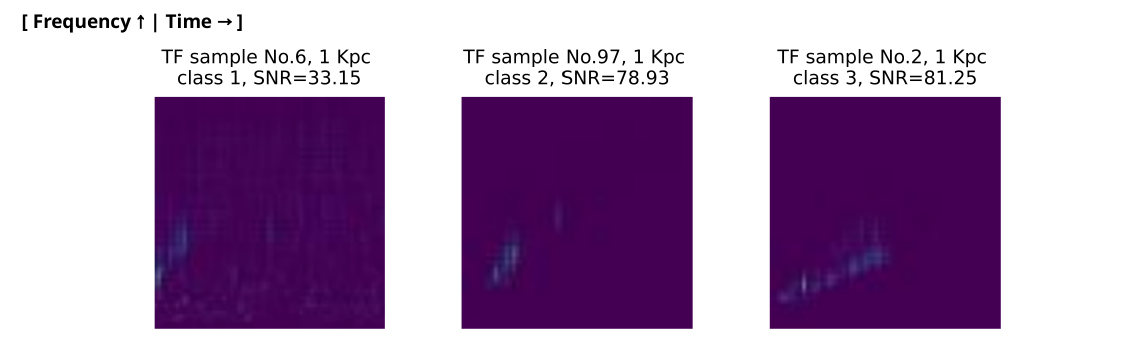} \\
  \includegraphics[width=12.9cm]{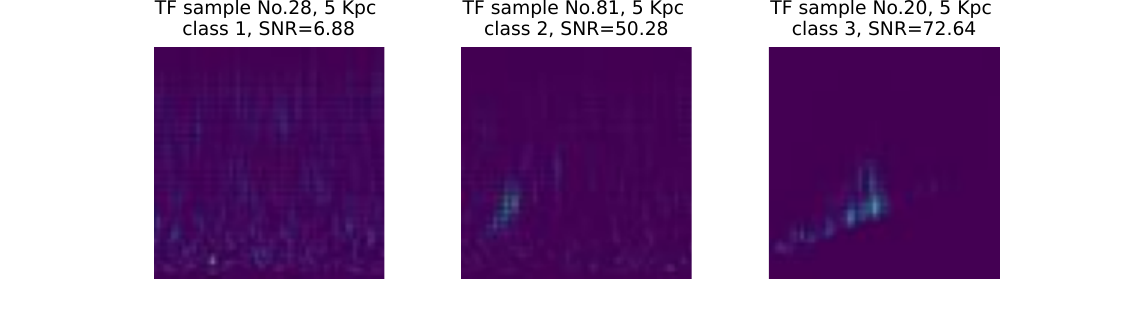} \\
  \includegraphics[width=12.9cm]{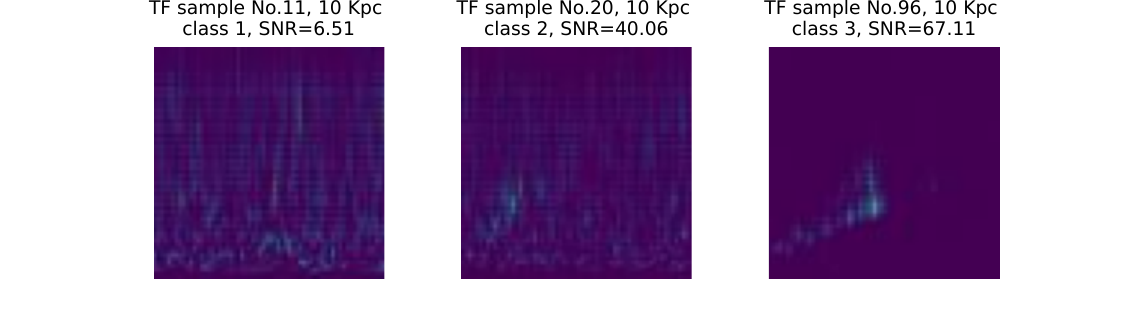}
  \caption{\label{fig:sample_images_genrel}Representative TF scalogram image samples after the RGB pixelization, showing the actual 64×64×3 pixel arrays input to the ResNet50 classifier. These consist of L1 noise plus numerical simulated CCSN GW signals. Arrows indicate the correspondence between image dimensions and physical quantities: horizontal = time, vertical = frequency. The visibility of the HFF varies depending on the noise realization and the nature of (and distance from) the emitting CCSN. The ResNet50 processes these images as array of pixels without awareness of their physical interpretation.}
\end{center}
\end{figure}

Identifying the main trends in the SNR histograms provides insight into the behavior of the ResNet50 predictions. We firstly see that the greater the distance, count of samples tend to be displaced at lower SNR values than those at shorter distances. This behavior is consistent with Fig.~\ref{fig:sample_images_genrel}: for the Morozova and Andresen models, the HFF becomes less visible at greater distances, meaning that noise realization dominates over the appearing of those GW signals. The TF image samples shown in the figure are representative; through a random exploration of dozens of samples, we found that samples with a lower (or even null) visibility of the HFF are more common among those generated at larger distances, both for Morozova and Andresen models. The case of the Cerd\'a-Dur\'an GW signal is different because, even though we observe the same statistical behavior of low SNR values at greater distances ($5$ kpc and $10$  kpc), the HFF is still visible. After all, the magnitude of the GW signal is large enough to be unaffected by the noise amplitude.

Another observable trend is the overlap among SNR distributions at different distances, considering a specific model injected into specific interferometric noise data (either L1, H1, or V1). Here we observe that, as the GW model has a smaller magnitude strain and/or the noise has a smaller sensitivity, histograms at larger distances ($5$ kpc and $10$ kpc) tend to be very overlapped in the low SNR regime. This behavior comes from the fact that a GW strain scales as $1/\text{distance}$, giving a very small variation at larger distances. This overlapping trend is really clear with the Andresen model being injected into the noise data of the three detectors, and with the Morozova model being injected into the V1 noise data
. The SNR distribution at $1$ kpc has little to no overlap with the SNR distributions for the other two distances. All SNR distributions associated with the Cerd\'a-Dur\'an GW model, across all chosen distances and detectors, exhibit minimal overlap.

Finally, let us focus on the distance relative dispersion of the distributions. If we consider the three left plots in Fig.~\ref{fig:wf_snr_gr_distribution} for Andresen GW model, we can see that at $1$ kpc the SNR distribution is approximately $3$ times more dispersed than distributions at $5$ kpc and $10$ kpc. This is not the case for the Morozova GW model and the Cerd\'a-Dur\'an GW model, in which their SNR distributions, for all distances, have similar dispersions (of just over a dozen SNR units). The relative dispersion of the SNR distributions of a model at different distances does not vary with respect to the interferometric noise. Moreover, given a distance and a single-interferometer data, SNR distribution is unique to its particular CCSN GW model. Hence, differences among the SNR distributions of distinct models suggest that some classes are intrinsically easier (or harder) to recognize than others, and this aspect will be more evident when working with larger distances.

A clarification: The resulting high SNR values shown in Fig.~\ref{fig:wf_snr_gr_distribution} are a deliberate feature of our study design, consistent with the $1/\text{distance}$ scaling of gravitational wave amplitude and standard methodologies used in proofs-of-concept in GW signals detectability (e.g.,~\cite{mS21} and~\cite{hC14}). That is to say, this choice allows us to first validate our classification methodology in a high-fidelity regime of rare, nearby events before assessing its performance, in future work, on the more challenging signals of low SNR values expected from extragalactic distances.

\subsubsection{Testing the optimized ResNet50 model}\label{sec:test_num_wf}

\begin{figure*}[htb]
\begin{center}
  \includegraphics[width=3.9cm]{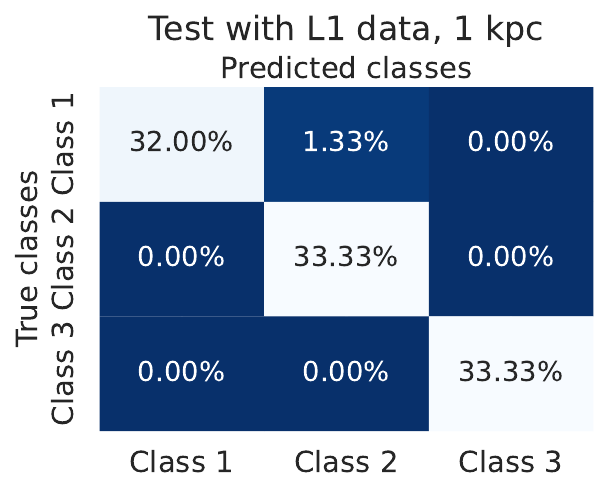} ~
  \includegraphics[width=3.9cm]{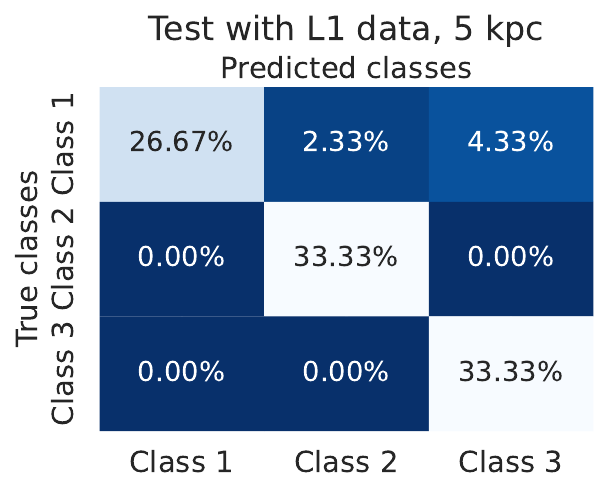} ~
  \includegraphics[width=3.9cm]{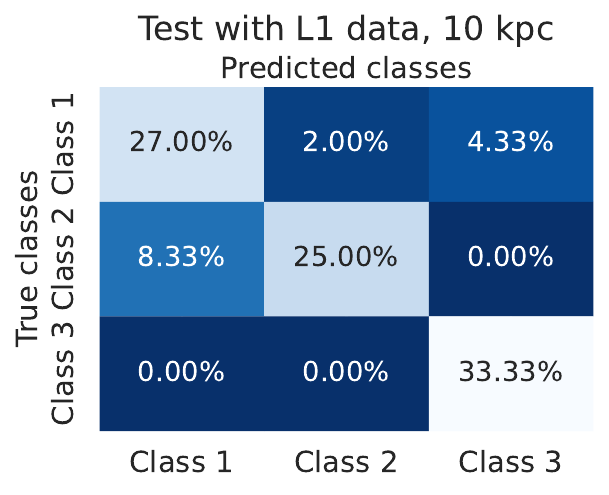}
  \vspace{.2cm} \\ 
  \includegraphics[width=3.9cm]{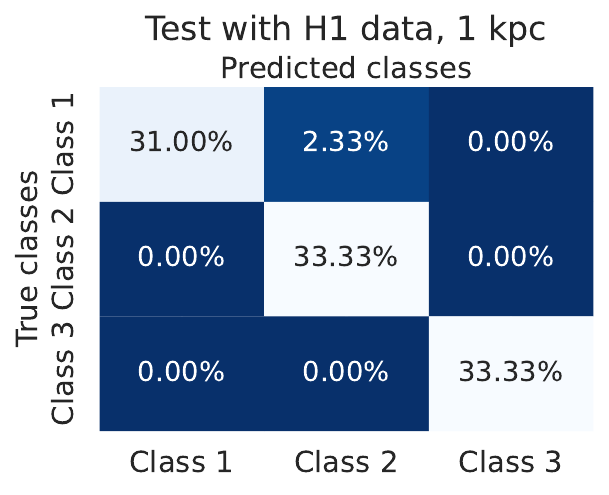} ~
  \includegraphics[width=3.9cm]{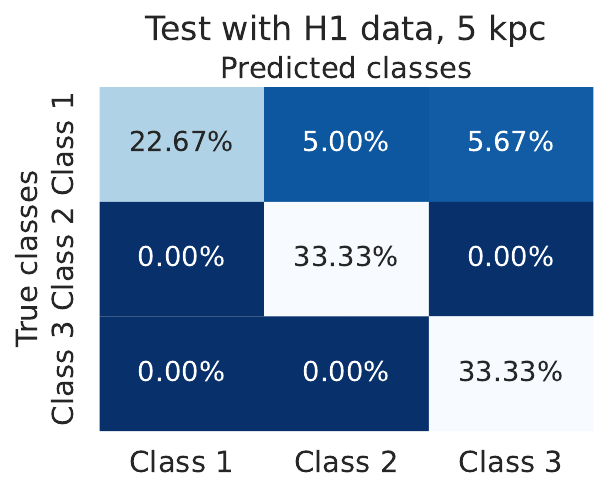} ~
  \includegraphics[width=3.9cm]{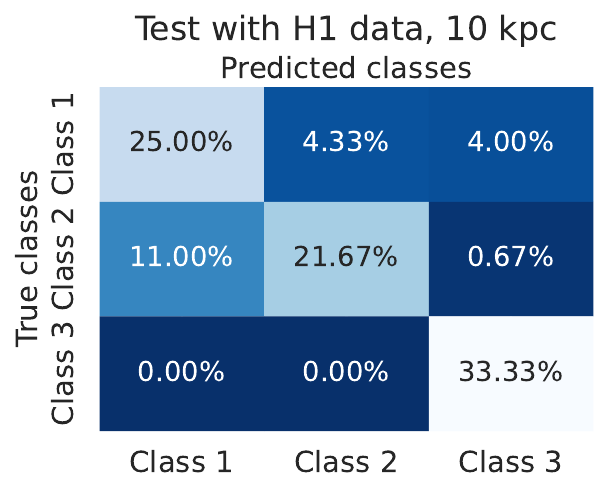}
  \vspace{.2cm} \\
  \includegraphics[width=3.9cm]{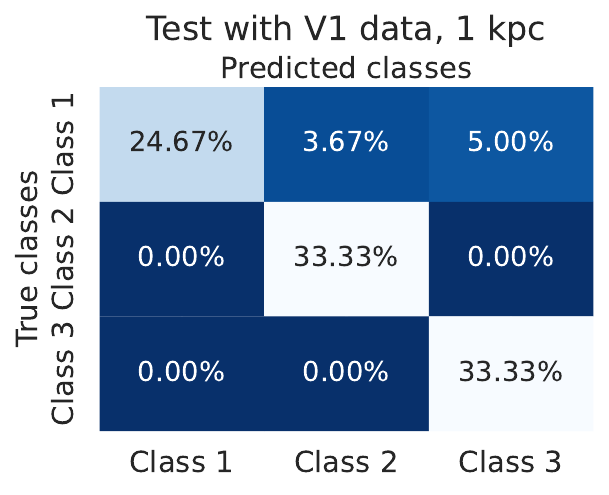} ~
  \includegraphics[width=3.9cm]{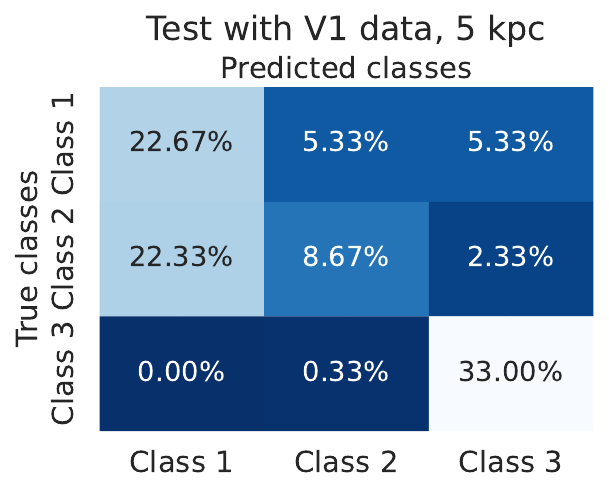} ~
  \includegraphics[width=3.9cm]{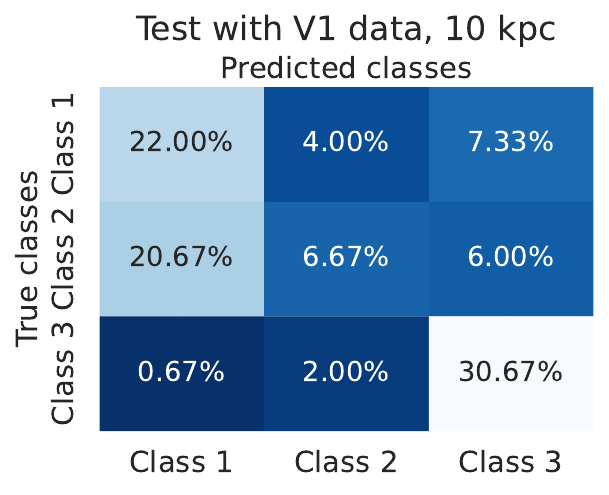}
  \vspace{.2cm}
  \caption{\label{fig:confu_matrix_genrel_test}Confusion matrices obtained from applying the optimized ResNet50 algorithm to classify the HFF in TF image samples containing single-interferometric noise plus numerical simulated CCSN GW signals. All matrices use overall normalization (sum of all elements = $100\%$). To appropriately understand these results, it is crucial to analyze the input data, which is codified in the SNR distributions of the training and testing datasets (shown in Figs.~\ref{fig:wf_phen_distribution} and ~\ref{fig:wf_snr_gr_distribution}, respectively).}
\end{center}
\end{figure*}

Now we present results obtained from testing our optimized ResNet50 model, working with samples with noise plus numerical simulated CCSN GW signals. Table~\ref{tab:genrelwf_test} shows test accuracies and test losses which show that, independent of the noise, the performance of the model decreases as we increase the distance. Moreover, we find that the best performances are obtained using L1 and the worst performances using V1. Given the sensitivity of detectors, it is expected that the lowest performances occur with V1 noise data. However, it would have to be explained why at $5$ kpc and $10$ kpc, test accuracy drastically decreases by about $30\%$ to $40\%$, pointing out that predictions for one of the three classes are, for the most part, incorrect. Then, to expand the predictive results of Table~\ref{tab:genrelwf_test} (particularly to elucidate the ability of ResNet50 to recognize the HFF depending on the sample class), we present in Fig.~\ref{fig:confu_matrix_genrel_test} confusion matrices for all the tests performed. Here we adopt a convention in which the sum of all elements is $100\%$ (overall normalization); therefore, an ideal classification gives a diagonal matrix with $33\%$ elements. According to these confusion matrices, we have that, in most cases (L1 and H1 at $1$ kpc and $5$ kpc, and V1 at $1$ kpc), class 1 samples (Andresen) are the most difficult to recognize, with a classification percentage varying from $22.67\%$ to $32.00\%$. These are followed by class 2 samples (Morozova) and class 3 samples (Cerd\'a-Dur\'an), which are fully recognized in $33.33\%$ of the predictions per class.

\begin{table}[t]
\caption{\label{tab:genrelwf_test}Evaluation metrics for the tests performed with the optimized ResNet50 model and samples containing numerical simulated CCSN waveforms. Independent of the noise, performance decreases as we increase the distance from the CCSN. Best performances are achieved with L1 data and worst performances with V1 data.}
\captionsetup{width=1.0\textwidth}
\centering
\small
\begin{tabular}{|c|c|c|c|c|} 
 \hline
  \textbf{Noise data} & \textbf{Distance (kpc)} & \textbf{Test accuracy} & \textbf{Test loss} \\
 \hline\hline
 L1 & $1$  & $0.9867$ & $0.04617$ \\
    & $5$  & $0.9333$ & $0.4856$ \\
    & $10$ & $0.8533$ & $0.8829$ \\
 \hline
 H1 & $1$ & $0.9767$ & $0.08379$ \\
    & $5$  & $0.8933$ & $0.6116$ \\
    & $10$ & $0.8000$ & $1.413$ \\
 \hline
 V1 & $1$  & $0.9133$ & $0.5717$ \\
    & $5$  & $0.6433$ & $3.235$ \\
    & $10$ & $0.5933$ & $3.189$ \\
 \hline 
 \end{tabular}
\end{table}

At this point, it is useful to remember what was pointed out in works~\cite{tG19,mM21}, namely that any detection machine learning algorithm not only inputs image samples as single templates (as is the case with Matched Filter) but also the distribution of these samples, that is to say, all the information about the dataset. Because of this, we can take advantage of the fact that the SNR populations of our input datasets give us valuable statistical information to understand, to some extent, the predictions of the ResNet50 algorithm. Therefore, under this framework, we can observe that predictions of the ResNet50, including the mistakes, are consistent with SNR populations of Fig.~\ref{fig:wf_snr_gr_distribution}. In all SNR distributions with the Andresen model, being located in ranges of lower SNR than those of samples with the Morozova and the Cerd\'a-Dur\'an models, are spaced enough to have almost zero overlap with the SNR distributions of those other models.

On the other hand, tests performed with L1 and H1 noise data at $10$ kpc show a slightly different trend. Here, the class 2 samples (Morozova) are the most misclassified, followed by the class 1 samples (Andresen), with a percentage discrepancy of $2\%$ to $3.33\%$ (Cerd\'a-Dur\'an samples maintain the full correct classifications of $33.33\%$). If we again contrast these results with the SNR populations in Fig.~\ref{fig:wf_snr_gr_distribution}, they would seem counter-intuitive; the SNR distributions for the Andresen and Morozova models, at $10$ kpc injected into L1 and H1 data, are still spaced enough with zero overlapping. That is to say, in this situation, the SNR values with Andresen models slightly surpass $10$, and SNR values with Morozova models are not lower than $35$, approximately.

To explain the above predictive behavior, we need to consider not only the nature of the input test dataset, but also the nature of the training input dataset, that is, the dataset of image samples containing L1, H1, and V1 data noise plus phenomenological waveforms. From histogram shown in the bottom panel of Fig.~\ref{fig:wf_phen_distribution}, we see that the SNR distribution for class 2 samples begins approximately at SNR$=35$, with a few training samples that gradually increase as we have move to greater SNR values. This means that if we input class 2 test samples of SNR very close to the aforementioned critical value (which is the case, as Fig.~\ref{fig:wf_snr_gr_distribution} shows for histograms of Morozova waveforms at distance of $10$ kpc), it will be more difficult for the ResNet50 to correctly classify them because the learning was based on a reduced number of train exemplars. This difficulty is more about the insufficiency of learning samples (to correctly associate common characteristics present in training samples and test samples) than merely working with lower SNR values.

The above explanation becomes even clearer if we focus on the most extreme and interesting failure cases, namely the tests with V1 data at $5$ kpc and $10$ kpc. In these tests we have a curious result: most of the class 2 samples are predicted as class 1 samples: at $5$ kpc, $23.33\%$ of the class 2 test data were misclassified as class 1, and at $10$ kpc, $20.67\%$ of class 2 test data were misclassified as class 1. These results can be interpreted by two interrelated facts: Firstly, testing with V1 data at $5$ kpc and $10$ kpc implies that we are working with the lowest SNR values, as shown in the bottom panel of Fig.~\ref{fig:wf_snr_gr_distribution}, and this decreases the visibility of the HFF slope as we observed in samples of Fig.~\ref{fig:sample_images_genrel}. Under this setting, it is expected that in a significant number of class 1 and class 2 samples, the magnitude of background noise is dominant over the magnitude of the GW strain, and the ResNet50 algorithm has to make a decision about these, even if wrong. Then, this fact explains why most of the $33.33\%$ images of class 2 were wrongly predicted. But now the question about why these samples were misclassified as class 1 (and not as class 3) arises, and the answer brings us to the second fact, which has to do, again, with the nature of the input training set. Regarding the bottom panel of Fig.~\ref{fig:wf_phen_distribution}, we mentioned that the SNR distribution of the training class 1 samples is shifted to lower SNR values than those of the SNR distributions of class 2 and 3 samples. In fact, according to the histograms, we have that approx. at SNR$<25$ the ResNet50 algorithm was trained only with class 1 samples; even in the region $25<$SNR$<35$, the number of class 1 samples is considerably larger than the class 2 and class 3 samples (more than $300$ samples vs. a hundred samples combining class 2 and 3). Therefore, when we input class 2 image samples of SNR$<35$ in the test (which is the case working with V1 data at $5$ kpc and $10$ kpc, as it is shown in the bottom panel of Fig.~\ref{fig:wf_snr_gr_distribution}), the ResNet50 favors class 1 predictions because it was trained mostly with those samples in this low SNR regime. Like tests with L1 and H1 data at $10$ kpc, the training dataset
did not provide enough information for the ResNet50 to correctly recognize the HFF slope in class 2 samples in this SNR regime, and it takes a decision based on what it knows.

Results shown in confusion matrices suggest that, undoubtedly, in the low SNR regime, the viability of our methodology is affected. However, this is not due to a malfunction of the ResNet50 architecture itself; this, as a standard computer vision model, is still very powerful (to the point that in the low SNR regime, it knew how to make what was the best decision \textit{given the training dataset}). But rather, the quality of our training dataset in this regime is not the best to ensure an appropriate classification. Therefore, here we stress that taking into account the SNR distributions of input (training and test) datasets is truly crucial for interpreting the predictive behavior of the optimized ResNet50 algorithm. And, more important is that by generating a high-quality dataset in terms of their SNR values (i.e., broad distributions that properly inform the learning process in all SNR regimes), we will have better conditions to perform predictions. The implementation of a systematic methodology to improve input image samples in terms of their SNR values is a relevant topic for future work.\\

{\bf Statistical validation in the low SNR regime}\\

To rigorously validate our interpretation that the optimized ResNet50's performance is limited by the SNR distribution of the training data, we designed a controlled test with samples of pure noise and samples of noise plus extremely low-SNR numerical simulated CCSN GW signals. It is important to note that our analysis, by theoretical justification, assumes the HFF is the dominant emission in CCSN gravitational-wave signals and, subsequently, is visible in the vast majority of detected CCSN GW signals undergoing analyses in post-processing. Therefore, this can be seen as an edge input case considering a rare scenario. 

Our ResNet50 model was not trained with pure noise samples. Then, these can be treated as belonging to a ``ghost class'' to explain predictions if they are inputted. This analysis was designed in three steps:

\begin{itemize}
    \item[a.] Generate two datasets. The first contains $1,800$ TF samples of pure noise: 600 for each detector, L1, H1, and V1. The other contains $1,800$ TF samples of noise plus a numerical simulated waveform: again, $600$ for each detector, where $200$ samples contain the HFF of the Andresen (class 1) waveform, $200$ the HFF of the Morozova (class 2) waveform, and $200$ contain the HFF of the Cerd\'a-Dur\'an (class 3) waveform. All waveforms were located at $50$ kpc, i.e., in a very low SNR regime.
    
    \vspace{.15cm}
    \item[b.] Obtain the test predictions for the aforementioned datasets. Each $i$-th prediction is given by the triad $P_i\left(c| \left\{\theta \right\}\right)$ ($c=1,2,3$), depicting the probabilities (scores) that the $i$-th TF sample belongs to the class $1$, $2$, or $3$, conditioned by the already learned model parameters $\left\{ \theta \right\}$. These probabilities are outputted by the final softmax layer of the ResNet50 and satisfies $\sum_{c=1}^{c=3}P_i\left(c| \left\{\theta \right\}\right)=1$.

    \vspace{.15cm}
    \item[c.] For each triad of probabilistic scores, select the maximum, to build two 1D distributions of $1,800$ elements: $\mathcal{R}$ for the pure noise samples and $\mathcal{S}$ for the samples of noise plus waveforms at $50$ kpc. Finally, the goal is to compare the distributions by performing a standard Kolmogorov–Smirnov (KS) test.
\end{itemize}

The KS test is a nonparametric test that is aimed at quantifying the similarity between $\mathcal{R}$ and $\mathcal{S}$. The KS statistic is defined as:

\begin{equation}
D_{\mathcal{R},\mathcal{S}} = \sup_{x} |F_{\mathcal{R}}(x) - F_{\mathcal{S}}(x)| ~, \label{eq:D_RS}
\end{equation}
where $F_{\mathcal{R}}$ and $F_{\mathcal{S}}$ are the cumulative distribution functions of $\mathcal{R}$ and $\mathcal{S}$, respectively. In particular, the null hypothesis (i.e., $H_0$) postulates that two samples, extracted from $\mathcal{R}$ and $\mathcal{S}$, respectively, come from the same distribution or, in other words, $\mathcal{R}$ and $\mathcal{S}$ are indistinguishable. For this hypothesis test, a $p$-value $> 0.05$ avoids rejecting $H_0$, pointing out that $D_{\mathcal{R},\mathcal{S}} \approx 0$ because of Eq.~(\ref{eq:D_RS}).

\begin{figure}[htp]
\begin{center}
  \includegraphics[width=6.3cm]{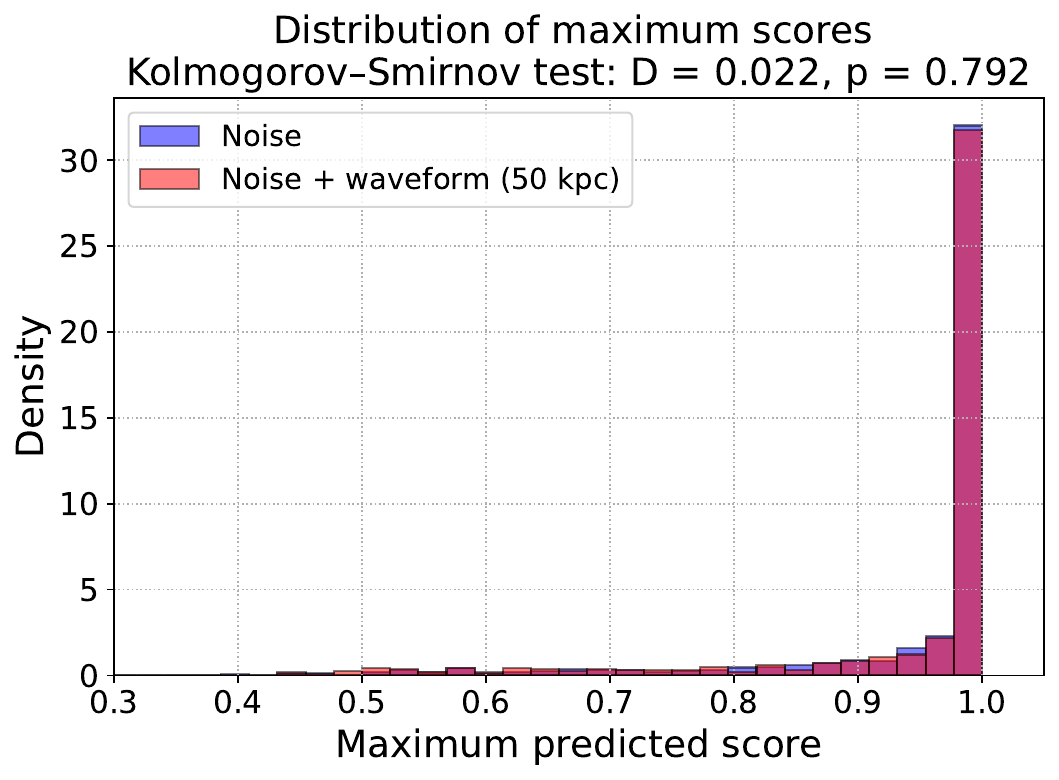}
  ~
  \includegraphics[width=6.3cm]{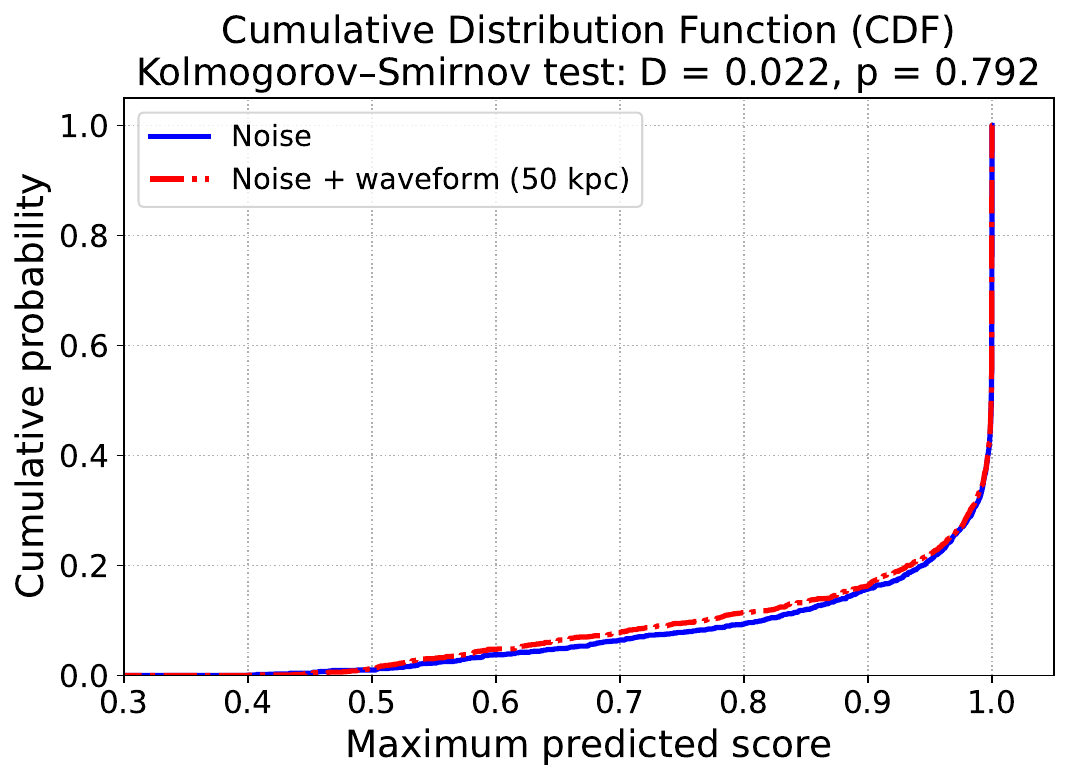}
  \caption{\label{fig:ks-test}Results of the Kolmogorov–Smirnov nonparametric test, inputting samples of noise only, and samples of noise with a numerical simulated CCSN waveform. We found that distributions of predictions $\mathcal{R}$ for noise only and $\mathcal{S}$ for noise plus waveforms at $50$ kpc (low SNR regime) are statistically equivalent or indistinguishable. Moreover, given that maximum scores tend to $1.0$, we have that there is a pathological overconfidence because of the training set (whose SNR values are shown in Fig.~\ref{fig:wf_phen_distribution}).}
\end{center}
\end{figure}

The results of the analysis are shown in Fig.~\ref{fig:ks-test}. In the upper panel, the distributions of maximum scores for the dataset of noise only and noise plus numerical simulated waveforms at $50$ kpc are shown. The trend is clear: maximum scores tend to $1$. Moreover, results of the KS test are excellent: $D=0.022$ shows there is an insignificant difference between distributions, and $p=0.792$ is well above the threshold of $0.05$. The resulting $p$-value can be interpreted as there is a $79.2\%$ probability of observing distributions $\mathcal{R}$ and $\mathcal{S}$ if they were identical. Then, we cannot reject the null hypothesis $H_0$. As a more rigorous visualization to support these results, the bottom panel of Fig.~\ref{fig:ks-test} shows the cumulative distribution functions, showing that $\mathcal{R}$ and $\mathcal{S}$, as defined in Eq.~(\ref{eq:D_RS}), are statistically indistinguishable.

\begin{figure}[ht]
\begin{center}
  \includegraphics[width=13.0cm]{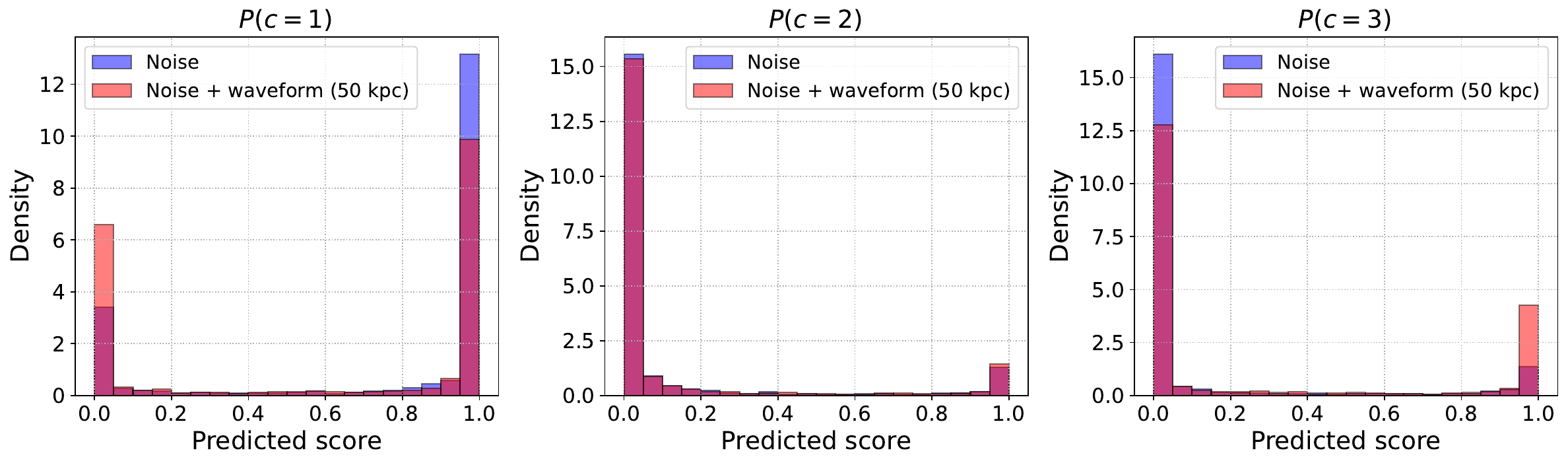}
  \caption{\label{fig:prob_max_classes}Distributions of maximum test scores, separated by class, for the Kolmogorov–Smirnov test. From the left panel, we observed that the pathological overconfidence of the ResNet50 comes from the predictive behavior of the class 1 samples. Moreover, class 2 and class 3 scores tend to $0$, as shown in the central and left panels. In summary, when working in the low SNR regime, the ResNet50 prioritizes the statistically plausible classes and refuses inconsistent options, given the training dataset with the distribution of SNR values shown in Fig.~\ref{fig:wf_phen_distribution}}
\end{center}
\end{figure}

Now, given the results, the natural question that arises is why maximum scores tend to $1.0$ and not to $0.3$ (take into account that a trend towards $0.3$ would depict total randomness). To explain this behavior, Fig.~\ref{fig:prob_max_classes} is illuminating. It shows probabilistic scores for the two datasets, separated by classes. Only for class 1, scores mostly tend to $1$, indicating that there is a pathological overconfidence, independent of the real presence of a CCSN GW signal. This result has to do with the fact that the training set, as shown in the bottom panel of Fig.~\ref{fig:wf_phen_distribution}, contains solely class 1 samples in the low SNR regime. Therefore, during tests with pure noise samples, the ResNet50 prioritizes the statistically plausible class. Moreover, we have that scores for classes 2 and 3 tend to $0$, indicating that the ResNet50 suppresses improbabilities given that, in the training set shown in Fig.~\ref{fig:wf_phen_distribution}, samples of these classes are nonexistent in the low SNR regime. In summary, the optimized ResNet50 does not 'guess' between classes in the low SNR regime but explicitly rejects inconsistent options that are implausible given the known training data. Anyway, taking an operational point of view, these results are good news. They mean that the ResNet50 takes the best decision given the training dataset, discarding problems such as overfitting of artifacts, data leakage, insufficient regularization in learning, etc. However, for future research, it would be important to implement alternative methodologies to generate training datasets such that, in the low SNR regime, all classes were present.

\subsubsection{Evaluation for ordinal classes}Given that target classes were defined by taken three ranges for the HFF slope that are adjacent to each other, it should be explored alternative metrics that take into account this adjacency and penalize more heavily misclassifications that are further. For this, we considered two additional test metrics. Firstly, the mean absolute error (MAE) which is computed as follows:
\begin{eqnarray}
    \mathrm{MAE} = \frac{1}{N_\mathrm{s}} \sum_{i=1}^{N_s} | y^i - \hat{y}^i| ~~,
\end{eqnarray}
where $N_\mathrm{s}$ is the number of samples, and $y^i$ and $\hat{y}^i$ the real and the predicted target classes, respectively. In addition, we drew on the Quadratic Weighted Kappa (QWK), which is generalization of the standard Cohen Kappa and it is computed as:
\begin{eqnarray}
    \mathrm{QWK} = 1 - \frac{D_\mathrm{o}}{D_\mathrm{e}} ~~,~~
    D_{o} = \sum_{j=1}^{C} \sum_{k=1}^{C} w_{jk} O_{jk} ~~,~~
    D_{e} = \sum_{j=1}^{C} \sum_{k=1}^{C} w_{jk} E_{jk} ~~,~~
\end{eqnarray}
where $C$ is the number of classes, $w_{jk}=\frac{(j-k)^2}{(C-1)^2}$ the quadratic weight, $O_{jk}$ the $jk$ element of the observed agreement matrix (that is to say, the usual confusion matrix), and $E_{jk} = \frac{\sum_{a=1}^C O_{ja} \sum_{b=1}^{C} O_{bk}}{N_{\mathrm{s}}}$ the $jk$ element of the expected agreement matrix.

Table~\ref{tab:genrelwf_test_ordinal} shows the results of evaluating with MAE and QWK metrics. Notice that here we have the same trend found based on results in Table~\ref{tab:genrelwf_test}, namely: best performance with L1 noise data, worst performance with V1 data, and about $30\%$ to $40\%$ decrease in test QWK when moving from $1$ kpc to $5$ kpc and $10$ kpc. These results allow us to conclude that even without assuming ordinal target labels, we can representatively summarize the performance of the ResNet50 with the current data.

\begin{table}[t]
\caption{\label{tab:genrelwf_test_ordinal}Evaluation of the ResNet50 classification algorithm taking into account the adjacency of the target labels. To penalize further missclassifications, we drew on the quadratic weighted kappa (QWK) and the mean absolute error (MAE).}
\captionsetup{width=1.0\textwidth}
\centering
\small
\begin{tabular}{|c|c|c|c|c|} 
 \hline
  \textbf{Noise data} & \textbf{Distance (kpc)} & \textbf{Test QWK} & \textbf{Test MAE} \\
 \hline\hline
 L1 & $1$  & $0.9899$ & $0.01333$ \\
    & $5$  & $0.8499$ & $0.1100$ \\
    & $10$ & $0.8019$ & $0.1900$ \\
 \hline
 H1 & $1$  & $0.9822$ & $0.02333$ \\
    & $5$  & $0.7844$ & $0.1633$ \\
    & $10$ & $0.7725$ & $0.2400$ \\
 \hline 
 V1 & $1$ & $0.8175$ & $0.1367$ \\
    & $5$  & $0.6608$ & $0.4100$ \\
    & $10$ & $0.5800$ & $0.4867$ \\
 \hline

 \end{tabular}
\end{table}

\subsubsection{Comparison with a post-hoc binned regression}\label{sec:binned_regression}

A pertinent question is whether a regression-based approach, followed by a binning of the continuous predictions, could yield equivalent or superior performance to our end-to-end classification model. To empirically address this question, we designed a post-hoc analysis. Taking advantage of our optimized ResNet50 model as a feature extractor, we consider its penultimate layer, namely a feature vector of 2048 components, as input to a multilayer perceptron (MLP) regressor of 5 layers:

\begin{enumerate}
    \item Fully connected layer with 512 neurons and ReLU activation,
    \item Dropout layer with rate 0.3,
    \item Fully connected layer with 256 neurons and ReLU activation function,
    \item Dropout layer with drouput rate 0.2
    \item Fully connected layer with 1 neuron and no activation.
\end{enumerate}

This MLP was trained to predict the continuous HFF slope value from the samples with noise plus phenomenological waveforms. Then, its predictions on the numerical simulated waveforms were binned into the three predefined slope classes (Class 1: Steep, Class 2: Moderate, Class 3: Low) for a direct comparison with our primary classification results. The performance of this regression-binning approach is summarized in Table~\ref{tab:regression-binning}, where its accuracy is contrasted with that of the original ResNet50 classifier across all detectors and distances.

\begin{table}[t]
\caption{\label{tab:regression-binning}Comparative test accuracy of the optimized ResNet50 classifier versus a post-hoc regression-then-binning approach using the ResNet50 as a feature extractor in addition to a multilayer perceptron of 5 layers. These tests were conducted by using samples containing real interferometric noise plus numerical simulated waveforms.}
\captionsetup{width=1.0\textwidth}
\centering
\small
\begin{tabular}{|c|c|c|c|} 
 \hline
  \textbf{Noise data} & \textbf{Distance (kpc)} & \textbf{\makecell{Classification\\test accuracy}} & \textbf{\makecell{Binned regression\\test accuracy}} \\
 \hline\hline
 L1 & $1$  & $0.9867$ & $0.5467$ \\
    & $5$  & $0.9333$ & $0.8500$ \\
    & $10$ & $0.8533$ & $0.7000$ \\
 \hline
 H1 & $1$  & $0.9767$ & $0.5833$ \\
    & $5$  & $0.8933$ & $0.8133$ \\
    & $10$ & $0.8000$ & $0.6767$ \\
 \hline 
 V1 & $1$  & $0.9133$ & $0.7567$ \\
    & $5$  & $0.6433$ & $0.6067$ \\
    & $10$ & $0.5933$ & $0.5933$ \\
 \hline
 \end{tabular}
\end{table}

\begin{figure*}[htb]
\begin{center}
  \includegraphics[width=3.9cm]{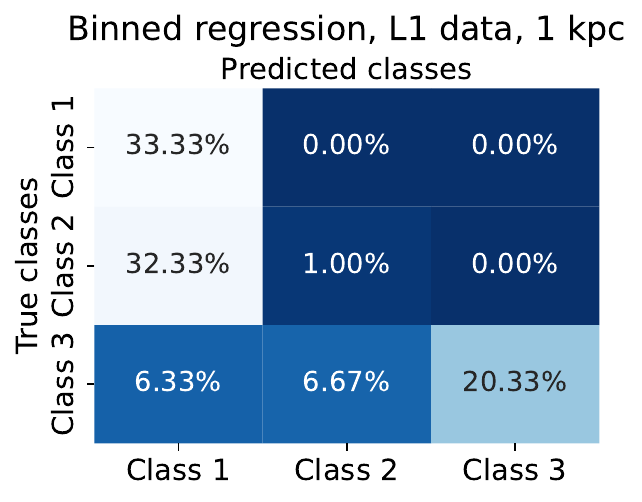} ~
  \includegraphics[width=3.9cm]{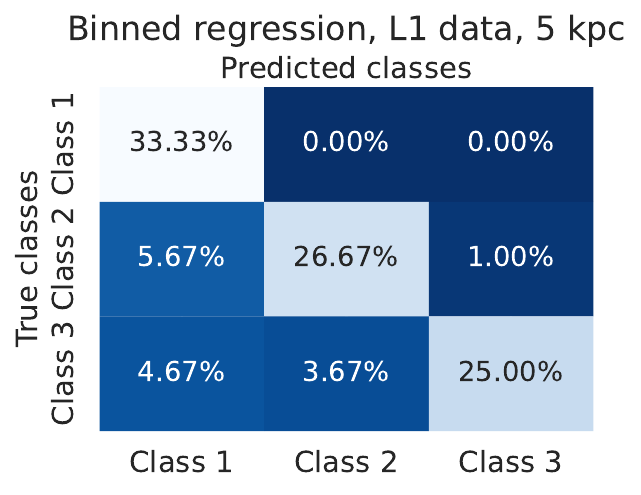} ~
  \includegraphics[width=3.9cm]{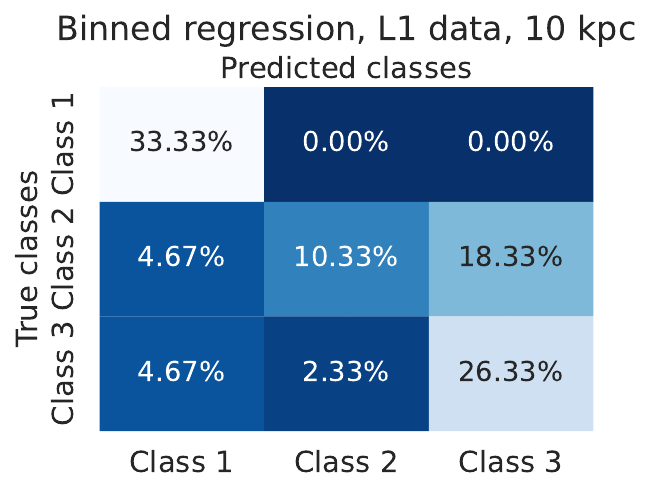}
  \vspace{.2cm} \\
  \includegraphics[width=3.9cm]{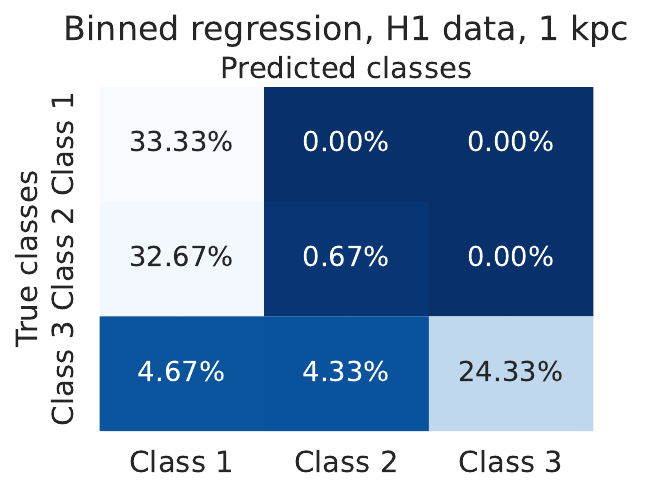} ~
  \includegraphics[width=3.9cm]{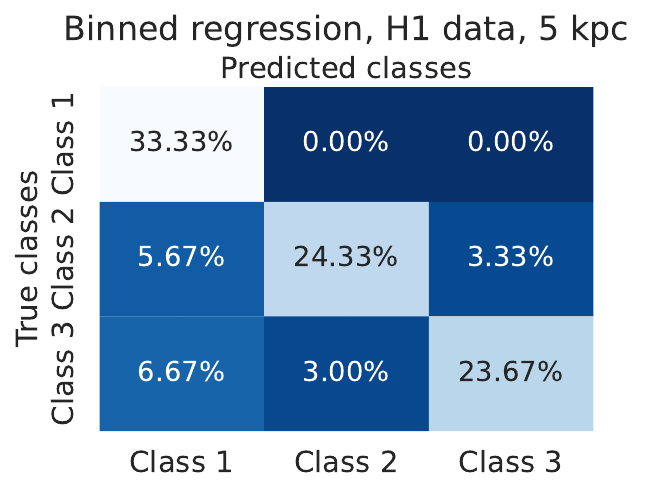} ~
  \includegraphics[width=3.9cm]{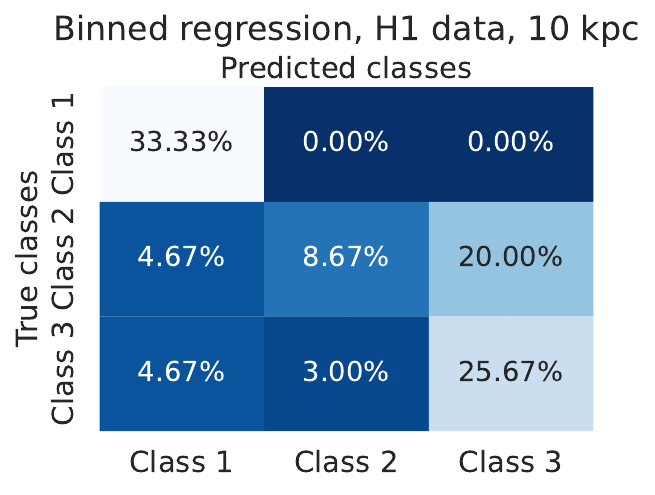}
  \vspace{.2cm} \\
  \includegraphics[width=3.9cm]{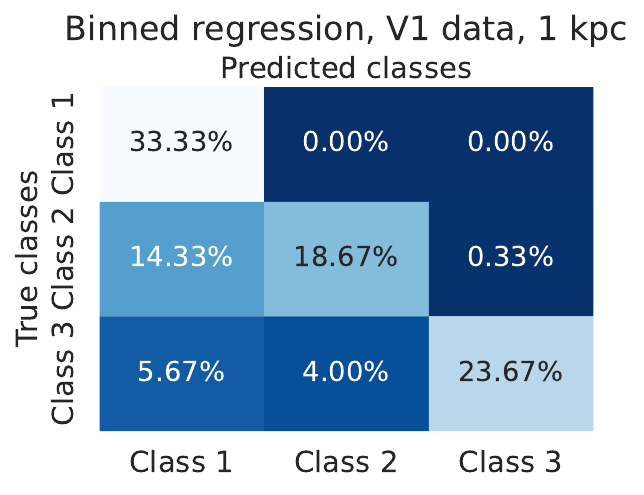} ~
  \includegraphics[width=3.9cm]{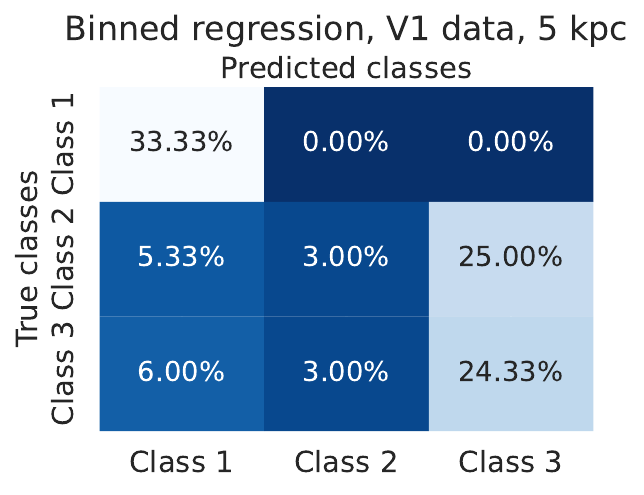} ~
  \includegraphics[width=3.9cm]{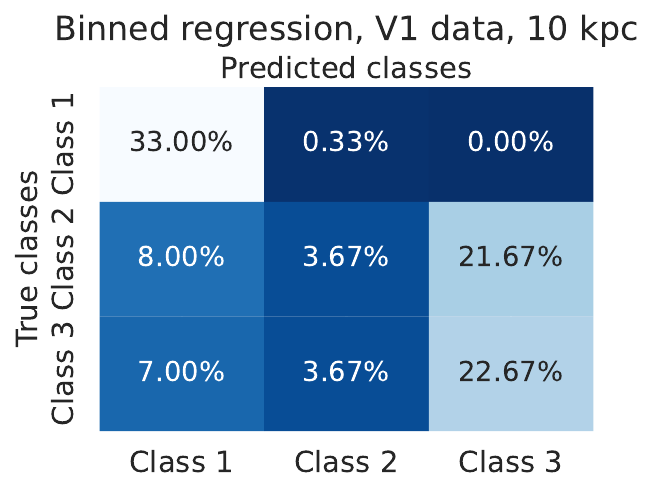}
  \vspace{.2cm}
  \caption{\label{fig:cfn_matrix_bin-regr}Confusion matrices for a binned regression, using overall normalization (sum of all elements = $100\%$). These results are consistent with the accuracy degradation of the regression model that was shown in Table~\ref{tab:regression-binning} and validate in a more detailed form that our original classification approach outperforms the regression-based approach.}
\end{center}
\end{figure*}

The results are unequivocal: the classification approach significantly outperforms the regression approach across the majority of the test scenarios. Misclassifications are particularly severe at the closest distance (1 kpc), where accuracy on H1 and L1 data (0.58 and 0.55, respectively) is little better than random guessing (0.33). Analysis of the confusion matrices from the binned regression reveals the source of this error. As it is shown in Fig.~\ref{fig:cfn_matrix_bin-regr}, at $1$ kpc with L1 and H1 data, the regression model exhibits a systematic failure to distinguish between Class 2 (Morozova, moderate HFF slope) and Class 1 (Andresen, steep HFF slope) TF image samples.

The aforementioned results suggest that the regression model failed to learn the precise morphological features that define the boundary between class 1 and class 2. Instead, it likely relied on coarser heuristics, such as duration or overall signal, which are not robust across different waveform types.

Notice that distribution of duration for training waveforms (upper right panel of Fig.~\ref{fig:wf_phen_distribution}) lacks values greater than $0.95$ s. Moreover, class 1 waveforms are the only ones in the short-duration region (from $0.3$ s to $0.6$ s), generating a strong correlation with high slopes in that region and also a bias towards these slope values. Then, as a result of this training setting, when long-duration waveforms (such as Morozova, with duration longer than $1$ s) are inputted, the rationale of the model breaks down, and probably the model addresses that uncertainty in consistency with the bias towards higher HFF slope values. This is a worse effect than that found in the optimized ResNet50 classifier, in which significant misclassifications of class 2 Morozova samples happened at larger distances ($5$ and $10$ kpc) with V1 data, as was discussed regarding confusion matrices of Fig.~\ref{fig:confu_matrix_genrel_test}.

Regarding the overall performance of the regression-binning approach, this exhibits a non-monotonic trend, improving from $1$ kpc to $5$ kpc and declining from $5$ kpc to $10$ kpc, in contrast to the stable degradation with distance of the original classifier. We posit this occurs because the MLP model, trained on idealized phenomenological waveforms, is confused by non-linear characteristics of high SNR signals at $1$ kpc. At $5$ kpc, noise could obscure these non-linearities, allowing the model to better approximate the slope. At $10$ kpc, the signal weakens beyond reliable estimation. At the end, this result highlights the key advantage of the original classification approach: it recognize robust features that generalize effectively across all SNR levels.

In conclusion, this comparative analysis robustly validates our choice of the classification methodology. It is fundamentally better optimized for the specific task of morphological categorization, learning highly robust decision boundaries that are superior to a simple binning of regression outputs. The significant performance gap shows that the classification approach is not merely a discretized form of regression, but a more powerful and appropriate framework for this particular problem.

\section{Conclusions}\label{sec:conclusions}

In this work, we present the first application of an optimized ResNet50 model, a state-of-the-art architecture in computer vision, to classify the HFF present in Morlet wavelet scalograms of CCSN GW signals embedded in real interferometric noise, which are input as time-frequency (TF) pixelated image samples. This task is a relevant step toward advancing the understanding of HFF morphology. We found 
the performance of the classification varies considerably depending on the noise realization, the nature of (and distance from) the emitting CCSN model, and the input datasets given the distribution-learning property of the ResNet50 algorithm.

We showed that our optimized ResNet50 algorithm can recognize the HFF by inputting TF image samples containing interferometric noise plus phenomenological and numerical simulated CCSN GW signals. When tested on samples containing numerical simulated CCSN GW signals, the resulting accuracies were consistent with the properties of the input datasets. From a practical standpoint, these properties are codified in the SNR distributions. Then, it was possible to interpret the predictions by analyzing these distributions.

We optimized the ResNet50 using TF image samples combining noise data from all LIGO-Virgo detectors (L1, H1, and V1) plus phenomenological waveforms. For the hyperparameter tuning, we used a \texttt{GridSearchCV} methodology and we found that the best hyperparameter combination is: batch size $b_\mathrm{s}=75$, number of training epochs $n_\mathrm{e}=40$, learning rate $\alpha=0.01$, momentum $m=0.6$, and optimizer RMSprop. Then, we trained the model with the best hyperparameter combination, and, after the testing, we reached excellent results: an accuracy of $0.994178$ and a (categorical cross-entropy) loss of $0.0303579$.

Then, we tested the ResNet50 algorithm using image samples of noise plus numerical simulated waveforms obtained from CCSN multidimensional simulations. These waveforms are more relevant for astrophysical purposes. Here, the performance varied depending on the distance from (and nature of the) CCSN emitting source and the input datasets. With L1 and H1 data at $1$ kpc and $5$ kpc, and with V1 data at $1$ kpc we reach excellent results. Image samples containing the Morozova model (class 2) and the Cerd\'a-Dur\'an model (class 3) are fully detected, $33\%$ of the test dataset per class; and an important percentage of samples containing the Andresen model (class 1) were recognized, from $22.67\%$ to $32.00\%$ of the test dataset. In this setting, the viability of our methodology is clearly shown.

On the other hand, working at $10$ kpc using L1 and H1 data and, more evidently, with V1 data at $5$ kpc and $10$ kpc, the situation is different. With L1 and H1 data, the ability to recognize image samples was adversely affected, even though results still show good statistical performance. With the Morozova model, $25\%$ and $21.67\%$ of the test dataset were detected, respectively; and with the Andresen model, $27\%$ and $25\%$ of the test dataset were recognized, respectively. However, when using V1 data and the furthest distances, it is unviable to recognize the HFF morphology of the Morozova model (and also for the Andresen model, as detailed below), because most of the samples containing this model were misclassified as Andresen samples: $22.33\%$ and $20.67\%$ of the test dataset were detected, for $5$ kpc and $10$ kpc, respectively. Samples containing the Cerd\'a-Dur\'an model, even in these last configurations, are robustly recognized by the ResNet50 algorithm: $33\%$ for $5$ kpc and $30.67\%$ for $10$ kpc, which is still excellent.

We interpret these undesirable misclassifications, especially in the worst cases of V1 data at $5$ kpc and $10$ kpc with Morozova and Andresen models. These misclassifications are attributable to the nature of the input training and test datasets. This nature is codified in their SNR distributions. By observing these distributions, we found that the ResNet50 algorithm was trained on a dataset with a significant imbalance in the low-SNR regime: it contained predominantly class 1 samples, with very few examples from class 2 or 3.

This, complemented by the fact that the noise realization dominates over the GW signals (i.e., the visibility of the HFF is lost) in a high percentage of Andresen and Morozova samples, explains why most samples with the Morozova model are mostly misclassified as samples with the Andresen model. Consequently, our methodology is unviable under these specific conditions.

Above explanation for the low SNR regime was quantitatively demonstrated by the Kolmogorov-Smirnov test, exhibiting a predictable bias towards class 1 due to the training set composition. This confirms that the primary limitation is not the model architecture, but the scope of the training data.

Our classification strategy to recognize the morphology of the HFF in relation to observational samples can be explained in terms of several factors, such as the nature of (and the distance from) the CCSN model and the noise realization. Therefore, it has implications for specific parameter estimation efforts. Predicting the specific dynamical behavior of the HFF (for instance, by estimating their slope by regression methodologies) involves a significant difficulty in interpreting wrong results; hence, a first broader, more interpretable, and actionable approach, to make fast desicions, is desirable.

Using only single-interferometer data in our study shows the strength of our methodology and results: we give minimal information to the ResNet50 to learn (only one TF image sample per detection), and still we reach excellent results. Nevertheless, our implementation can be readily adapted to a complete process of CCSN GW signals detection in a network of three detectors and HFF characterization. In that case, for instance, we could input simultaneous TF image samples along the three RGB channels without any modification in the architecture of the ResNet50, reducing the HFF characterization to a simple transfer learning procedure.

Our methodology has also limitations, particularly related to predictions in low SNR regime. In that sense, we also conclude that the viability of the ResNet50 classification model will vary depending on the particular configuration that we have: GW CCSNe model, noise realization, and input (training and testing) datasets given the learning-distribution property of the ResNet50 algorithm. In particular, how to improve the quality of the input datasets to reach good performance results with further distances and larger noise realizations is an aspect that deserves to be addressed in future works. We would expect that training datasets with wider SNR distributions (allowing for more exemplars in the low SNR regime), in conjunction with enhanced procedures to increase the sensitivity for detecting GW signals embedded in the noise (for instance, wavelet transformations with different mother wavelets, autoencoders to denoise data, etc.), would improve the ResNet50 performance. Other relevant aspects is the source orientation, and the inclusion of data from a network of detectors is also a matter that would need to be addressed in prospective research.

To finish, we comment on distinct advantages of this approach in comparison to other machine learning methodologies for analyses of CCSN GW signals. We know these signals are mainly stochastic and really expensive to generate with numerical simulations, implying that we have limited training data. The good news is that ResNet50 (like other deep residual networks) is one of the best deep learning approaches to enable real-time analysis with limited data. They facilitate transfer learning, which enhances generalization and adaptability and reduces computational costs. For more general analyses (for instance, multi-feature detections), we could start with a ResNet50 model pre-trained with less expensive and less realistic signals, then apply fast fine-tuning (i.e., making small adjustments to their hyperparameters) using more realistic multidimensional CCSN GW signals.

As mentioned, the core innovation of ResNet50 is the residual units, which prevent accuracy degradation when working with a lot of layers while achieving higher predictive performance. Moreover, these units solve the vanishing gradient problem because residual units bypass complex non-linear stacked layers and preserve, or minimally modify, the gradient flow during the backpropagation learning (indeed, when the gradient flow is modified in convolutional residual blocks, batch normalization stabilizes that modification). Take into account that residual networks are also usually used as feature extractors in advanced models for object detection and segmentation. Besides, they have very good hardware compatibility, thanks to optimized implementations for GPUs, TPUs, and devices with pruning/quantization. Then, working with these architectures in CCSN GW data analyses puts us in a good position for thinking about (and designing) more complex analyses.


\bmhead{Acknowledgements}

This work was supported by CONAHCYT Frontiers Science project No. 376127 {\it Sombras, lentes y ondas gravitatorias generadas por objetos compactos astrofísicos}. 
M.D.M. acknowledge the support of CONAHCYT postdoctoral project No. 3751010, CONAHCYT Frontiers Science project No. 376127, and PROSNI-UDG. C.M. wants to thank CONAHCYT and PROSNI-UDG. 
This research has used data or software obtained from the Gravitational Wave Open Science Center (gwosc.org), a service of the LIGO Scientific Collaboration, the Virgo Collaboration, and KAGRA. This material is based upon work supported by NSF's LIGO Laboratory which is a major facility fully funded by the National Science Foundation, as well as the Science and Technology Facilities Council (STFC) of the United Kingdom, the Max-Planck-Society (MPS), and the State of Niedersachsen/Germany for support of the construction of Advanced LIGO and construction and operation of the GEO600 detector. Additional support for Advanced LIGO was provided by the Australian Research Council. Virgo is funded, through the European Gravitational Observatory (EGO), by the French Centre National de Recherche Scientifique (CNRS), the Italian Istituto Nazionale di Fisica Nucleare (INFN) and the Dutch Nikhef, with contributions by institutions from Belgium, Germany, Greece, Hungary, Ireland, Japan, Monaco, Poland, Portugal, Spain. KAGRA is supported by Ministry of Education, Culture, Sports, Science and Technology (MEXT), Japan Society for the Promotion of Science (JSPS) in Japan; National Research Foundation (NRF) and Ministry of Science and ICT (MSIT) in Korea; Academia Sinica (AS) and the National Science and Technology Council (NSTC) in Taiwan.

\section*{Declarations}

\subsection*{Funding}
This research has been possible thanks to the Consejo Nacional de Humanidades, Ciencias y Tecnologías (CONAHCYT) Frontiers Science grant No. 376127 {\it Sombras, lentes y ondas gravitatorias generadas por objetos compactos astrofísicos}, CONAHCYT postdoctoral grant No. 3751010, and PROSNI of Universidad de Guadalajara.

\subsection*{Conflict of interest}
The authors declare no competing interests.

\subsection*{Code availablity}
Following the \href{https://www.unesco.org/en/open-science/about?hub=686}{UNESCO Recommendation of Open Science}, all the codes implemented in this work are available, under a GPL-3.0 license, on the GitHub repositories:

\begin{itemize}
    \item \href{https://github.com/ManuelDMorales/datagen-sngw-phen}{{\bf datagen-sngw-phen}} - dataset generator of strain samples with LIGO-Virgo noise plus phenomenological waveforms.
    \item \href{https://github.com/ManuelDMorales/datagen-sngw-genrel}{{\bf datagen-sngw-genrel}} - dataset generator of strain samples with LIGO-Virgo noise plus numerical simulated waveforms.
    \item \href{https://github.com/ManuelDMorales/resnet50-sngw-hff}{{\bf resnet50-sngw-hff}} - converter of samples from strain data to pixelized scalogram images, and ResNet50 to classify the high-frequency emission in CCSNe GW.
\end{itemize}

\subsection*{Author contributions}
Study conception and design: M. D. Morales and J. M. Antelis. Data collection: M. D. Morales. Software implementation: M. D. Morales. Analysis and interpretation of results: M. D. Morales. Draft manuscript preparation: M. D. Morales. Funding acquisition: M. D. Morales and C. Moreno. Supervision: J. M. Antelis and C. Moreno. Review: M. D. Morales, J. M. Antelis and C. Moreno. All authors have read and agreed to the published version of the manuscript.






\bibliography{sn-bibliography}

@article{bA16,
      author        = {B.P. Abbott and R. Abbott and T.D. Abbott and M.R. Abernathy and F. Acernese and K. Ackley and others},
      title         = {Observation of {G}ravitational {W}aves from a {B}inary {B}lack {H}ole {M}ergers},
      journal       = {Phys. Rev. Lett.},
      volume        = {116},
      pages         = {061102},
      doi           = {},
      year          = {2016}
}

@article{bA17,
      author        = {B.P. Abbott and R. Abbott and T.D. Abbott and F. Acernese and K. Ackley and C. Adams and others},
      title         = {{GW}170817: {O}bservation of {G}ravitational {W}aves from a {B}inary {N}eutron {S}tar {I}nspiral},
      journal       = {Phys. Rev. Lett.},
      volume        = {119},
      pages         = {161101},
      doi           = {},
      year          = {2017}
}

@article{bA17_2,
      author        = {B.P. Abbott and R. Abbott and T.D. Abbott and F. Acernese and K. Ackley and C. Adams and others},
      title         = {Gravitational {W}aves and {G}amma-{R}ays from a {B}inary {N}eutron {S}tar {M}erger: {GW}170817 and {GRB 170817A}},
      journal       = {ApJL},
      volume        = {848},
      pages         = {2},
      doi           = {},
      year          = {2017}
}

@article{mS21,
      author        = {M.J. Szczepa\'nczyk and J.M. Antelis and M. Benjamin and M. Cavagli\'a and D. Gondek-Rosi\'nska and T. Hansen and others},
      title         = {Detecting and reconstructing gravitational waves from the next galactic core-collapse supernova in the advanced detector era},
      journal       = {Phys. Rev. D},
      volume        = {104},
      pages         = {102002},
      doi           = {},
      year          = {2021}
}

@article{hC14,
      author        = {H.-Y. Chen and D. E. Holz},
      title         = {The {L}oudest {G}ravitational {W}ave {E}vents},
      journal       = {arXiv:1409.0522},
      volume        = {},
      pages         = {},
      doi           = {},
      year          = {2014}
}

@article{hB90,
      author        = {H.A. Bethe},
      title         = {Supernova mechanisms},
      journal       = {Rev. Mod. Phys.},
      volume        = {62},
      pages         = {801},
      doi           = {},
      year          = {1990}
}

@article{hJ12,
      author        = {H.-T. Janka},
      title         = {Explosion {M}echanisms of {C}ore-{C}ollapse {S}upernovae},
      journal       = {Annu. Rev. Nucl. Part. Sci.},
      volume        = {62},
      pages         = {407--451},
      doi           = {},
      year          = {2012}
}

@article{sS08,
      author        = {S. Scheidegger and T. Fischer and S.C. Whitehouse and M. Liebendörfer},
      title         = {Gravitational waves from 3D MHD core collapse simulations},
      journal       = {A\&A},
      volume        = {490},
      pages         = {231--241},
      doi           = {},
      year          = {2008}
}

@article{bM13,
      author        = {B. Müller and H.-T. Janka and A. Marek},
      title         = {A {N}ew {M}ulti-{D}imensional {G}eneral {R}elativistic {N}eutrino {H}ydrodynamics {C}ode of {C}ore-{C}ollapse {S}upernovae {III}. {G}ravitational {W}ave {S}ignals from {S}upernova {E}xplosion {M}odels},
      journal       = {ApJ},
      volume        = {766},
      pages         = {43},
      doi           = {},
      year          = {2013}
}

@article{tK16,
      author        = {T. Kuroda and K. Kotake and T. Takiwaki},
      title         = {A {N}ew {G}ravitational-{W}ave {S}ignature from {S}tanding {A}ccretion {S}hock {I}nstabilities in {S}upernovae},
      journal       = {ApJL},
      volume        = {829},
      pages         = {L14},
      doi           = {},
      year          = {2016}
}

@article{dV23,
      author        = {D. Vartanyan and A. Burrows and T. Wang and M.S.B. Coleman and C.J. White},
      title         = {Gravitational-wave signature of core-collapse supernovae},
      journal       = {Phys. Rev. D},
      volume        = {107},
      pages         = {103015},
      doi           = {},
      year          = {2023}
}

@article{iH09,
      author        = {I.S. Heng},
      title         = {Rotating stellar core-collapse waveform decomposition: a principal component analysis approach},
      journal       = {Class. Quantum Grav.},
      volume        = {26},
      pages         = {105005},
      doi           = {},
      year          = {2009}
}

@article{sS19,
      author        = {S. Suvorova1 and J. Powell and A. Melatos},
      title         = {Reconstructing gravitational wave core-collapse supernova signals with dynamic time warping},
      journal       = {Phys. Rev. D },
      volume        = {99},
      pages         = {123012},
      doi           = {},
      year          = {2019}
}

@article{sS24,
      author        = {S. Sasaoka and Y. Sakai and D. Dominguez and K. Somiya and K. Saka and K. Oohara and others},
      title         = {Parameter estimation of protoneutron stars from gravitational wave signals using the {H}ilbert-{H}uang transform},
      journal       = {Phys. Rev. D},
      volume        = {110},
      pages         = {104020},
      doi           = {},
      year          = {2024}
}

@article{yY24,
      author        = {Y. Yuan and X.-L. Fan and H.-J. Lü and Y.-Y. Sun and K. Lin},
      title         = {Waveform reconstruction of core-collapse supernova gravitational waves with ensemble empirical mode decomposition},
      journal       = {MNRAS},
      volume        = {529(4)},
      pages         = {3235--3243},
      doi           = {},
      year          = {2024}
}

@article{kK06,
      author        = {K. Kotake and K. Sato and K. Takahashi},
      title         = {Explosion mechanism, neutrino burst and gravitational wave in core-collapse supernovae},
      journal       = {Rep. Prog. Phys.},
      volume        = {69},
      pages         = {971},
      doi           = {},
      year          = {2006}
}

@article{zL23,
      author        = {Z. Lin and A. Rijal and C. Lunardini and M.D. Morales and M. Zanolin},
      title         = {Characterizing a supernova’s standing accretion shock instability with neutrinos and gravitational waves},
      journal       = {Phys. Rev. D},
      volume        = {107},
      pages         = {083017},
      doi           = {},
      year          = {2023}
}

@article{tB23,
      author        = {T. Bruel and M.-A. Bizouard and M. Obergaulinger and P. Maturana-Russel and A. Torres-Forné and P. Cerdá-Durán and others},
      title         = {Inference of protoneutron star properties in core-collapse supernovae from a gravitational-wave detector network},
      journal       = {Phys. Rev. D},
      volume        = {107},
      pages         = {083029},
      doi           = {},
      year          = {2023}
}

@article{aM24,
      author        = {A. Mezzacappa and M. Zanolin},
      title         = {Gravitational {W}aves from {N}eutrino-{D}riven {C}ore {C}ollapse {S}upernovae: {P}redictions, {D}etection, and {P}arameter {E}stimation},
      journal       = {arXiv:2401.11635},
      volume        = {},
      pages         = {},
      doi           = {},
      year          = {2024}
}

@article{pA18,
      author        = {P. Astone and P. Cerd\'a-Dur\'an and I. Di Palma and M. Drago and F. Muciaccia and C. Palomba and F. Ricci},
      title         = {New method to observe gravitational waves emitted by core collapse supernovae},
      journal       = {Phys. Rev. D},
      volume        = {98},
      pages         = {122002},
      doi           = {},
      year          = {2018}
}

@article{pC25,
      author        = {P. Cerd\'a-Dur\'an and M. L\'opez and A. Favali and I. Di Palma and M. Drago and F. Ricci},
      title         = {Phenomenological gravitational waveforms for core-collapse supernovae},
      journal       = {arXiv:2501.11401},
      volume        = {},
      pages         = {},
      doi           = {},
      year          = {2025}
}

@article{cT23,
      author        = {C\'esar Eduardo Tiznado Alonso and Claudia Moreno and Manuel D. Morales and Mauricio Antelis},
      title         = {Detector de ondas gravitacionales fenomenol\'ogicas de supernovas basado en aprendizaje supervisado},
      journal       = {Research in Computing Science},
      volume        = {152(6)},
      pages         = {259--271},
      doi           = {},
      year          = {2023}
}

@article{aC23,
      author        = {A. Casallas-Lagos and J.M. Antelis and C. Moreno and M. Zanolin and A. Mezzacappa and M. J. Szczepa\'nczyk},
      title         = {Characterizing the temporal evolution of the high-frequency gravitational wave emission for a core collapse supernova with laser interferometric data: {A} neural network approach},
      journal       = {Phys. Rev. D},
      volume        = {108},
      pages         = {084027},
      doi           = {},
      year          = {2023}
}

@article{ac24,
      author        = {R.D. Murphy and A. Casallas-Lagos and A. Mezzacappa and M. Zanolin and R.E. Landfield and E.J. Lentz and others},
      title         = {Dependence of the reconstructed core-collapse supernova gravitational wave high-frequency feature on the nuclear equation of state in real interferometric data},
      journal       = {Phys. Rev. D},
      volume        = {110},
      pages         = {083006},
      doi           = {},
      year          = {2024}
}

@article{vM18,
      author        = {V. Morozova and D. Radice and A. Burrows and D. Vartanyan},
      title         = {The gravitational wave signal from core-collapse super-novae},
      journal       = {Astrophys. J.},
      volume        = {861},
      pages         = {10},
      doi           = {},
      year          = {2018}
}

@article{hA19,
      author        = {H. Andresen and E. Müller and H.-Th. Janka and A. Summa and K. Gill and M. Zanolin},
      title         = {Gravitational waves from 3D core-collapse supernova models: {T}he impact of moderate progenitor rotation},
      journal       = {MNRAS},
      volume        = {486},
      pages         = {2238–2253},
      doi           = {},
      year          = {2019}
}

@article{pC13,
      author        = {P. Cerd\'a-Dur\'an and N. DeBrye and M.A. Aloy and J.A. Font and M. Obergaulinger},
      title         = {Gravitational wave signatures in black hole forming core collapse},
      journal       = {Astrophys. J. Lett.},
      volume        = {779},
      pages         = {L18},
      doi           = {},
      year          = {2013}
}

@article{rK87,
      author        = {R. Kronland-Martinet and J. Morlet and A. Grossmann},
      title         = {Analysis of sound patterns through wavelet transforms},
      journal       = {Int. J. Patt. Recogn. Art. Intell.},
      volume        = {1},
      pages         = {273–302},
      doi           = {},
      year          = {1987}
}

@article{mM21,
      author        = {M.D. Morales and J.M. Antelis and C. Moreno and A.I. Nesterov},
      title         = {Deep {L}earning for {G}ravitational-{W}ave {D}ata {A}nalysis: {A} {R}esampling {W}hite-{B}ox {A}pproach },
      journal       = {Sensors},
      volume        = {21(9)},
      pages         = {3174},
      doi           = {},
      year          = {2021}
}

@article{kH16,
      author        = {K. He and X. Zhang and S. Ren and J. Sun},
      title         = {Deep {R}esidual {L}earning for {I}mage {R}ecognition},
      journal       = {IEEE Conference on Computer Vision and Pattern Recognition (CVPR)},
      volume        = {},
      pages         = {770-778},
      doi           = {},
      year          = {2016}
}

@article{yL98,
      author        = {Y. Lecun and L. Bottou and Y. Bengio and P. Haffner},
      title         = {Gradient-based learning applied to document recognition},
      journal       = {Proc. IEEE},
      volume        = {86(11)},
      pages         = {2278-2324},
      doi           = {},
      year          = {1998}
}

@article{aK17,
      author        = {A. Krizhevsky and I. Sutskever and G.E. Hinton},
      title         = {ImageNet classification with deep convolutional neural networks},
      journal       = {CACM},
      volume        = {60(6)},
      pages         = {84-90},
      doi           = {},
      year          = {2017}
}

@article{kS15,
      author        = {K. Simonyan and A. Zisserman},
      title         = {Very {D}eep {C}onvolutional {N}etworks for {L}arge-{S}cale {I}mage {R}ecognition},
      journal       = {The 3rd {I}nternational {C}onference on {L}earning {R}epresentations (ICLR2015)},
      volume        = {},
      pages         = {},
      doi           = {},
      year          = {2015}
}

@article{sI15,
      author        = {S. Ioffe and C. Szegedy},
      title         = {Batch normalization: accelerating deep network training by reducing internal covariate shift},
      journal       = {Proceedings of the 32nd International Conference on International Conference on Machine Learning}, 
      volume        = {37},
      pages         = {448-456},
      doi           = {},
      year          = {2015}
}

@article{vN10,
      author        = {V. Nair and G.E. Hinton},
      title         = {Rectified {L}inear {U}nits {I}mprove {R}estricted {B}oltzmann {M}achines},
      journal       = {Proceedings of the 27th International Conference on Machine Learning}, 
      volume        = {},
      pages         = {807-814},
      doi           = {},
      year          = {2010}
}

@article{dK14,
      author        = {D.P. Kingma and J. Ba},
      title         = {Adam: {A} {M}ethod for {S}tochastic {O}ptimization},
      journal       = {Proceedings of the 3rd International Conference on Learning Representations (ICLR)}, 
      volume        = {},
      pages         = {},
      doi           = {},
      year          = {2014}
}

@article{pN23,
      author        = {P. Nousi and A. E. Koloniari and N. Passalis and P. Iosif and N. Stergioulas and A. Tefas},
      title         = {Deep residual networks for gravitational wave detection},
      journal       = {Phys. Rev. D},
      volume        = {108},
      pages         = {024022},
      doi           = {},
      year          = {2023}
}

@article{tS24,
      author        = {Tian-Yang Sun and Chun-Yu Xiong and Shang-Jie Jin and Yu-Xin Wang and  Jing-Fei Zhang and Xin Zhang},
      title         = {Efficient parameter inference for gravitational wave signals in the presence of transient noises using temporal and time-spectral fusion normalizing flow},
      journal       = {Chinese Phys. C},
      volume        = {48(4)},
      pages         = {045108},
      doi           = {},
      year          = {2024}
}

@article{sN24,
      author        = {S. Nunes and G. Escrig and O.G. Freitas and J.A. Font and T. Fernandes and A. Onofre and A. Torres-Forné},
      title         = {Deep-learning classification and parameter inference of rotational core-collapse supernovae},
      journal       = {Phys. Rev. D},
      volume        = {110},
      pages         = {064037},
      doi           = {},
      year          = {2024}
}

@article{mS22,
      author        = {M. Shafiq and Z. Gu},
      title         = {Deep {R}esidual {L}earning for {I}mage {R}ecognition: {A} {S}urvey},
      journal       = {Appl. Sci.},
      volume        = {12(18)},
      pages         = {8972},
      doi           = {},
      year          = {2022}
}

@article{xG10,
      author        = {X. Glorot and Y. Bengio},
      title         = {Understanding the difficulty of training deep feedforward neural networks},
      journal       = {13th Int. Conf. Mach. Learn. ICML 2010},
      volume        = {9},
      pages         = {249–256},
      doi           = {},
      year          = {2010}
}

@article{rA23,
      author        = {R. Abdulkadirov and P. Lyakhov and N. Nagornov},
      title         = {Survey of {O}ptimization {A}lgorithms in {M}odern {N}eural {N}etworks},
      journal       = {Mathematics},
      volume        = {11(11)},
      pages         = {2466},
      doi           = {},
      year          = {2023}
}

@article{tG19,
      author        = {T.D. Gebhard and N. Kilbertus and I. Harry and B. Schölkopf},
      title         = {Convolutional neural networks: {A} magic bullet for gravitational-wave detection{?}},
      journal       = {Phys. Rev. D},
      volume        = {100},
      pages         = {063015},
      doi           = {},
      year          = {2019}
}

@book{Th09,
  title     = {The {E}lements of {S}tatistical {L}earning: {D}ata {M}ining, {I}nference, and {P}rediction, 2nd ed.},
  author    = {T. Hastie and R. Tibshirani and J. Friedman},
  year      = {2009},
  publisher = {Springer},
  address   = {Berlin/Heidelberg}
}

@incollection{eA22,
   author    = {Abdikamalov, E. and Pagliaroli, G. and Radice, D.},
   title     = {Gravitational {W}aves from {C}ore-{C}ollapse {S}upernovae},
   editor    = {Bambi, C. and Katsanevas, S. and Kokkotas, K.D.},
   booktitle = {Handbook of {G}ravitational {W}ave {A}stronomy},
   year      = {2022},
}

@misc{pyCBC,
   author    = {A. Nitz and I. Harry and D. Brown and C.M. Biwer and J. Willis and T.D. Canton and others},
   title     = {gwastro/pycbc: v2.4.0 release of {P}y{CBC}},
   doi = {10.5281/zenodo.10013996},
   year      = {2023},
}

@misc{TensorFlow,
   author    = {T. GoogleBrain},
   title     = {TensorFlow},
   doi = {10.5281/zenodo.10126399},
   year      = {2023},
}

@misc{Keras,
  title={Keras},
  author={F. Chollet and others},
  year={2015},
  url={https://keras.io},
}

@misc{Pillow,
  title={Pillow (PIL Fork)},
  author={Alex Clark},
  year={2015},
  url={https://python-pillow.org/}
}

\end{document}